\def\square{\hbox{\vrule\vbox{\hrule\phantom{o}\hrule}\vrule}}

\documentclass[12pt]{article}
\usepackage{graphicx}

\newcommand{\drm}{{\rm d}}

\newcommand{\epsilonbf}{\mbox{\boldmath $\epsilon$}}
\newcommand{\Gammabf}{\mbox{\boldmath $\Gamma$}}
\newcommand{\nc}{\newcommand}

\nc{\varPhi}{\Phi} \nc{\bi}{\bibitem} \nc{\til}{\tilde}
\nc{\Lc}{{\cal L}} \nc{\om}{\omega} \nc{\bb}{\begin{equation}}
\nc{\ee}{\end{equation}} \nc{\vecna}{\mbox{\boldmath $\nabla$}}
\nc{\ga}{\gamma} \nc{\erm}{{\rm e}} \nc{\ib}{{\bf i}}
\nc{\sig}{\sigma} \nc{\longra}{\longrightarrow} \nc{\al}{\alpha}
\nc{\vare}{\varepsilon} \nc{\C}{I\!\!\!C} \nc{\hb}{\hbar}
\nc{\1}{1\!\!1} \nc{\lan}{\langle} \nc{\ran}{\rangle}
\nc{\Om}{\Omega} \nc{\R}{I\!R} \nc{\wide}{\widehat}
\nc{\Psit}{\widetilde{\Psi}} \nc{\Rt}{\widetilde{R}}
\nc{\psib}{\overline{\psi}} \nc{\psit}{\widetilde{\psi}} \nc{\bt}{\beta}
\nc{\und}{\underline} \nc{\sta}{\stackrel} \nc{\cent}{\centerline}
\nc{\vs}{\vspace*} \nc{\zb}{\overline{\psi}} \nc{\doz}{\dot{\psi}}
\nc{\ddoz}{\ddot{\psi}} \nc{\dozb}{\dot{\overline{\psi}}}
\nc{\ddozb}{\ddot{\overline{\psi}}} \nc{\dox}{\dot{x}} \nc{\dopi}{\dot{\pi}}
\nc{\dopsi}{\dot{\psi}} \nc{\Vbf}{\mbox{\boldmath $V$}}
\nc{\sbf}{\mbox{\boldmath $s$}} \nc{\Sigbf}{\mbox{\boldmath $\Sigma$}}
\nc{\ombf}{\mbox{\boldmath $\omega$}} \nc{\gabf}{\mbox{\boldmath $\gamma$}}
\nc{\omp}{\omega_{{\rm \psi}}}
\nc{\vbf}{\mbox{\boldmath $v$}} \nc{\imp}{\mbox{\boldmath $p$}}
\nc{\vo}{{\wide {\vbf}}} \nc{\po}{{\wide {p}}} \nc{\Po}{{\wide {P}}}
\nc{\impo}{\wide {\imp}} \nc{\ii}{\`{\i \ }} \nc{\ug}{\; = \;}

\begin{document}

\baselineskip 0.7cm

\cent{\bf Introduction of a Quantum of Time (``chronon")}

\cent{\bf and its Consequences for Quantum Mechanics$^{(\dagger)}$}

\footnotetext{$^{(\dagger)}$ Work partially supported by CAPES, CNPq, FAPESP
and by INFN, MURST, CNR.}

\vs{0.5cm}

\begin{center}
{Ruy H. A. FARIAS }

\vs{0.1 cm}

{\em LNLS - Laborat\'orio Nacional de Luz S\'{\i}ncrotron, Campinas, S.P.,
Brazil}

\vs{0.2cm}

and

\vs{0.2cm}

{Erasmo RECAMI }

\vs{0.1cm}

{\em Facolt\`a di Ingegneria, Universit\`a statale di Bergamo, Italy, {\rm and}}

{\em INFN--Sezione di Milano, Milan, Italy; {\rm recami@mi.infn.it}}

\end{center}

\vs{1.0cm} \baselineskip 0.6cm

{\bf Abstract} --- \ We discuss the consequences of the introduction of a
quantum of time $\tau_0$
in the formalism of non-relativistic quantum mechanics, by referring
ourselves, in particular, to the theory of the {\em chronon} as proposed
by P.Caldirola. Such an interesting ``finite difference" theory, forwards
---at the classical level--- a solution for the motion of a particle endowed
with a non-negligible charge in an external electromagnetic field, overcoming
all the known difficulties met by Abraham--Lorentz's and Dirac's approaches
(and even allowing a clear answer to the question whether a free falling
charged particle does or does not emit radiation), and ---at the quantum
level--- yields a remarkable mass spectrum for leptons.

After having briefly reviewed Caldirola's approach, our first aim is to
work out, discuss, and compare to one another the {\em new}
representations of Quantum Mechanics (QM) resulting from it, in the
Schr\"odinger, Heisenberg and density--operator (Liouville--von Neumann)
pictures, respectively.

Moreover, for each representation, three ({\em retarded, symmetric} and
{\em advanced}) {\em formulations} are possible, which refer either to
times $t$ and $t-\tau_0$, or to times $t-\tau_0/2$ and $t+\tau_0/2$, or to
times $t$ and $t+\tau_0$, respectively. \ It is interesting to notice that,
when the chronon tends to zero, the ordinary QM is obtained as the limiting
case of the ``symmetric" formulation only; while the ``retarded" one does
naturally appear to describe QM with friction, i.e., to describe
{\em dissipative} quantum systems (like a particle moving in an absorbing
medium). In this sense, {\em discretized} QM is much richer than the
ordinary one.

We also obtain the (retarded) finite--difference Schr\"odinger equation
within the Feynman path integral approach, and study some of its relevant
solutions. We then derive the time--evolution operators of this discrete
theory, and use them to get the finite--difference Heisenberg equations.

When discussing the mutual compatibility of the various
pictures listed above, we find that they {\em can} be written down
in a form such that they result to be equivalent (as it happens in
the ``continuous" case of ordinary QM), even if the Heisenberg
picture cannot be derived by ``discretizing" directly the ordinary
Heisenberg representation.

Afterwards, some typical applications and examples are studied, as the free
particle, the harmonic oscillator and the hydrogen atom; and various cases
are pointed out, for which the predictions of discrete QM differ from those
expected from ``continuous" QM.

At last, the density matrix formalism is applied to the solution of the
{\em measurement problem} in QM, with very interesting results, as for
instance a natural explication of ``decoherence", which reveal the power
of dicretized (in particular, {\em retarded\/}) QM.

\newpage

\section{Introduction}
\mbox{}

The idea of a discrete temporal evolution is not a new
one and, as almost all the physical ideas, has from time to
time been recovered from oblivion.\footnote{Historical aspects
related to the introduction of a fundamental interval of time in Physics
can be found in F.Casagrande, ref.\cite{CAS}.} \ For instance,
in classical Greece this idea came to light as part of the
atomistic thought. \ In the Middle Age, belief in the discontinuous
character of time was at the basis of the ``theistic atomism''
held by the Arabic thinkers of the Kal\={a}m\cite{JAMM}. \ In Europe,
discussions about the discreteness of space and time can
be found for example in the writings of Isidore of Sevilla,
Nicolaus Boneti and Henry of Harclay, who discussed the
nature of continuum. \ In more recent times, the idea of the
existence of a fundamental interval of time was
rejected by Leibniz, since it was incompatible with his
rationalistic philosophy. Within modern physics, however,
Planck's famous work on black body radiation inspired a new
view of the subject. In fact, the introduction of the quanta
opened a wide range of new scientific possibilities
regarding the way the physical world can be conceived.
Including considerations, like those in the present paper, on the
``discretization" of time within the framework of Quantum Mechanics (QM).

\indent
       In the early years of our century, Mach regarded the concept of
continuum to be a consequence of our physiological limitations:
$<<$...le temps et l'espace ne repr\'{e}sentent, au point de vue
physiologique, qu'un continue apparent, qu'ils se composent tr\`{e}s
vraisemblablement d'elements discontinus, mais qu'on ne peut distinguer
nettement les uns des autres.$>>$\cite{ARZE} \ Also Poincar\'{e} took
into consideration the possible existence of what he called an ``atom of
time": the minimum amount of time which allows to distinguish between two
states of a system.\cite{POIN1} \ Finally, in the twenties,
J.J.Thomson\cite{THOM} suggested the electric force to act in a discontinuous
way, producing finite increments of momentum separated by finite intervals
of time.  Such a seminal work has ever since inspired a series of papers on
the existence of a fundamental interval of time, named {\em chronon\/}; \ even
if the repercussions of all that work was small, at that time. \  A  further
seminal article was the one by Ambarzumian and Ivanenko\cite{AMBIV}, appeared
in 1930, which assumed space--time as being discrete and also stimulated a
large number of subsequent papers.

        It is important to stress that, in principle, time discretization
can be introduced in two distinct (and completely different) ways:\\

\noindent
(1) by attributing to time a discrete structure, i.e., by regarding time
not as a continuum, but as a one-dimensional ``lattice";\\

\noindent
(2) by considering time as a continuum, in
which events can take place (discontinuously) only at discrete instants
of time.\\

\indent
        Almost all the attempts to introduce a discretization of
time followed the first way, generally as part of a more extended procedure
in which the space-time as a whole is considered as intrinsically discrete
(four-dimensional lattice).  Recently, also T.D.Lee\cite{LEE} introduced
a time discretization on the basis of the {\em finite} number
of experimental measurements performable in any finite interval of time.
For an early approach in this direction, see e.g. T.Tati\cite{TATI} and
references therein.

        The second approach was first adopted in the twenties
(e.g., by Levi\cite{LP} and by Pokrowski\cite{LP}),
after Thomson's work, and resulted in the first real example of a theory
based on the existence of a fundamental interval of time: the one
set forth by Caldirola, in the fifties\cite{CALREV,CALgen}.  Namely, Caldirola
formulated a theory for the {\em classical} electron, with the aim of
providing a consistent (classical) theory for its motion in an
electromagnetic field. In the late seventies, Caldirola extended its
procedure to non-relativistic quantum mechanics.\\

It is known that the classical theory of the electron in an electromagnetic
field (despite the efforts by Abraham,\cite{ABRA} Lorentz,\cite{LORE}
Poincar\'{e}\cite{POIN} and Dirac,\cite{DIRA} among
others) actually presents many serious problems; {\em except ---of course---
when the field of the particle is neglected}.\footnote{It is interesting to
note that all those problems have been ---necessarily--- tackled by
Yaghjian in his recent book,\cite{YAGH} when he faced the question of the
relativistic motion of a charged, macroscopic sphere in an external
electromagnetic field.} \ By replacing Dirac's differential
equation by two {\em finite--difference equations}, Caldirola developed
a theory in which the main difficulties of Dirac's theory were overcome.
As we shall see, in his relativistically invariant formalism the chronon
characterizes the changes suffered by the dynamical state of the electron
when it is submitted to external forces. So that the electron
will be regarded as an (extended-like) object, which is point-like only at
discrete positions $x_{n}$ (along its trajectory) such that the electron
takes a quantum of proper time to travel from one position to the following
one [or, rather, two chronons: see the following]. \ It is tempting to
examine extensively the generalization of such a theory to the quantum
domain; and this will be performed in the present work. \ Let us recall
that one of the most interesting aspects of the {\em discretized}
Schr\"{o}dinger equations is that the mass of the {\em muon} and of the
{\em tau} lepton followed as corresponding to the two levels of the first
(degenerate) excited state of the electron.\\

In conventional QM there is a perfect equivalence among its various pictures:
Schr\"{o}dinger's, Heisenberg's, density matrix's . When discretizing the
evolution equations, we shall succeed in writing down those pictures in a
form such that they result to be still equivalent. \ However, in order to
be compatible with the Schr\"{o}dinger representation, our Heisenberg equations
{\em cannot}, in general, be obtained by a {\em direct} discretization of the
continuous Heisenberg equation.

The plan of this work is the following. \ In Chapt.\ref{sec:2} we present a
brief review of the main classical theories of the electron, including
Caldirola's. \ In Chapt.\ref{sec:3} we introduce the
three discretized {\em forms} ("retarded", "advanced", "symmetrical") of the
Schr\"{o}dinger equation, analyze the main characteristics of such
formulations, and derive the retarded one from Feynman's path integral
approach. \ In Chapt.\ref{sec:4}, our discrete theory is applied to some
simple quantum systems, such as the harmonic oscillator, the free particle
and the hydrogen atom. The possible experimental deviations
from the predictions of ordinary QM are investigated. \ In Chapt.\ref{sec:5},
a new derivation of the discretized Liouville--von Neumann equation, starting
from the {\em coarse grained} hypothesis, is presented.  Such a
representation is then adopted for tackling the {\em measurement problem}
in QM, with rather interesting results. \ Finally, a discussion on the
possible interpretation of our discretized equations can be found in
Chapt.\ref{sec:6}.

\section{The Introduction of the Chronon in the Classical Theory of
the Electron}
\label{sec:2}

\mbox{}
\indent
        Almost a century after its discovery, the electron continues to be
an object waiting for a convincing description (cf., e.g., \cite{RECSAL}),
both in classical and quantum electrodynamics. As Schr\"odinger put it, the
electron is ---still--- a stranger in electrodynamics.  Maxwell's
electromagnetism is a field theoretical approach in which no reference is
made to the existence of material corpuscles. Thus, one may say that one of
the most controversial questions of the twentieth century physics, the
wave--particle paradox, is not characteristic of QM only.
In the electron classical theory, matching the description of the
electromagnetic fields (obeying Maxwell equations) with the
existence of charge carriers like the electron is still a challenging
task.

The hypothesis that electric currents could be associated with charge
carriers was already present in the early ``particle electrodynamics"
formulated in 1846 by Fechner and Weber.  But such an idea was taken
into consideration again only a few decades later, in 1881, by Helmholtz.
Up to that time, electrodynamics had been developed on the hypothesis of an
electromagnetic {\em continuum\/}\cite{COULBLOCK}
and of an {\em ether}.  In that same year, J.J.Thomson wrote his
seminal paper in which the electron mass was regarded as purely
electromagnetic in nature.  Namely, the energy and momentum associated
with the (electromagnetic) fields produced by an electron were held entirely
responsible for the energy and momentum of the electron itself.\cite{BELLONI}

Lorentz's electrodynamics, which described the particle--particle interaction
via electromagnetic fields by the famous force law

\begin{equation}
\bf{f} = \rho \left({\bf{E} + \frac{1}{c}\bf{v}\wedge\bf{B}}\right) \ ,
\end{equation}

\noindent
$\rho$ being the charge density of the particle on which the fields act,
dates back to the beginning of the 1890 decade.  The electron was finally
discovered by Thomson in 1897, and in the following years various theories
appeared.  The famous (pre-relativistic) theories by
Abraham, Lorentz and Poincar\'{e} regarded it as an extended--type object,
endowed again with a purely electromagnetic mass. As well-known, in 1903
Abraham proposed the simple-minded (and questionable) model of a rigid sphere,
with a uniform electric charge density on its surface. The theory of Lorentz
(1904) was quite similar, trying to improve the situation with the mere
introduction of the effects resulting from the Lorentz--Fitzgerald
contraction.

\subsection{The theory of the electron by Abraham--Lorentz}

A main difficulty for an accurate description of the electron motion was
the inclusion of the radiation reaction, i.e., of the effect produced on such
a motion by the fields radiated by the particle itself. In the model proposed
by Abraham--Lorentz the assumption of a purely electromagnetic structure
for the electron implied that

\begin{equation}
\bf{F}_{\rm p}+\bf{F}_{\rm ext}=0 \ ,
\label{eq:al}
\end{equation}

\noindent
where $\bf{F}_{\rm p}$ is the self-force due to the self-fields of the
particle, and $\bf{F}_{\rm ext}$ is the external force.  According to
Lorentz's law, the self-force was given by

$$\bf{F}_{\rm p} = \int \rho \left({\bf{E}_{\rm p} + \frac{1}{c}\bf{v}\wedge
\bf{B}_{\rm p}}\right) \drm^3{\bf r} \ ,$$

\noindent
where $\bf{E}_{\rm p}$ and $\bf{B}_{\rm p}$ are the fields produced by the
charge density $\rho$ itself, according to the Maxwell--Lorentz equations.
For the radiation reaction force, Lorentz obtained the following expression:

\begin{equation}
{\bf F}_{\rm p}=-\frac{4}{3c^2}W_{\rm el} {\bf a}+\frac{2}{3}\frac{k e^2}{c^3}
\dot{\bf a}
-\frac{2e^2}{3c^3}\sum_{n=2}^{\infty}\frac{(-1)^n}{n!}\frac{1}{c^n}
\frac{\drm^n {\bf a}}{\drm t^n} {\rm O}(R^{n-1}) \ ,
\label{eq:lor}
\end{equation}

\noindent
where $k \equiv (4 \pi \epsilon_0)^{-1}$ \ [in the following, whenever
convenient, we shall assume units such that numerically $k = 1$], \ and where

$$W_{\rm el} \equiv \frac{1}{2} \int \int \frac{\rho ({\bf r}) \:
\rho ({\bf r}')}
{\left|{{\bf r}-{\bf r}'}\right|} \drm^3{\bf r} \: \drm^3{\bf r}' $$

\noindent
is the electrostatic self-energy of the considered charge distribution, and
$R$ is the radius of the electron. \ All the terms in the sum are
structure dependent; i.e., they depend on $R$ and on the charge
distribution. \ By identifying the electromagnetic mass of the particle with
its electrostatic self-energy

$$m_{\rm el}=\frac{W_{\rm el}}{c^2} \ ,$$

\noindent
it was possible to write eq.(\ref{eq:al}) as

\begin{equation}
\frac{4}{3}m_{\rm el}\dot{\bf v}-\Gammabf = {\bf F}_{\rm ext} \ ,
\end{equation}

\noindent
so that one got:

\begin{equation}
\Gammabf = \frac{2}{3}\frac{e^2}{c^3}\dot{\bf a}\left({1+{\rm O}(R)}\right) \ ,
\label{eq:react}
\end{equation}

\noindent
which was the equation of motion in the Abraham--Lorentz model. \ Quantity
$\Gammabf$ is the radiation reaction force, namely, the reaction force
acting on the electron. \ A problem with equation (4) was constituted by the
factor $\frac{4}{3}$. In fact, if the mass is supposed to be of
electromagnetic origin only, then the total momentum of the electron would
be given by

\begin{equation}
{\bf p} = \frac{4}{3}\frac{W_{\rm el}}{c^2} {\bf v} \ ,
\end{equation}

\noindent
which is {\em not} invariant under Lorentz transformations.  That model,
therefore, was non-relativistic. \ Finally, we can observe from equation
(\ref{eq:lor}) that the structure dependent terms are functions of higher
derivatives of the acceleration; even more, the resulting differential
equation is {\em of the third order}, so that initial position and initial
velocity are not enough to single out a solution.  In order to suppress the
structure terms, one ought to reduce the electron to a point, \
($R \rightarrow 0$), \ but in this case the self-energy $W_{\rm el}$ and
mass $m_{\rm el}$ would diverge!

After the emergence of the special theory of relativity or, rather, after
the publication by Lorentz in 1904 of his famous transformations, some
attempts were made to adapt the model to the new
requirements.\cite{LAUE,SCHOTT,PAGE} \ Abraham
himself (1905) succeeded in deriving the following generalization of the
radiation reaction term (\ref{eq:react}):

\begin{equation}
\Gamma_{\mu}=\frac{2}{3}\frac{e^2}{c}\left({\frac{\drm^2 u_{\mu}}{\drm s^2}
+\frac{u_{\mu}u^{\nu}}{c^2} \frac{\drm ^2u_{\nu}}{\drm s^2}}\right)  .
\label{eq:abragen}
\end{equation}

A solution for the problem of the electron momentum non-covariance
was proposed by Poincar\'e in 1905, by the  addition of cohesive
forces of non-electromagnetic character, which ---however--- made the
electron no longer purely electromagnetic in nature.

On the other hand, electrons could not be considered pointlike, due to
the obvious divergence of their energy when $R \rightarrow 0$; thus, a
description of the electron motion could not forget about the structure
terms. \ Only Fermi succeeded to show, later, that the correct relation for
the momentum of a purely electromagnetic electron could be obtained without
Poincar\'{e}'s cohesive forces.\cite{FERMI}

\subsection{Dirac's theory of the classical electron}

Notwithstanding its inconsistencies, the theory by Abraham--Lorentz was
the most accepted theory of the electron, until the publication of Dirac's
theory in 1938. During the long period in between, as well as afterwards,
however, various further attempts to solve the problem were set forth,
either by means of extended-type models (Mie, Page, Schott,
etc.\cite{MODELS}), or by trying again to treat the electron as a pointlike
particle (Fokker, Wentzel, etc.\cite{ROHR}).

Dirac's approach\cite{DIRA} is the best known attempt to describe the
classical electron.  It by-passed the critical problem of the previous
theories of Abraham and Lorentz by working out for the pointlike electron
a trick which {\em avoided} divergences! \ By using the conservation laws of
energy and momentum, and Maxwell equations, Dirac calculated the flux of the
energy--momentum 4-vector through a tube of radius $\epsilon \ll R$
(quantity $R$ being the radius of the electron at rest) surrounding
the world line of the particle, and obtained:

\begin{equation}
m \frac{\drm u_{\mu}}{\drm s} = F_{\mu} + \Gamma_{\mu} \ ,
\label{eq:dirrel}
\end{equation}

\noindent
where $\Gamma_{\mu}$ is the Abraham 4-vector (\ref{eq:abragen}), that is, the
reaction force acting on the electron itself;  and
$F_{\mu}$ is the 4-vector that represents the external field acting on
the particle:

\begin{equation}
F_{\mu}=\frac{e}{c} F_{\mu \nu} u^{\nu} \; .
\label{eq:eqnove}
\end{equation}

According to such a model, the rest mass $m_0$ of the electron is the
limiting, finite value obtained as the difference of two quantities tending to
infinity when $R \rightarrow 0$:

$$m_0 = \lim_{\epsilon \rightarrow 0} \left({\frac{1}{2}
\frac{e^2}{c^2\epsilon}-k(\epsilon)}\right) ,$$

\noindent
the procedure followed by Dirac being an early example of elimination of
divergences by means of a subtractive method.

\noindent
At the non-relativistic limit, Dirac's equation goes into
the one previously obtained by Abraham--Lorentz:

\begin{equation}
m_0\frac{\drm{\bf v}}{\drm t}-\frac{2}{3}\frac{e^2}{c^3}\frac{\drm^2{\bf v}}
{\drm t^2}=
e\left({{\bf E}+\frac{1}{c} {\bf v}\wedge {\bf B}}\right) \ ,
\label{eq:dirnrel}
\end{equation}

\noindent
except for the fact that in Abraham--Lorentz's approach $m_0$ diverged.  \
The last equation shows that the reaction force equals \ ${2 \over 3} \;
{e^2 \over c^3} \; {\drm^2 {\bf v} \over {\drm} t^2}$.

Dirac's dynamical equation (\ref{eq:dirrel}) was later reobtained also
from different, improved models.\cite{SHEMB} \
Wheeler and Feynman, for example, rederived eq.(\ref{eq:dirrel}) by basing
electromagnetism on an action principle applied to particles only, via their
absorber hypothesis.\cite{WF} \
However, eq.(\ref{eq:dirrel}) also presents many troubles, related to the
infinite many non-physical solutions that it possesses. \ Actually, as
already mentioned, it is a
third--order differential equation, requiring three initial conditions for
singling out one of its solutions. \ In the description of a {\em free} electron,
e.g., it even yields ``self-accelerating" solutions ({\em runaway
solutions\/}),\cite{ELIE1} for which velocity and acceleration increase
spontaneously and indefinitely.\cite{ROHR} \ Moreover, for an electron
submitted to an
electromagnetic pulse, further non-physical solutions appear, related this
time to {\em pre-accelerations}\cite{ASHA}. \  If the electron comes from
infinity with a uniform velocity $v_0$ and, at a certain instant of time
$t_0$, is submitted to an electromagnetic pulse, then it starts accelerating
{\em before} $t_0$. \ Drawbacks like these motivate further attempts to find
out a coherent model for the classical electron.

\subsection{Caldirola's theory for the classical electron}

Among the various attempts to formulate a more satisfactory theory, we want
to focus our attention on the one proposed by P.Caldirola.  Like Dirac's,
Caldirola's theory is also Lorentz invariant.  Continuity, in fact,  is not
an assumption required by Lorentz invariance.\cite{SNYD} \ The theory
postulates the existence of a universal interval $\tau_0$ of {\em proper}
time, even if time flows continuously as in the ordinary theory.  When an
external force acts on the electron, however, the reaction of the particle
to the applied force is not continuous: The value of the electron velocity
$u_\mu$ is supposed to jump from $u_\mu(\tau - \tau_0)$ to $u_\mu(\tau)$
only at certain positions $s_{n}$ along its world line; these discrete
positions being such that the electron takes a time $\tau_0$ for travelling
from one position $s_{\rm{n} - 1}$ to the next one $s_{n}$.

In this theory\cite{CALREV} the electron, in principle, is still considered
as pointlike, but the Dirac relativistic equations for the classical
radiating electron are replaced: \ (i) by a corresponding
{\em finite--difference} (retarded) equation in the velocity $u^\mu(\tau)$

\begin{eqnarray}
{{m_0} \over {\tau_0}}\left\{ {u_\mu \left( \tau  \right)-u_\mu \left(
{\tau -\tau_0} \right)+{{u_\mu \left( \tau  \right)
u_\nu \left( \tau  \right)} \over {c^2}}\left[ {u_\nu \left( \tau
\right)-u_\nu \left( {\tau -\tau_0} \right)} \right]} \right\} \ =\nonumber \\
= \ {e \over c}F_{\mu \nu}\left( \tau  \right)u_\nu \left( \tau  \right) ,
\label{eq:retrel}
\end{eqnarray}

\noindent
which reduces to the Dirac equation (\ref{eq:dirrel}) when $\tau_{0}
\rightarrow 0$, but cannot be derived from it (in the sense that it cannot be
obtained by a simple discretization of the time derivatives
appearing in Dirac's original equation); \ and \ (ii) by a second
equation, connecting this time the ``discrete positions" $x^\mu(\tau)$
along the world line of the particle. In fact, the dynamical law above is
by itself unable to specify univocally the variables $u_{\mu}(\tau)$ and
$x_{\mu}(\tau)$ which describe the motion of the particle. Caldirola named
it the {\em transmission law}:

\begin{equation}
x_\mu \left( {n\tau_0} \right)-x_\mu \left[ {\left( {n-1} \right)\tau_0} \right]=
{{\tau_0} \over 2}\left\{ {u_\mu \left( {n\tau_0} \right)-u_\mu \left[ {\left( {n-1}
\right)\tau_0} \right]} \right\} ,
\label{eq:trel}
\end{equation}

\noindent
which is valid inside each discrete interval $\tau_{0}$, and describes the
{\em internal} or {\em microscopic} motion of the electron.

In these equations, $u^\mu(\tau)$ is the ordinary four-vector velocity
satisfying the condition

$$u_\mu(\tau) u^\mu(\tau) = -c^2 \ \ \ \ \ \ \ {\rm for} \ \tau = n \tau_0$$

\noindent
where $n = 0,1,2,...$ \ and \ $\mu,\nu = 0,1,2,3$; \ $F^{\mu \nu}$ is the
external (retarded) electromagnetic field tensor, \ and the quantity

\begin{equation}
{\tau_0 \over 2} \equiv \theta_0 = {2 \over 3}{{k e^2} \over {m_0 c^3}} \simeq
6.266 \times 10^{-24} \; {\rm s}
\label{chronon}
\end{equation}

\noindent
is defined as the chronon associated with the electron (as it will be
justified below). 
The chronon $\theta_0 = \tau_0 / 2$ depends on the particle
(internal) properties: namely, on its charge $e$ and rest mass $m_0$.

\indent
       As a result, the electron happens to appear eventually as an
extended--like\cite{SALREC} particle, with internal structure, rather than as
a pointlike object (as initially assumed).  For instance, one may imagine
that the particle does not react instantaneously to the action of an external
force because of its finite extension (the numerical value of the chronon
is of the same order as the time spent by light to travel along an electron
classical diameter). \ As already said, eq.(\ref{eq:retrel}) describes the
motion of an object that happens to be pointlike only at discrete positions
$s_{n}$ along its trajectory\cite{CALREV}; even if both position and velocity
are still continuous and well-behaved functions of the parameter $\tau$, since
they are differentiable functions of $\tau$.

It is essential to notice that a discrete character is given to the electron
merely by the introduction of the fundamental quantum of time, without
any need of a ``model" for the electron.  Actually it is well-known that many
difficulties are met not only by the strictly pointlike models, but also by
the extended-type particle models (``spheres", ``tops", ``gyroscopes", etc.).
In A.O.Barut's words, \ ``If a spinning particle is not quite a point particle,
nor a solid three dimensional top, what can it be?". \ We deem the answer
stays with a third type of models, the ``extended-like" ones, as the present
theory; or as the (related) theoretical approach\cite{SALREC,RECSAL} in which
the center of the {pointlike} charge is spatially distinct from the particle
center-of-mass. \ Anyway, it is not necessary to recall that the worst
troubles met in quantum field theory (e.g., in quantum electrodynamics), like
the presence of divergencies, are due to the pointlike character still
attributed to (spinning) particles; since the problem of a suitable
model for elementary particles was transported, {\em unsolved}, from classical
to quantum physics. One could say that problem to be the most important in
modern particle physics.

\indent
        Equations (\ref{eq:retrel}) and (\ref{eq:trel}) together provide a
full description of the
motion of the electron. \ Notice that the global, ``macroscopic" motion can
be the same for different solutions of the transmission law. \ The behaviour
of the electron under the action of external electromagnetic fields is
completely described by its macroscopic motion.

As in Dirac's case, the equations above are invariant under Lorentz
transformations; but, as we are going to see, are {\em free} from
pre-accelerations, self-accelerating solutions, and problems (that had
raised great debates in the first half of the century) with the hyperbolic
motion.

\indent
        In the {\em non-relativistic limit} the previous (retarded) equations
reduced to the form

\begin{equation}
{{m_0} \over {\tau_0}}\left[ {{\bf v}\left( t \right)-{\bf v}\left(
{t-\tau_0} \right)} \right]= e \left[ {{\bf E}\left( t \right)+{1 \over c}
{\bf v}\left( t \right)\wedge {\bf B}\left( t \right)} \right] ,
\label{eq:retnrel}
\end{equation}

\begin{equation}
\bf r\left( t \right)-\bf r\left( {t-\tau_0} \right)={{\tau_0} \over 2}\left[ {{\bf v}\left( t \right)
-{\bf v}\left( {t-\tau_0} \right)} \right] ,
\label{eq:tnrel}
\end{equation}

\noindent
which {\em can} be obtained ---this time-- from eq.(\ref{eq:dirnrel}), by
directly replacing the time derivatives by the corresponding finite--difference
expressions.  The macroscopic equation (\ref{eq:retnrel}) had already been
obtained also by other authors\cite{SCHOTT,PAGE,BW,ELIE2} for the dynamics of
extended--type electrons.

\indent
       The important point is that eqs.(\ref{eq:retrel}),(\ref{eq:trel}), \
or eqs.(\ref{eq:retnrel}),(\ref{eq:tnrel}), \ allow to overcome the
difficulties
met with the Dirac classical equation (\ref{eq:dirrel}). \ In fact, the
electron {\em macroscopic} motion is completely determined once velocity and
initial position are given. \ Solutions of the relativistic equations
(\ref{eq:retrel}),(\ref{eq:trel}) for the radiating electron  ---or of the
corresponding non-relativistic equations (\ref{eq:retnrel}),(\ref{eq:tnrel})---
were obtained for several problems, the resulting motions never presenting
unphysical behaviour; so that the following questions can be regarded as
having been solved\cite{CALREV}: \ A) {\em exact relativistic solutions\/}: \
1) free
electron motion; \ 2) electron under the action of an electromagnetic
pulse;\cite{CIREL} \ 3) hyperbolic motion\cite{LANZ}; \ \ B) {\em
non-relativistic approximate solutions\/}: \ 1) electron under the action of
time-dependent forces; \ 2) electron in a constant, uniform magnetic
field\cite{PETTI}; \ 3) electron moving along a straight line under the action
of an elastic restoring force\cite{CCP}. \\

\indent
       Before going on, let us briefly study the electron radiation properties as
deduced from the finite-difference relativistic equations (\ref{eq:retrel}),
(\ref{eq:trel}), with the aim of showing the advantages of the present formalism
with respect to the Abraham--Lorentz--Dirac one. \ Such equations can be
written\cite{LANZ,CALREV}

\begin{equation}
{\Delta Q_\mu (\tau) \over \tau_0} + R_\mu (\tau) + S_\mu (\tau) \; = \;
{e \over c} F_{\mu \nu} (\tau) u^\nu (\tau) \ ,
\label{eq:retreln}
\end{equation}

\noindent
where:

\begin{equation}
\Delta Q_\mu \; \equiv \; {m_0} \left[ {u_\mu (\tau) -
u_\mu (\tau -\tau_0)} \right] \ ;
\label{eq:retdef1}
\end{equation}

\begin{equation}
R_\mu (\tau) \; \equiv \; - {m_0 \over {2 \tau_0}} \ \left\{
{{u_\mu (\tau) u_\nu (\tau)} \over c^2} \; \left[ u^\nu (\tau + \tau_0) +
u^\nu (\tau - \tau_0) - 2 u^\nu (\tau) \right] \; \right\} \ ;
\label{eq:retdef2}
\end{equation}

\begin{equation}
S_\mu (\tau) \; = \;  - {m_0 \over {2 \tau_0}} \ \left\{
{{u_\mu (\tau) u_\nu (\tau)} \over c^2} \; \left[ u^\nu (\tau + \tau_0) -
u^\nu (\tau - \tau_0) \right] \; \right\} \ .
\label{eq:retdef3}
\end{equation}

\noindent
In eq.(\ref{eq:retreln}), the first term $\Delta Q_0 / \tau_0$ represents the
variation per unit of proper time (in the interval $\tau - \tau_0$  to $\tau$)
of the particle energy--momentum vector. \ The second one, $R_\mu (\tau)$, is
a dissipative term, since it contains only even derivatives of the velocity
as one can prove by expanding $u^\nu (\tau + \tau_0)$ and
$u^\nu (\tau - \tau_0)$ in terms of $\tau_0$; furthermore, it is never
negative,\cite{CALREV,LANZ} and can therefore represent the energy-momentum
{\em radiated} by the electron in the unit of proper time. \ The third term,
$S_\mu (\tau)$, is conservative and represents the rate of change in proper
time of the electron {\em reaction} energy-momentum.

\indent
       The time component ($\mu = 0$) of eq.(\ref{eq:retreln}) writes:

\begin{equation}
{{T(\tau) - T(\tau - \tau_0)} \over \tau_0} + R_0 (\tau) + S_0 (\tau) \; = \;
 P^{\rm ext}(\tau) \ ,
\label{eq:retreln0}
\end{equation}

\noindent
quantity $T(\tau)$ being the kinetic energy

\begin{equation}
T(\tau) \ = \ m_0 c^2 \left( {1 \over {\sqrt{1 - \beta^2}}} - 1 \right)
 \ ; \nonumber
\end{equation}

\noindent
so that in eq.(\ref{eq:retreln0}) the first term replaces the proper-time
derivative of the kinetic energy, the second one is the energy radiated by
the electron in the unit of proper time,  $S_0(\tau)$ is the variation rate
in proper time of the electron reaction energy (radiative correction), and
$P^{\rm ext}(\tau)$ is the work done by the external forces in the unit of
proper time.

\indent
       We are now ready to show, e.g., that eq.(\ref{eq:retreln0}) yields a
clear explanation for the origin of the so-called ``acceleration energy"
(Schott energy), appearing in the energy--conservation relation for the
Dirac equation. \ In fact, by expanding in power series with respect to
$\tau_0$ the l.h. sides of eqs.(\ref{eq:retreln}),(\ref{eq:retdef1}),
(\ref{eq:retdef2}),(\ref{eq:retdef3}) for $\mu = 0$, and keeping only the
first-order terms, one gets

\begin{equation}
{{T(\tau) - T(\tau - \tau_0)} \over \tau_0} \; \simeq \; {{\drm T} \over
{\drm \tau}} - {2 \over 3} \, {e^2 \over c^2} \, {{\drm a_0} \over
{\drm \tau}} \ ;
\label{eq:retr1}
\end{equation}

\begin{equation}
R_0 (\tau) \; \simeq \; {1 \over {\sqrt{1 - \beta^2}}} \, {2 \over 3} \,
{e^2 \over c^3} \, a_\mu a^\mu \ ;
\label{eq:retr2}
\end{equation}

\begin{equation}
S_0 (\tau) \; \simeq \; 0 \ ;
\label{eq:retr3}
\end{equation}

\noindent
where $a^\mu$ is the four-acceleration

$$a^\mu \; \equiv \; {{drm u^\mu} \over {drm \tau}} \; = \; \gamma \,
{{drm u^\mu} \over {drm t}}$$

\noindent
quantity $\gamma$ being the Lorentz factor. \  Therefore
eq.(\ref{eq:retreln0}), to the first order in $\tau_0$, becomes:

\begin{equation}
{{\drm T} \over {\drm \tau}} - {2 \over 3} \, {e^2 \over c^2} \,
{{\drm a_0} \over {\drm \tau}} + {2 \over 3} \, {e^2 \over c^3} \,
{{a_\mu a^\mu} \over {\sqrt{1 - \beta^2}}} \ \simeq \ P^{\rm ext}(\tau)
\label{eq:retr4}
\end{equation}

\noindent
or, passing from the proper time $\tau$ to the observer's time $t$:

\begin{equation}
{{\drm T} \over {\drm t}} - {2 \over 3} \, {e^2 \over c^2} \,
{{\drm a_0} \over {\drm t}} + {2 \over 3} \, {e^2 \over c} \,
{a_\mu a^\mu} \ \simeq \ P^{\rm ext}(\tau) \ {{\drm \tau} \over {\drm t}} \ .
\label{eq:retr5}
\end{equation}

\noindent
The last relation is identical with the energy conservation law found by Fulton
and Rohrlich\cite{FR} for the Dirac equation. In eq.(26) the derivative of
$(2e^2 / 3c^2) a_0$ appears, which is just the {\em acceleration energy}. Our
approach
clearly shows that {\em it arises only by expanding in a power series of}
$\tau_0$ the kinetic energy increment suffered by the electron during the
fundamental proper-time interval $\tau_0$; \ whilst such a Schott energy (as
well as higher-order energy terms) does not show up when adopting the
full formalism of finite--difference equations. \ We shall come back to this
important point in Subsection 2.4.

\indent
       Let us finally observe\cite{CALREV} that, when setting

\begin{equation}
{m_0 \over {e c \tau_0}} \; \left[ u_\mu (\tau) u_\nu (\tau - \tau_0) -
u_\mu (\tau - \tau_0) u_\nu (\tau) \right] \ \equiv \ F^{\rm self}_{\mu \nu} \ ,
\label{eq:retr6}
\end{equation}

\noindent
the relativistic equation of motion (\ref{eq:retrel}) reads:

\begin{equation}
{e \over c} \; \left( F^{\rm self}_{\mu \nu} + F^{\rm ext}_{\mu \nu} \right)
u^\nu \ = \ 0 \ ,
\label{eq:retrelnn}
\end{equation}

\noindent
confirming that $F^{\rm self}_{\mu \nu}$ represents the (retarded) self-field
associated with the moving electron.

\subsection{The three alternative formulations of Caldirola's theory}

Two more (alternative) formulations are possible of Caldirola's equations,
based on different discretization procedures. In fact, equations
(\ref{eq:retrel}) and (\ref{eq:trel}) describe an intrinsically radiating
particle.  And, by  expanding equation (\ref{eq:retrel})
in terms of $\tau_0$, a radiation reaction term appears.  Caldirola called
those equations the {\em retarded} form of the electron equations of motion.

\indent
       By rewriting the finite--difference equations, on the contrary, in
the form:

\begin{eqnarray}
{{m_0} \over {\tau_0}}\left\{ {u_\mu \left( {\tau +\tau_0} \right)-u_\mu
\left( \tau \right)+{{u_\mu \left( \tau  \right)u_\nu \left( \tau  \right)}
\over {c^2}}\left[ {u_\nu \left( {\tau +\tau_0} \right)-u_\nu \left( \tau
\right)} \right]} \right\} \ =\nonumber \\
 = \ {e \over c}F_{\mu
\nu}\left( \tau  \right)u_\nu \left( \tau  \right) ,
\label{eq:advrel}
\end{eqnarray}

\begin{equation}
x_\mu \left[ {\left( {n+1} \right)\tau_0} \right]-x_\mu \left( {n\tau_0}
\right)=\tau_0 u_\mu \left( {n\tau_0} \right) ,
\label{eq:tarel}
\end{equation}

\noindent
one gets the {\em advanced} formulation of
the electron theory, since the motion ---according to
eqs.(\ref{eq:advrel}),(\ref{eq:tarel})--- is now determined by advanced
actions.  In contrast with the retarded formulation, the advanced one
describes an electron which absorbs energy from the external world.

\indent
        Finally, by adding together retarded and advanced actions, Caldirola
wrote down the {\em symmetric} formulation of the electron theory:

\begin{eqnarray}
{{m_0} \over {2\tau_0}}\left\{ {u_\mu \left( {\tau +\tau_0} \right)-u_\mu \left(
{\tau -\tau_0} \right)+{{u_\mu \left( \tau  \right)u_\nu \left( \tau  \right)} \over {c^2}}
\left[ {u_\nu \left( {\tau +\tau_0} \right)-u_\nu \left( {\tau -\tau_0} \right)} \right]}
\right\} \ =\nonumber \\
= \ {e \over c}F_{\mu \nu}(\tau)u_\nu(\tau) ,
\end{eqnarray}

\begin{equation}
x_\mu \left[ {\left( {n+1} \right)\tau_0} \right]-x_\mu \left( {\left( {n-1} \right)\tau_0}
 \right)=2\tau_0u_\mu \left( {n\tau_0} \right) ,
\end{equation}

\noindent
which does not include any radiation reactions, and describes
a non radiating electron.\\

\indent
       Before closing this brief introduction to Caldirola's theory, it is
worthwhile to present two more relevant results derived from it, one of
them following below, in the next Subsection. \ If we  consider a free
particle and look for the ``internal solutions" of the
equation (\ref{eq:tnrel}),  we get  ---for a periodical solution of the type

$$\dot{x}=-\beta_0 \; c \; \sin\left({\frac{2 \pi \tau}{\tau_0}}\right)$$
$$\dot{y}=-\beta_0 \; c \; \cos\left({\frac{2 \pi \tau}{\tau_0}}\right)$$
$$\dot{z}=0$$

\noindent
(which describes a uniform circular motion) and by imposing the kinetic energy
of the internal rotational motion to equal the intrinsic energy $m_0c^2$ of
the particle---  that the amplitude of the oscillations is given by
$\beta_0^2=\frac{3}{4}$. \ Thus, the magnetic moment corresponding to this
motion is exactly the {\em anomalous magnetic moment} of the
electron,\cite{CAL0} obtained here in a purely classical context:

$$\mu_a=\frac{1}{4 \pi} \; \frac{e^3}{m_0c^2} \ . $$

This shows that the anomalous magnetic moment is an intrinsically classical,
and not quantum, result; and the absence of $\hbar$ in the last expression
is a remarkable confirmation of this fact.

\subsection{Hyperbolic motions}

In a review paper on the theories of electron including radiation--reaction
effects, Erber\cite{ERBER} criticized Caldirola's theory for its
results in the case of hyperbolic motion.

\indent
       Let us recall that the opinion of Pauli and von Laue (among others)
was that a charge
performing uniformly accelerated motions ---e.g., an electron in free fall---
could not emit radiation. That opinion was strengthened by the
invariance of Maxwell equations under the group of conformal
transformations,\cite{BATE} which in particular includes transformations
from rest to uniformly accelerated motions. \ Since the first decades of the
twentieth century, this had been ---however--- an open question, as the works
by Born and Schott had on the contrary suggested a radiation emission in such
a case. \ In 1960, Fulton and Rohrlich\cite{FR}
{\em demonstrated} that from Dirac's equation for the classical electron
the emission of radiation during the hyperbolic motion follows.

\indent
       A solution of this paradox is possible within Caldirola's theory, and
it was worked out in ref.\cite{LANZ} \  By analysing the energy conservation
law for an electron submitted to an external force and following a procedure
similar to that of Fulton and Rohrlich, Lanz obtained the
expression (\ref{eq:retreln0}) above. \ By expanding it in terms of $\tau$
and keeping only the first order terms, he arrived at
expression (\ref{eq:retr4}), identical to the one obtained by Fulton and
Rohrlich, in which ---let us repeat--- the Schott energy appears: \ a term
that Fulton and Rohrlich (having obtained it from Dirac's expression for
the radiation reaction) interpreted as a part of the internal energy of the
charged particle.

\indent
       For the particular case of {\em hyperbolic motion}, it is

$$a_{\mu}a^{\mu} = \frac{\drm a_0}{\drm \tau}$$

\noindent
so that there is {\em no} radiation reaction [cf. eqs.(\ref{eq:retr4}) or
(\ref{eq:retr5})]. \ However, neither the acceleration energy,
nor the energy radiated by the charge per unit of proper time,
$\frac{2}{3}e^2a_{\mu}a^{\mu}$, are zero!

\indent
       The difference is that in the discrete case this acceleration energy
does not exist as such, being a term proceeding from the discretized
expression for the charged particle kinetic energy variation. As we have
seen in eq.(\ref{eq:retr1}), the Schott term appears when the variation of
the kinetic energy during the fundamental interval of proper time is
expanded in powers of $\tau_0$:

$$\frac{T(\tau)-T(\tau-\tau_0)}{\tau_0} \; \cong \; \frac{\drm }{\drm \tau}T
- \frac{2}{3} \, \frac{e^2}{c^2} \, \frac{\drm }{\drm \tau}a_0 .$$

\noindent
This is an interesting result, since it was not easy to understand the
physical meaning of the Schott acceleration energy.  With the introduction of
the fundamental interval of time, as we know, the changes in the kinetic
energy are no longer continuous, and the Schott term merely expresses,
to first order, the variation of the kinetic energy when passing from one
discrete instant of time to the subsequent one.

\indent
       In eqs.(\ref{eq:retr1}) and (\ref{eq:retr4}), the derivative
${\drm T} / {\drm\tau}$ is a point function, forwarding the kinetic energy
slope {\em at} the instant $\tau$. \ And the dissipative term
$\frac{2}{3}e^2a_{\mu}a^{\mu}$ is just a relativistic generalization
of the Larmor radiation law: then, if there is acceleration, there is also
radiation emission.

\indent
       For the hyperbolic motion, however, the energy dissipated (because of
the acceleration) has just the same magnitude as the energy gain due to the
kinetic energy increase.  We are not forced to resort to `pre-accelerations'
in order to justify the origin of such energies...\cite{PLASS} \ Thus,
the present theory provides a clear picture of the physical processes
involved in the uniformly accelerated motion of a charged particle.

\section{The Hypothesis of the Chronon in Quantum\hfill\break
Mechanics}
\label{sec:3}
\mbox{}

\indent
        Let us pass to the main topic of the present paper: the chronon in
Quantum Mechanics. The speculations about the discreteness of time (on the basis of
of possible physical evidences) in QM go back to the first decades of this
century, and various theories have been proposed developing QM
on a space-time lattice (cf., e.g., refs.\cite{COLE}).  This is not the case
with the hypothesis of the chronon, where we do {\bf not} actually
have a discretization of the time coordinate.
In the twenties, for example, Pokrowski\cite{LP} suggested
the introduction of a fundamental interval of time, starting from an
analysis of the shortest wavelengths
detected at that time in cosmic radiation. More recently, for instance,
Ehrlich\cite{EHRL} proposed a quantization of the
elementary particle lifetimes, suggesting the value
$4.4 \times 10^{-24} \; {\rm s}$ for the {\em quantum} of time.  But a
time discretization is suggested by the very
foundations of QM. There are physical limits that
prevent the distinction of arbitrarily close successive
states in the time evolution of a quantum system.
Basically, such limitations result from the Heisenberg
relations: in such a way that, if a discretization is to be introduced in the
description of a quantum system, it cannot possess a universal value,
since those limitations depend on the
characteristics of the particular system under consideration. In other
words, the value of the fundamental interval of time has to change a priori
from system to system. All these points make
the extension of Caldirola's procedure to QM justifiable.

        In the seventies, Caldirola extended the introduction of the
chronon to QM, following the same guidelines
that had led him to his theory of the electron. So, time is still a
continuous variable, but the evolution of the system
along its world line is discontinuous. As
for the electron theory in the non-relativistic limit,
one has to substitute the corresponding finite--difference expression for the
time derivatives;  e.g.:

\begin{equation}
{{\drm f\left( t \right)} \over {\drm t}}\to {{f\left( t \right)-f\left( {t-\Delta t} \right)}
\over {\Delta t}}
\end{equation}

\noindent
where proper time is now replaced by the local time $t$.
Such a procedure was then applied to obtain the finite--difference
form of the Schr\"{o}dinger equation. As for the
electron case, there are three different ways to
perform the discretization, and three ``Schr\"{o}dinger
equations" can be obtained\cite{CALMON79}:

\begin{equation}
i{\hbar  \over \tau }\left[ {\Psi \left( {\bf x,t} \right)-\Psi \left( {\bf x,t-\tau } \right)} \right]
= \hat{H}\Psi \left( {\bf x,t} \right) ,
\label{eq:ret}
\end{equation}

\begin{equation}
i{\hbar  \over {2\tau }}\left[ {\Psi \left( {\bf x,t+\tau } \right)-\Psi \left(
{\bf x,t-\tau } \right)} \right]=\hat{H}\Psi \left( {\bf x,t} \right),
\label{eq:symm}
\end{equation}

\begin{equation}
i{\hbar  \over \tau }\left[ {\Psi \left( {\bf x,t+\tau } \right)-\Psi \left( {\bf x,t}
\right)} \right] = \hat{H}\Psi \left( {\bf x,t} \right),
\label{eq:adv}
\end{equation}

\noindent
which are, respectively, the {\em retarded}, {\em symmetric} and
{\em advanced} Schr\"{o}dinger equations, all of them transforming into the (same)
continuous equation when the fundamental interval of time (that can now be
called just $\tau$) goes to zero.
It can be immediately observed that the
symmetric equation is of the second order, while the other two are
first order equations. As in the continuous case, for a finite--difference
equation of order $n$ a single and complete solution
requires $n$ initial conditions to be specified.

\indent
        As the equations are different, the solutions they provide
are also fundamentally different. In order to study
the properties of such equations there are two basic
procedures. For some special cases, they can be solved by one
of the various existing methods of solving finite-difference equations,
or by means of an attempt solution, an
{\em ansatz}. The other way is to find a new Hamiltonian $\tilde{H}$ such
that a new continuous Schr\"{o}dinger equation

\begin{equation}
i\hbar {{\partial \Psi \left( {\bf x,t} \right)} \over {\partial t}}=
\tilde{H}\Psi \left( {\bf x,t} \right)
\end{equation}

\noindent
reproduces, at the points $t=n\tau$, the same results
obtained from the discretized equations. As has been shown
by Casagrande and Montaldi\cite{CASA}, it is always possible to
find a continuous generating function which makes it
possible to obtain a differential equation equivalent to the
original finite-difference one, such that in every point of interest their
solutions are identical. This procedure is useful since it is
generally very difficult to work with the finite--difference
equations on a qualitative basis. Except for some very
special cases, they can be only numerically solved. This
equivalent Hamiltonian $\tilde{H}$ is, however, non-hermitian
and it is frequently very difficult to be obtained.
For the special case where the Hamiltonian is time independent,
the equivalent Hamiltonian is quite easy to calculate.
For the symmetric equation, e.g., it is given by:

\begin{equation}
\tilde{H}={\hbar  \over \tau }\sin^{-1}
 ( {{\tau  \over \hbar }\hat{H}} ) .
\end{equation}

\noindent
 As expected, $\tilde{H}\rightarrow \hat{H}$ when $\tau \rightarrow 0$.
 Caldirola\cite{CALquan} used the
{\em symmetric} equation to describe the non-radiating electron
(bound electron) since for Hamiltonians explicitly
independent of time its solutions are always of oscillating
character:

$$\Psi \left( {\bf x,t} \right)=\exp \left( {-i{t \over \tau }\sin^{-1}
 ( {{\tau  \over t}\hat{H}} )} \right)f\left( {\bf x} \right) .$$

\noindent
In the classical theory of electron, the symmetric equation
also represents a non-radiating motion. It provides only an
approximate description of the motion without taking into
account the effects due to the self fields of the electron.
However, in the quantum theory it plays a fundamental
role. In the discrete formalism, it is the only way to describe
a bound non-radiating particle.

\indent
        The solutions of the {\em advanced} and {\em retarded} equations
show completely different behaviour. For a hamiltonian
explicitly independent of time the solutions have a general
form given by

$$\Psi \left( {\bf x,t} \right)=\left[ {1+i{\tau  \over \hbar }\hat{H}} \right]^{{{-t}
\mathord{\left/ {\vphantom {{-t} \tau }} \right. \kern-\nulldelimiterspace} \tau }}f\left(
{\bf x} \right)$$

\noindent
and, expanding $f(x)$ in terms of the eigenfunctions of $\hat{H}$,

$$\hat H u_n\left( {\bf x} \right)=W_nu_n\left( {\bf x} \right)$$

$$f\left( {\bf x} \right)=\sum\limits_n {c_nu_n}\left( {\bf x} \right)$$

\noindent
with

$$\sum\limits_n {\left| {c_n} \right|^2}=1 ,$$

\noindent
it can be obtained that

$$\Psi \left( {\bf x,t} \right)=\sum\limits_n {c_n}\left[ {1+i{\tau
\over \hbar }W_n} \right]^{{{-t} \mathord{\left/ {\vphantom {{-t} \tau }}
\right. \kern-\nulldelimiterspace} \tau }}u_n\left( {\bf x} \right) .$$

\noindent
In particular, the norm of this solution is given by

$$\left| {\Psi \left( {\bf x,t} \right)} \right|^2=\sum\limits_n {\left|
{c_n} \right|}^2\exp \left( {-\gamma_nt} \right)$$

\noindent
with

$$\gamma_n={1 \over \tau }\ln \left( {1+{{\tau ^2} \over {\hbar ^2}}W_n^2}
\right)={{W_n^2} \over {\hbar ^2}}\tau +O\left( {\tau ^3} \right).$$

\noindent
The presence of a damping factor depending critically on the
value $\tau$ of the chronon must be remarked.

\indent
        This {\em dissipative} behaviour originates from the {\em retarded}
character of the equation. The analogy with the electron
theory also holds and the retarded equation possesses
intrinsically dissipative solutions representing a radiating
system. The Hamiltonian has the same status as in the
continuous case: it is an observable since it is a hermitian
operator and its eigenvectors form a basis of the state
space. However, due to the damping term, the norm of the state
vector is not constant anymore. An opposite behaviour is
observed for the solutions of the advanced equation, in the
sense that they increase exponentially.

Before going on, let us at least mention that the discretized QM (as well as
Caldirola et al.'s approach to ``QM with friction") can find room within
the theories\cite{SANTIL} based on the so-calles Lie-admissible
algebras. A lot of related work (not covered in the present review) can be
found e.g. in refs.\cite{JANlieadm,MIGN,SANTIL}; see
also\cite{JANheisen,JANetal,JANdirac,MONZAN}.

For a different approach to decaying states, see e.g.\cite{AGO}.

\subsection{The mass of the muon}

        The most impressive achievement due to the introduction of
the chronon hypothesis in the realm of QM
is obtained in the description of a bound electron using the new
formalism. Bound states are described by the symmetric
Schr\"{o}dinger equation and a Hamiltonian that
does not depend explicitly on time. A general solution can
be obtained by using a convenient {\em ansatz}:

$$\Psi \left( {\bf x,t} \right)=\sum\limits_n {u_n}\left( {\bf x} \right)\exp
\left( {-i\alpha_nt} \right),$$

\noindent
where $\hat{H}u_{n}\left( \bf{x}\right)=E_{n}u_{n}\left( \bf{x}\right)$
gives the spectrum of eigenvalues of the
Hamiltonian. If the fundamental interval of
time $\tau$ corresponds to the chronon $\theta_{0}$ associated with the
classical electron, it can be straightforwardly obtained that

$$\alpha_n={1 \over {\theta_0}}\sin^{-1}
 \left( {{{E_n\theta_0} \over \hbar }} \right).$$

This solution gives rise to an upper limit for the
eigenvalues of the Hamiltonian due to the condition

$$\left| {{{E_n\theta_0} \over \hbar }} \right|\le 1.$$

\noindent
 Since $\theta_{0}$ is finite, there is a maximum value for
the energy of the electron given by

$$E_{\max }={\hbar  \over {\theta_0}}={2 \over 3}{{\hbar m_0c^3} \over
{e^2}}\approx 105.04 \; \; {\rm MeV}.$$

\noindent
Now, including the rest energy of the electron, we finally
get

$$E=E_{\max }+E_0^{{\rm electron}}\approx 105.55 \; \; {\rm MeV}.$$

\noindent
which is very close (an error of 0.1\%) to the measured
value of the rest mass of the {\bf muon}. Using the equivalent
hamiltonian method it is possible to extend the basis of
eigenstates out of the critical limit. However, for the
eigenvalues above the critical limit the corresponding
eigenstates are unstable and decay in time:

$$\Psi \left( {\bf x,t} \right)=\sum\limits_n {c_nu_n}\left( {\bf x} \right)\exp
\left( {-i\gamma_nt} \right) \exp\left( {-k_nt} \right),$$

\noindent
As for the retarded equation, the norm of the state vector is
not constant and decays exponentially with time for those
eigenstates out of the stability range. Caldirola\cite{CALquan}
interpreted this norm as giving the probability of the
existence of the particle in its original Hilbert space and
associated a mean lifetime with these states.

\indent
The considerations regarding the muon as an excited state of the
electron can be traced back
to the very days of its discovery. Particularly, it  has already
been observed that the ratio between the masses of the two particles
is almost exactly $3/(2 \alpha)$, where $\alpha$ is the fine structure
constant.\cite{NAMB}  It has already been remarked that
${\textstyle{2 \over 3}}\alpha $ is just the coefficient of the
radiative reaction term in Dirac's equation for the classical
electron.\cite{ROSEN} \ Bohm and Weinstein\cite{BW} put
forward the hypothesis that various kinds of ``mesons"
could be excited states of the electron. Dirac\cite{DIRAC62}
even proposed a specific model for an extended electron
so as to interpret the muon as an excited state of the
electron (on this point, cf. also refs.\cite{BAR78}).

Caldirola\cite{CAL78} observed that by means of the Heisenberg
uncertainty relations it is possible to associate the existence of the
muon as an excited state of the electron with the introduction of
the chronon in the theory of electron. The relation

$$\Delta \tau \; \Delta E \geq \hbar /2 $$

\noindent
imposes limitations in the determination, at a certain instant $\tau$, of the
energy {\em E} associated with the internal motion of the electron. If
excited states of the particle  corresponding to larger values of mass
exist, then it is possible only to speak of an ``electron with rest mass
$m_{0}$" when $\Delta E \leq (\mu_{0}-m_{0})c^{2}$, where $\mu_{0}$ is
the rest mass of the internal excited state. Such internal states could only
be excited in the presence of sufficiently strong interactions. From the
uncertainty relation we have that

$$\Delta\tau \geq \frac{\hbar}{2\left({\mu_{0}-m_{0}}\right) c^{2}} ,$$

\noindent
and, supposing the muon as an excited state, we get

$$\left({\mu_{0}-m_{0}}\right) c^{2} \cong \frac{3}{2} \frac{\hbar c}{e^{2}}
m_{0} c^{2}  .$$

Thus, it can be finally obtained that

$$\Delta \tau \geq \frac{1}{3} \frac{e^{2}}{m_{0}c^{2}} = \frac{\tau_{0}}{2}  ,$$

\noindent
i.e., that the value of the rest mass of an interacting electron can be taken
only inside an interval of the proper time larger than half a chronon. So,
when we  take into account two successive states, each one endowed with
the same uncertainty $\Delta \tau$,
they must then be separated by a time interval of at least $2 \: \Delta \tau$,
which corresponds exactly to the chronon $\tau_{0}$.

\subsection{The mass spectrum of leptons}

In order to obtain the mass of another particle, a possibility to be considered
is to take the symmetric equation as describing the muon. According to
this {\em na\"{\i}ve}
argumentation, the equation also foresees a maximum limit for the energy of
the eigenstates of the muon. By assuming the equation as successively describing
the particles corresponding to these maxima, an expression can be set up for
the various limit values, given by

\begin{equation}
E_{0}^{\left( n \right)}=m_{0}c^{2} \left[ {{3 \over 2}{{\hbar c} \over {e^2}}+1}
\right]^n=m_0c^2\left[ {{3 \over 2}{1 \over \alpha }+1} \right]^n \ ,
\end{equation}

\noindent
such that, for\vspace{0.5cm}
\begin{center}
\begin{tabular}{lclc}
$     n=0$ & $\rightarrow$ & $E^{(0)}=0.511 \;$ {\rm MeV} & ({\rm electron}) \\
$     n=1$ & $\rightarrow$ & $E^{(1)}=105.55 \;$ {\rm MeV} & ({\rm muon}) \\
$     n=2$ & $\rightarrow$ & $E^{(2)}=21801.54 \;$ {\rm MeV} & ({\rm heavy lepton?}) \ , \\
\end{tabular}
\end{center}
\vspace{0.5cm}
\noindent
the masses for the first excited states can be obtained, including a possible
heavy lepton which, according to the experimental results up to now,
does not seem to exist.

\indent
Following a suggestion of Barut\cite{BAR79}, according to which it should
be possible to obtain the excited states of the electron from the coupling
of its intrinsic magnetic moment with its self field,
Caldirola\cite{CAL80,BENZA},
considering a model of the extended electron as a micro-universe,
succeeded in evaluating also the mass of the lepton $\tau$ .

Caldirola took into account, for the electron, a model of a point-object
moving around in a 4-dimensional {\em de Sitter}  micro-universe
characterized by

$$c^2t^2-x^2-y^2-z^2=c^2\tau_{0}^2  ,$$

\noindent
where $\tau_{0}$ is the chronon associated with the electron and the radius
of the micro-universe is given by $a=c\tau_{0}$. Considering the spectrum
of excited states obtained from the na\"{\i}ve argumentation above, we
find that each excited state determines a characteristic radius for the
micro-universe. Thus, for each particle, the trajectory
of the point-object  is confined to a spherical shell defined by its
characteristic radius and by the characteristic radius of its excited state.
For the electron, e.g., the point-object moves around, inside the spherical
shell defined by its corresponding radius and by the one associated with its
excited state: the muon. Such radii are given by

\begin{equation}
a^{(n)}=\tau_{0}c\left[{\frac{3}{2}\frac{1}{\alpha}+1}\right]^{-n} .
\end{equation}

According to the model ---supposing that the intrinsic energy of the
lepton $e^{(n)}$ is given by $m^{(n)}c^2$--- the lepton moves in
its associated micro-universe along a circular trajectory with
a velocity $\beta=\frac{\sqrt{3}}{2}$, to which corresponds an
intrinsic magnetic moment

\begin{equation}
\mu_a^{(n)}=\frac{1}{4 \pi}\frac{e^2}{m^{(n)}c^2}  .
\end{equation}

Starting from Barut's suggestion, Caldirola obtained for the lepton $e^{(n)}$
an extra self-energy given by

$$E^{(n,p)}=(2p)^4 m^{(n)}c^2  .$$

The condition set down on the trajectory of the point-object, so that it remains
confined to its corresponding spherical shell, is given by

$$E^{(n,p)} \leq \left[{\frac{3}{2}\frac{1}{\alpha}+1}\right] m_0c^2  ,$$

\noindent
and the values attainable by $p$ are: $p=0$ for $n=0$, and $p=0,1$
for $n \neq 0$. The spectrum of mass is then finally given by

\begin{equation}
m^{(n,p)}=\left[{1+(2p)^4}\right] m^{(n)} = m_{0} \left[{1+(2p)^4}\right]
\left[{\frac{3}{2}\frac{1}{\alpha}+1}\right]^n .
\end{equation}

Thus, for different values of $n$ and $p$ we have:\vspace{0.5cm}
\begin{center}
\begin{tabular}{|c|c|c|c|}  \hline \hline
{\bf n} & {\bf p} & ${\bf m}^{(n)}$  &  \\ \hline
0 & 0 & 0.511 {\rm MeV} & {{\rm electron}}   \\ \hline
1 & 0 & 105.55 {\rm MeV} & {{\rm muon}} \\
  & 1 & 1794.33 {\rm MeV} & {{\rm tau}} \\ \hline \hline
\end{tabular}
\end{center}
\vspace{0.5cm}

It must be remarked that the {\em tau} appears as an internal excited state
of the muon and its mass is in fair agreement\cite{HIKA} with the experimental values:
$m_{\tau}\approx \:1784 \; {\rm MeV}$. The difference between these values
is less than 1\%. This is quite amazing if we consider the simplicity of the
model. The model foresees the existence of other excited states which do not
seem to exist. This is to some extent justifiable once the muon is obtained
as an excited electron and the description of the electron does
not  anticipate the existence of any other state. In order to obtain the lepton {\em tau}
it was necessary to introduce in the formalism the coupling of the intrinsic
magnetic moment with the self-field of the electron.

\subsection{Feynman path integrals}

\begin{figure}
\begin{center}
 \scalebox{0.8}{\includegraphics{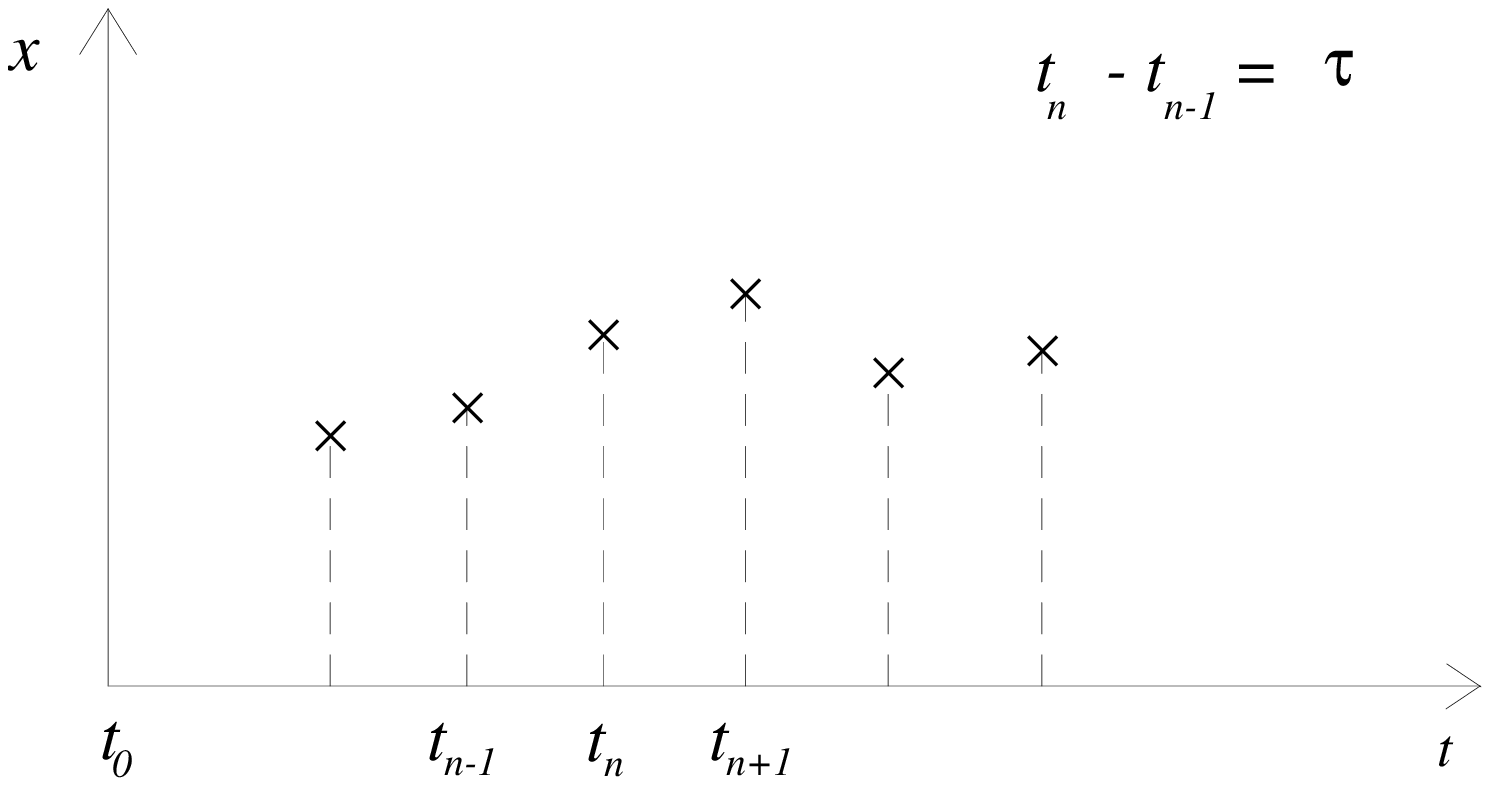}}
\caption{Discrete steps in the time evolution of the considered
system (particle).}
\end{center}
\label{fig:4-1}
\end{figure}

The discretized Schr\"{o}dinger equations can easily be obtained
using Feynman's path integral approach. This is
particularly interesting since it gives a clearer idea of the
meaning of these equations. According to the hypothesis
of a chronon, time is still a continuous variable and the
introduction of the fundamental interval of time is
connected ---let us recall-- only with the reaction of the system to the
action of a force. It is convenient to restrict the derivation to the
one-dimensional case, considering a particle under the
action of a potential $V(x,t)$. While the time coordinate is
continuous, we assume a discretization of the system (particle) position
corresponding to instants separated by time intervals $\tau$ [cf.
figure \ref{fig:4-1}].

        The transition amplitude for a particle going from an
initial point $(x_{1},t_{1})$ of the space-time to a final point
$(x_{n},t_{n})$ is given by the propagator

\begin{equation}
K\left( {x_n,t_n;x_1,t_1} \right)=\left\langle {{x_n,t_n}} \mathrel{\left | {\vphantom {{x_n,t_n}
{x_1,t_1}}} \right. \kern-\nulldelimiterspace} {{x_1,t_1}} \right\rangle.
\end{equation}

        In Feynman's approach this transition amplitude is
associated with a path integral, where the classical {\em action}
plays a fundamental role. It is convenient to introduce the
notation

\begin{equation}
S\left( {n,n-1} \right)\equiv \int\limits_{t_{n-1}}^{t_n} {\drm t \:{\cal L}
\left( x,\dot{x} \right)}
\label{eq:action}
\end{equation}

\noindent
such that ${\cal L} \left(x,\dot{x}\right)$  is the classical Lagrangian
and $S(n,n_{1})$ is the classical action. Thus, for two consecutive
instants of time, the propagator is given by

\begin{equation}
K\left( {x_n,t_n;x_{n-1},t_{n-1}} \right)={1 \over A}\exp \left( {{i \over \hbar }S
\left( {x_n,t_n;x_{n-1},t_{n-1}} \right)} \right).
\end{equation}

\indent
        The path integral has the meaning of a sum over all the
possible paths traversed by the particle and can be written as

\begin{equation}
\left\langle {{x_n,t_n}} \mathrel{\left | {\vphantom {{x_n,t_n} {x_1,t_1}}} \right.
\kern-\nulldelimiterspace} {{x_1,t_1}} \right\rangle=\mathop {\lim }\limits_{N\to \infty }
A^{-N}\int {\drm x_{N-1}\int {\drm x_{N-2}}\cdots }\int {\drm x_2\prod\limits_{n-2}^N {\exp
\left( {{i \over \hbar }S\left( {n,n-1} \right)} \right)}},
\end{equation}

\noindent
where $A$ is a normalization factor. In order to obtain the
discretized Schr\"{o}dinger equations we have to consider the
evolution of a quantum state between two consecutive
configurations $(x_{n-1},t_{n-1})$ e $(x_{n},t_{n})$. The state of the
system at $t_{n}$ is:

\begin{equation}
\Psi \left( {x_n,t_n} \right)=\int\limits_{-\infty }^{+\infty } {K\left( {x_n,t_n;x_{n-1},
t_{n-1}} \right)}\Psi \left( {x_{n-1},t_{n-1}} \right)\drm x_{n-1} .
\end{equation}

\indent
        On the other side, it follows from the definition of
the classical action (eq. \ref{eq:action}) that

\begin{equation}
S\left( {x_n,t_n;x_{n-1},t_{n-1}} \right)={m \over {2\tau }}\left( {x_n-x_{n-1}}
\right)^2-\tau V\left( {{{x_n+x_{n-1}} \over 2},t_{n-1}} \right).
\end{equation}

\noindent
Thus, the state at $t_{n}$ is given by

\begin{eqnarray}
\Psi \left( {x_n,t_n} \right)=\int\limits_{-\infty }^{+\infty } {\exp \left\{ {{{im} \over
 {2\hbar \tau }}\left( {x_n-x_{n-1}} \right)^2-i{\tau  \over \hbar }V\left( {{{x_n+x_{n-1}}
\over 2},t_{n-1}} \right)} \right\}}\nonumber \\
\Psi ( {x_{n-1},t_{n-1}} )\drm x_{n-1}.
\label{eq:psi1}
\end{eqnarray}

\indent
        When $\tau\approx 0$, for $x_{n}$ slightly different from $x_{n-1}$,
the integral due to the quadratic term is rather small. The contributions
are considerable only for $x_{n}\approx x_{n-1}$. Thus, we can
make the following approximation

$$x_{n-1}=x_n+\eta \;\; \rightarrow\;\; \drm x_{n-1}\equiv \drm \eta , $$

\noindent
such that

$$\Psi \left( {x_{n-1},t_{n-1}} \right)\cong \Psi \left( {x_n,t_{n-1}} \right)+\left( {{{\partial \Psi \left( {x_n,t_{n-1}} \right)} \over {\partial x}}} \right)\eta +\left( {{{\partial ^2\Psi } \over {\partial x^2}}} \right)\eta ^2.$$

By inserting this expression into equation~(\ref{eq:psi1}), supposing
that\footnote{The potential is supposed to vary slowly with $x$.}

$$V\left( {x+{\eta  \over 2}} \right)\approx V\left( x \right) ,$$

\noindent
and taking into account only the terms to the first order in $\tau$, we get

$$\Psi \left( {x_n,t_n} \right)={1 \over A}\exp \left( {-{i \over \hbar }\tau V\left(
{x_n,t_{n-1}} \right)} \right)\left( {{{2i\hbar \pi \tau } \over m}} \right)^{{1
\mathord{\left/ {\vphantom {1 2}} \right. \kern-\nulldelimiterspace} 2}}\left(
{\Psi \left( {x_n,t_{n-1}} \right)+{{i\hbar \tau } \over {2m}}{{\partial ^2\Psi }
\over {\partial x^2}}} \right)$$

Notwithstanding the fact that $\exp(-i \tau V(x_{n},t_{n})/\hbar)$  is
a function defined only for certain
well-determined values, it can be expanded in powers of $\tau$,
around an arbitrary position $(x_{n},t_{n})$. Choosing $A=(2i\hbar\pi\tau/m)^{-1/2}$,
such that $\tau \rightarrow 0$
in the continuous limit, we obtain

\begin{eqnarray}
\Psi \left( {x_n,t_{n-1}+\tau } \right)-\Psi \left( {x_n,t_{n-1}} \right)=-{i \over \hbar }\tau
V\left( {x_n,t_{n-1}} \right)\Psi \left( {x_n,t_{n-1}} \right) \ +\nonumber \\
 + \ {{i\hbar \tau } \over {2m}}
{{\partial ^2\Psi } \over {\partial x^2}}+ {\rm O}\left( {\tau ^2} \right)
\end{eqnarray}

\indent
By a simple reordering of terms we finally get

$$i{{\Psi \left( {x_n,t_{n-1}+\tau } \right)-\Psi \left( {x_n,t_{n-1}} \right)}
\over \tau }=\left\{ {-{{\hbar ^2} \over {2m}}{{\partial ^2} \over {\partial x^2}}+V\left(
{x_n,t_{n-1}} \right)} \right\}\Psi \left( {x_n,t_{n-1}} \right)$$

\indent
Following this procedure we obtain the {\bf advanced}
finite--difference Schr\"{o}dinger equation which describes
a particle performing a one-dimensional motion under the potential
$V(x,t)$.

\indent
The solutions of the advanced equation show
an amplification factor which may suggest that the
particle absorbs energy from the field described by the Hamiltonian
in order to evolve in time. In the classical domain
the {\em advanced} equation is interpreted as describing
a {\em positron}. However, in the realm of the non-relativistic QM,
it is more naturally interpreted as representing a system which takes
in energy from the environment.

\indent
In order to obtain the discrete Schr\"{o}dinger equation only the
terms to the first order in
$\tau$ have been taken into account. Since the limit $\tau \rightarrow 0$
has not been accomplished, the equation thus obtained is only
an approximation. The meaning of this fact may be related to another one
we are going to face below in this paper, when considering the measurement
problem in QM.

\indent
It is rather interesting to remark that in order to obtain
the {\em retarded} equation we have to regard the propagator as acting
backward in time. The conventional procedure used in the continuous case
always provides us with the {\em advanced} equation.
Therefore, the potential describes a mechanism for transferring  energy
from a field to the system.
The {\em retarded} equation  can be obtained only by
assuming an inversion of the time order, considering
the expression

\begin{equation}
\Psi \left( {x_{n-1},t_{n-1}} \right)=\int\limits_{-\infty }^{+\infty }
{{1 \over A}}\exp
\left\{ {{i \over \hbar }\int\limits_{t_n}^{t_{n-1}} {{\cal L}\drm t}}
\right\}\Psi \left( {x_n,t_n}
\right)\drm x_n  ,
\end{equation}

\noindent
which can be rigorously obtained by merely using the closure
relation for the eigenstates of the position operator and then
redefining the propagator in the inverse time order.
With this expression, it is possible to obtain the {\em retarded}
Schr\"{o}dinger equation.  The symmetric equation can easily be
obtained by a similar procedure.

\indent
An interesting characteristic related to these apparently
opposed equations is the impossibility of obtaining one from
the other by a simple time inversion. The time order in the propagators
must be related to the inclusion, in these propagators, of something like
the advanced and retarded potentials. Thus, in order to obtain
the retarded equation we have to take into account effects
that act backward in time. Considerations like these that led
to the derivation of the three discretized equations can supply
useful guidelines for the comprehension of their meaning.

\subsection{The Schr\"{o}dinger and Heisenberg pictures}
\label{sec:heisen}

In discrete QM, as well as in the ``continuous" one,
the utilization of discretized Heisenberg equations
is expected to be preferable for certain types of problems.
As it happens for the continuous case, the discretized
versions of the Schr\"{o}dinger and Heisenberg pictures are also
equivalent. However, we show below that the Heisenberg equations
{\bf cannot}, in general, be obtained by a direct discretization of the
continuous equations.

\indent
First of all, it is convenient to introduce the discrete time evolution
operator for the symmetric

\begin{equation}
\hat{U}\left( {t,t_0} \right)=\exp \left[ {-{{i\left( {t-t_0} \right)} \over \tau }\sin^{-1}
\left( {{{\tau \hat{H}} \over \hbar }} \right)} \right]
\end{equation}

\noindent
and for the retarded equation,

\begin{equation}
\hat{U}\left( {t,t_0} \right)=\left[ {1+{i \over \hbar }\tau \hat{H}} \right]^{{{-\left( {t-t_0}
\right)} \mathord{\left/ {\vphantom {{-\left( {t-t_0} \right)} \tau }} \right.
\kern-\nulldelimiterspace} \tau }}
\end{equation}

\indent
In order to simplify the equations, the following notation will be used
throughout this Section

\begin{equation}
\Delta f(t) \longleftrightarrow \frac{f(t+\tau)-f(t-\tau)}{2 \tau}
\end{equation}

\begin{equation}
\Delta_{R} f(t) \longleftrightarrow \frac{f(t)-f(t-\tau)}{\tau}
\end{equation}

\indent
For both the operators above it can easily be demonstrated that, if
the Hamiltonian $\hat{H}$ is a hermitian operator, the following equations
are valid:

\begin{equation}
\Delta\hat{U} \left({t,t_{0}}\right) = \frac{1}{i\hbar}
\hat{U} \left({t,t_{0}}\right) \hat{H} ,
\end{equation}

\begin{equation}
\Delta\hat{U}^{\dagger} \left({t,t_{0}}\right) = -\frac{1}{i\hbar}
\hat{U}^{\dagger} \left({t,t_{0}}\right) \hat{H} .
\end{equation}

\indent
In the Heisenberg picture the time evolution is transferred from the state
vector to the operator representing the observable according to the
definition

\begin{equation}
\hat{A}^{\rm H}\equiv \hat{U}^{\dagger}\left( {t,t_0=0} \right)
\hat{A}^{\rm S}\hat{U}\left( {t,t_0=0} \right)
\end{equation}

\indent
For the {\em symmetric} case, for a given operator $\hat{A}^{\rm S}$,
the time evolution of the operator $\hat{A}^{\rm H}(t)$ is given by

$$\Delta \hat{A}^{\rm H}\left( t \right)=\Delta \left[
{\hat{U}^{\dagger}\left( {t,t_0=0}
\right)\hat{A}^{\rm S}\hat{U}\left( {t,t_0=0} \right)} \right]$$

\begin{equation}
\Delta \hat{A}^{\rm H}\left( t \right)={1 \over {i\hbar }}\left[
{\hat{A}^{\rm H},\hat{H}} \right]
\end{equation}

\noindent
which has exactly the same form as the equivalent equation for the
continuous case. The important feature of the time evolution operator
which is used to derive the expression above is that
it is an {\em unitary} operator. This is true for the symmetric case.
For the retarded case, however, this property is not satisfied anymore.
Differently from the symmetric and continuous cases, the state
of the system is also time dependent in the retarded Heisenberg picture:

\begin{equation}
\left| {\Psi^{\rm H} (t)}  \right\rangle=\left[ {1+{{\tau ^2\hat{H}^2} \over
{\hbar ^2}}} \right]^{{{-\left( {t-t_0} \right)} \mathord{\left/ {\vphantom
{{-\left( {t-t_0} \right)} \tau }} \right. \kern-\nulldelimiterspace} \tau }}\left|
{\Psi^{\rm S} (t_{0})} \right\rangle
\end{equation}

\indent
By using the property $\left[\hat{A},f\left(\hat{A}\right)\right]=0$,
it is possible to show that the evolution law for the operators
in the retarded case is given by:

\begin{equation}
\Delta \hat{A}^{\rm H}\left( t \right)=\left\{ {{1 \over {i\hbar }}\left[
{\hat{A}^{\rm S}\left( t \right),\hat{H}^{\rm S}
\left( t \right)} \right]+\Delta \hat{A}^{\rm S}\left( t \right)} \right\}^{\rm H}
\end{equation}

\indent
In short, we can conclude that the discrete symmetric case and the continuous case
are formally very similar and the Heisenberg equation can be obtained
by a direct discretization of the continuous equation. For the retarded and
advanced cases, however, this does not hold. In the Appendices we analyse
the compatibility between the Heisenberg and Schr\"{o}dinger pictures.

Let us here mention that a lot of parallel work has been done by Jannussis
et al.; they, e.g., studied the retarded, dissipative case in the Heisenberg
representation, passing then to study in that picture the (normal or damped)
harmonic oscillator: see\cite{JANheisen} (cf. also\cite{JANetal,JANdirac}).

\subsection{Time-dependent Hamiltonians}
\label{time}

When we restricted the analysis of the discretized equations to the time
independent Hamiltonians this was made aiming at simplicity. When
the Hamiltonian is explicitly time dependent the situation is very
similar to the continuous case. It is always difficult to work with
such Hamiltonians but, as in the continuous case, the theory of small
perturbations can also be applied. For the symmetric equation,
when the Hamiltonian is of the form

\begin{equation}
\hat{H}=\hat{H}_{0}+\hat{V}(t) ,
\end{equation}

\noindent
such that  $\hat{V}$ is a small perturbation related to $\hat{H}_{0}$,
the resolution method turns out to be very similar to the usual one.
The solutions are equivalent to the continuous solutions
followed by an exponentially varying term. It is always possible to solve
this kind of problem using an appropriate ansatz.

\indent
However, another factor must be considered, related to the existence
of a limit beyond which $\hat{H}$ does not have stable eigenstates.
For the symmetric equation, the equivalent Hamiltonian is given by

\begin{equation}
\tilde{H}={\hbar  \over \tau }\sin^{-1} \left( {{\tau  \over \hbar }\hat{H}} \right) .
\end{equation}

\indent
Thus, as previously stressed, beyond the critical value the eigenvalues are
not real and the operator $\tilde{H}$ is not hermitian anymore. Below that limit,
$\tilde{H}$ is a densely defined and a self-adjoint operator in the ${\cal L} \subset L^{2}$
subspace defined by the eigenfunctions of $\tilde{H}$. When the limit value is
exceeded the system changes to an excited state and the previous state loses
physical meaning. In this way, it is convenient to restrict the observables
to self-adjoint
operators that keep invariant the subspace ${\cal L}$. The perturbation $\hat{V}$ is
assumed to satisfy this requirement.

\indent
In the usual QM it is convenient, in order to deal with time dependent
perturbations, to work with the interaction representation (Dirac's picture). In this
representation, the evolution of the state is determined by the time dependent potential
$ \hat{V}(t)$, while the evolution of the observable is determined by the stationary
part of the Hamiltonian $\hat{H}_{0}$. In the discrete formalism, the time evolution
operator defined for $\hat{H}_{0}$, in the symmetric case, is given by

\begin{equation}
\hat{U}_{0}\left( {t,t_0} \right)=\exp \left[ {-{{i\left( {t-t_0} \right)} \over \tau }\sin^{-1}
\left( {{{\tau \hat{H}_{0}} \over \hbar }} \right)} \right]
\end{equation}

\indent
In the interaction picture the vector state is defined, from the state in the
Schr\"{o}dinger picture, as

\begin{equation}
\left| {\Psi^{\rm I} (t)}  \right\rangle=\hat{U}^{\dagger}_{0}(t) \left|
{\Psi^{\rm S} (t_{0})} \right\rangle  ,
\end{equation}

\noindent
where $\hat{U}^{\dagger}_{0}(t)\equiv \hat{U}^{\dagger}_{0}(t,t_{0}=0)$.
On the other hand, the operators are defined as

\begin{equation}
\hat{A}^{\rm I}=\hat{U}_0^{\dagger}\left( t \right)\hat{A}^{{\rm S}}
\hat{U}_0\left( t \right)
\end{equation}

\indent
So, it is possible to show that, in the interaction picture, the evolution of the
vector state is determined by the equation

\begin{equation}
i\hbar \Delta \Psi ^{\rm I}\left( {\bf{x},t} \right)={{i\hbar } \over {2\tau }}\left[
{\Psi ^{\rm I}\left( {\bf{x},t+\tau } \right)-\Psi ^{\rm I}\left( {\bf{x},t-\tau }
 \right)} \right]=\hat{V}^{\rm I}\Psi ^{\rm I}\left( {\bf{x},t} \right) ,
\label{eq:evolint}
\end{equation}

\noindent
which is equivalent to a direct discretization of the continuous equation. For the
operators we get that

\begin{equation}
\Delta \hat{A}^{\rm I}\left( t \right)={{\hat{A}^{\rm I}\left( {t+\tau }
\right)-\hat{A}^{\rm I}\left( {t-\tau } \right)} \over {2\tau }}={1 \over {i\hbar }}
\left[ {\hat{A}^{\rm I},\hat{H}_0} \right] ,
\end{equation}

\noindent
which is also equivalent to the continuous equation.

\indent
Thus, for the symmetric case, the discrete interaction picture  keeps the same characteristics
of the continuous one for the evolution of the operators and state vectors once,
obviously, the eigenstates of $\hat{H}$  remain below the stability limit.
We can adopt, for the discrete case, a procedure  similar to that one
commonly used in QM to deal with small time dependent perturbations.

We consider, in the interaction picture, the same basis of eigenstates
associated with the stationary Hamiltonian $\hat{H}_{0}$, given
by $\left| {n}\right\rangle$. Then,

$$\left| {\Psi (t)} \right\rangle^{\rm I}=\sum\limits_n
{{\Psi (t)}\left\langle {n}
\mathrel{\left | {\vphantom {n {\Psi \left( t \right)}}} \right.
\kern-\nulldelimiterspace}
{{\Psi \left( t \right)}} \right\rangle^{\rm I}\left|{n}
\right\rangle=\sum\limits_n
{c_n\left( t \right)}\left|  {n}\right\rangle}$$

\noindent
is the expansion, over this basis, of the state of the system at a certain instant {\em t}.
It must be observed that the evolution of the state of the system is determined
once the coefficients $c_{n}(t)$ are known. Using the evolution equation (\ref{eq:evolint})
it can be obtained that

$$i\hbar \Delta \left\langle {n} \mathrel{\left | {\vphantom {n {\Psi
\left( t \right)}}} \right.
\kern-\nulldelimiterspace} {{\Psi \left( t \right)}} \right\rangle^{\rm I}=
\sum\limits_m {\left\langle n \right|\hat{V}^{\rm I}\left| {m}
 \right\rangle\left\langle {m} \mathrel{\left | {\vphantom {m {\Psi
\left( t \right)}}}
\right. \kern-\nulldelimiterspace} {{\Psi \left( t \right)}}
\right\rangle^{\rm I}} .$$

\noindent
Using the evolution operator to rewrite the perturbation $\hat{V}$ in the
Schr\"{o}dinger picture we get that

\begin{equation}
i\hbar \Delta c_n\left( t \right)=\sum\limits_m {c_m}\left( t \right)V_{nm}\left( t \right)\exp
\left( {i\omega_{nm}t} \right) ,
\label{eq:coef}
\end{equation}

\noindent
such that

$$\omega_{nm}={1 \over \tau }\left[ {\sin^{-1} \left( {{{\tau E_n} \over \hbar }}
 \right)\- \sin^{-1} \left( {{{\tau E_m} \over \hbar }} \right)} \right] ,$$

\noindent
and we obtain the evolution equation for the coefficients $c_{n}(t)$, the solution
of which gives the time evolution of the system.

As in the usual QM, it is also possible to work with the
interaction picture evolution operator, $\hat{U}^{\rm I} (t,t_{0})$,
which is defined as

$$\left|{\Psi (t)}\right\rangle^{\rm I}=\hat{U}^{\rm I}(t,t_{0})\left|{\Psi (t_{0})}
\right\rangle^{\rm I} ,$$

\noindent
such that (\ref{eq:evolint}) can be written as

\begin{equation}
i\hbar \Delta\hat{U}^{\rm I}(t,t_{0})=\hat{V}^{\rm I}(t) \hat{U}^{\rm I}(t,t_{0})  .
\end{equation}

\indent
The operator $\hat{U}^{\rm I} (t,t_{0})$ has to satisfy the initial condition
$\hat{U}^{\rm I} (t,t_{0}) = 0$. Given this condition, we
have for the finite--difference equation
above the solution

$$\hat{U}^{\rm I} (t,t_{0})=\exp\left[ {\frac{-i(t-t_{0})}{\tau}
\sin^{-1}\left({\frac{\tau \hat{V}^{\rm I}(t)}{\hbar}}\right)}\right] .$$

\indent
Differently from the continuous case, where the approximate evolution
operator turns out to be an infinite Dyson series, a well determined
expression is obtained. The solution of the problem is obtained by
correlating the elements of the matrix associated with such operator
to the evolution coefficients $c_{n}(t)$.

In general, the finite--difference equations are harder
to be analytically solved than the equivalent differential equations.
In particular, such difficulty is far more stressed for the system
of equations obtained from the formalism above.

An alternative approach is to use the equivalent
Hamiltonians.\cite{CAL78}
Once the equivalent Hamiltonian is found the procedure is exactly the same
as for the continuous theory. If the perturbation term $\hat{V}$ is
small the equivalent Hamiltonian can be written as

$$\tilde{H}=\frac{\hbar}{\tau}\sin^{-1}\left( {\frac{\tau}{\hbar}\hat{H}_{0}}\right)
+\hat{V}(t)=\hat{H}_{0}+\hat{V}(t) . $$

\indent
In the interaction picture, the state of the system is now
defined as

\begin{equation}
\left|{\Psi^{\rm I}(t)}\right\rangle=\exp{i\frac{\tilde{H}_{0}t}{\hbar}}
\left|{\Psi^{\rm S}(t)}\right\rangle ,
\label{eq:psi}
\end{equation}

\noindent
and the operators are given by

\begin{equation}
\hat{A}^{\rm I}=\exp\left({i\frac{\tilde{H}_{0}t}{\hbar}}\right)
\hat{A}^{\rm S} \exp\left({-i\frac{\tilde{H}_{0}t}{\hbar}}\right) .
\label{eq:hamilt}
\end{equation}

\indent
The states (\ref{eq:psi}) evolve according to the equation

\begin{equation}
i\hbar\frac{\partial}{\partial t}\left|{\Psi^{\rm I}(t)}\right\rangle=
\hat{V}^{\rm I} \left|{\Psi^{\rm I}(t)}\right\rangle ,
\label{eq:evolpsi}
\end{equation}

\noindent
where $\hat{V}^{\rm I}$ is obtained according to definition (\ref{eq:hamilt}).

Now, small time dependent perturbations can be dealt with by taking into account
the time evolution operator defined by

\begin{equation}
\left|{\Psi^{\rm I}(t)}\right\rangle=\hat{U}^{\rm I}(t,t_{0})\left|{\Psi^{\rm I}(t_{0})}\right\rangle .
\end{equation}

\noindent
According to the evolution law (\ref{eq:evolpsi}) we have

\begin{equation}
i\hbar\frac{\drm }{\drm t}\hat{U}^{\rm I}(t,t_{0})=\hat{V}^{\rm I}(t)
\hat{U}^{\rm I}(t,t_{0}) .
\end{equation}

\indent
Thus, once given that $\hat{U}^{\rm I}(t_{0},t_{0})=1$ it turns out
that the time evolution operator is given by

$$\hat{U}^{\rm I}(t,t_{0})=1-\frac{i}{\hbar}\int_{t_{0}}^{t} \hat{V}^{\rm I}(t')
\hat{U}^{\rm I}(t',t_{0}) \drm t' $$

\noindent
or

$$\hat{U}^{\rm I}(t,t_{0})=1+\sum_{n=1}\left({-\frac{i}{\hbar}}\right)^{n}
\int_{t_{0}}^{t}\drm t_{1}\int_{t_{0}}^{t_{1}}\drm t_{2} \cdots
\int_{t_{0}}^{t_{n-1}}\drm t_{n}
\hat{V}^{\rm I}(t_{1}) \hat{V}^{\rm I}(t_{2}) \cdots \hat{V}^{\rm I}(t_{n}) , $$

\noindent
where the evolution operator is obtained in terms of a Dyson series.

In order to draw a parallel between the elements of the matrix of the evolution
operator and the evolution coefficients $c_{n}(t)$ obtained from the continuous
equation equivalent to (\ref{eq:coef}), we have to use the basis of
eigenstates of the stationary Hamiltonian $\hat{H}_{0}$. If the
initial state of the system is an eigenstate $\left|{m}\right\rangle$ of
that operator then, at a subsequent time, we have

$$c_{n}(t)=\left\langle{n|\hat{U}^{\rm I}(t,t_{0})|m}\right\rangle .$$

\indent
The method of the equivalent Hamiltonian is simpler to use, since
it takes full advantage of the continuous formalism.

\section{Some Applications of the Discretized\hfill\break
Quantum Equations}
\label{sec:4}

\mbox{}
\indent
Turning back to more general questions, it is interesting to analyse the physical
consequences resulting from the introduction of the fundamental interval of
time in QM. In this Section we apply the discretized equations
to some typical problems.

\subsection{The simple harmonic oscillator}

The Hamiltonian that describes a simple harmonic oscillator  does not depend
explicitly
on time. The introduction of the discretization in the time coordinate does not
affect the outputs obtained from the continuous equation for the spatial
branch of the solution. This is always the case when the potential does not
have an explicit time dependence. For potentials like this, the solutions
of the discrete equations
are always  formally identical, with changes in the numerical values which
depend on the eigenvalues of the Hamiltonian considered and on the value of
the chronon associated with the system described. We have the same spectrum of
eigenvalues and the same basis of eigenstates but with the time evolution given
by a  different expression.

For the simple harmonic oscillator, the Hamiltonian is given by

\begin{equation}
\hat{H}=\frac{1}{2m}\hat{\bf P}^2+\frac{m\omega^2}{2}\hat{\bf X}^2  ,
\end{equation}

\noindent
to which the eigenvalue equation corresponds:

\begin{equation}
\hat{H} \left| {u_n}\right\rangle =E_n  \left| {u_n}\right\rangle  ,
\end{equation}

\noindent
so that $E_n$ gives the energy eigenvalue spectrum of the oscillator.

As mentioned previously, since this Hamiltonian does not  depends
explicitly on time, there is always an upper limit for the possible values
of  its energy eigenvalues. In the basis of eigenfunctions of $ \hat{H}$
a general state of the oscillator can be written as

$$\left| {\Psi(t)}\right\rangle = \sum_{n} c_n(0)\left|{u_n}\right\rangle
\exp{\left[ -i\frac{t}{\tau} \sin^{-1}{\left(  \frac{E_n \tau}{\hbar} \right) } \right] }  ,$$

\noindent
with $c_n(0)=\left\langle {u_n|\Psi(t=0)} \right\rangle$. Naturally, when $\tau \rightarrow 0$,
the solution above recovers the continuous expression with its time dependency
given by $\exp{\left( {\frac{-i E_n t}{\hbar}}\right)}$.  Therefore, there is only
a small phase difference between the two expressions. For the mean value of an arbitrary
observable,

$$\left\langle{\Psi(t)|\hat{A}|\Psi(t)}\right\rangle=\sum_{m=0} \sum_{n=0}
c_m^\ast(0) c_n(0) A_{mn} \exp{\left[{\frac{i}{\hbar}\left({E_m-E_n}\right)t}\right]} \; \cdot$$\hfill\break
$$\cdot \; \exp{\left[{i \left({E_m^3-E_n^3}\right)\frac{t\tau^2}{3! \hbar^3}}\right]}
 \ + \ {\rm O}(\tau^4) \; ,$$

\noindent
with $A_{mn}=\left\langle{u_m|\hat{A}|u_n}\right\rangle$, we obtain an
additional phase term which implies a small deviation of the resulting
frequencies when compared to the Bohr frequencies of the harmonic oscillator.
To first approximation, this  deviation is given by the term depending on $\tau^2$
in the expression above.

It must be emphasized that the restrictions imposed on the spectrum of eigenvalues
of $\hat{H}$  break the basis of eigenvectors: the number of eigenvectors becomes
{\em finite} and does not constitute a complete set and, therefore, does not form
a basis anymore. For eigenstates beyond the upper limit the states are unstable and
decay exponentially with time.

\begin{figure}
\begin{center}
 \scalebox{0.8}{\includegraphics{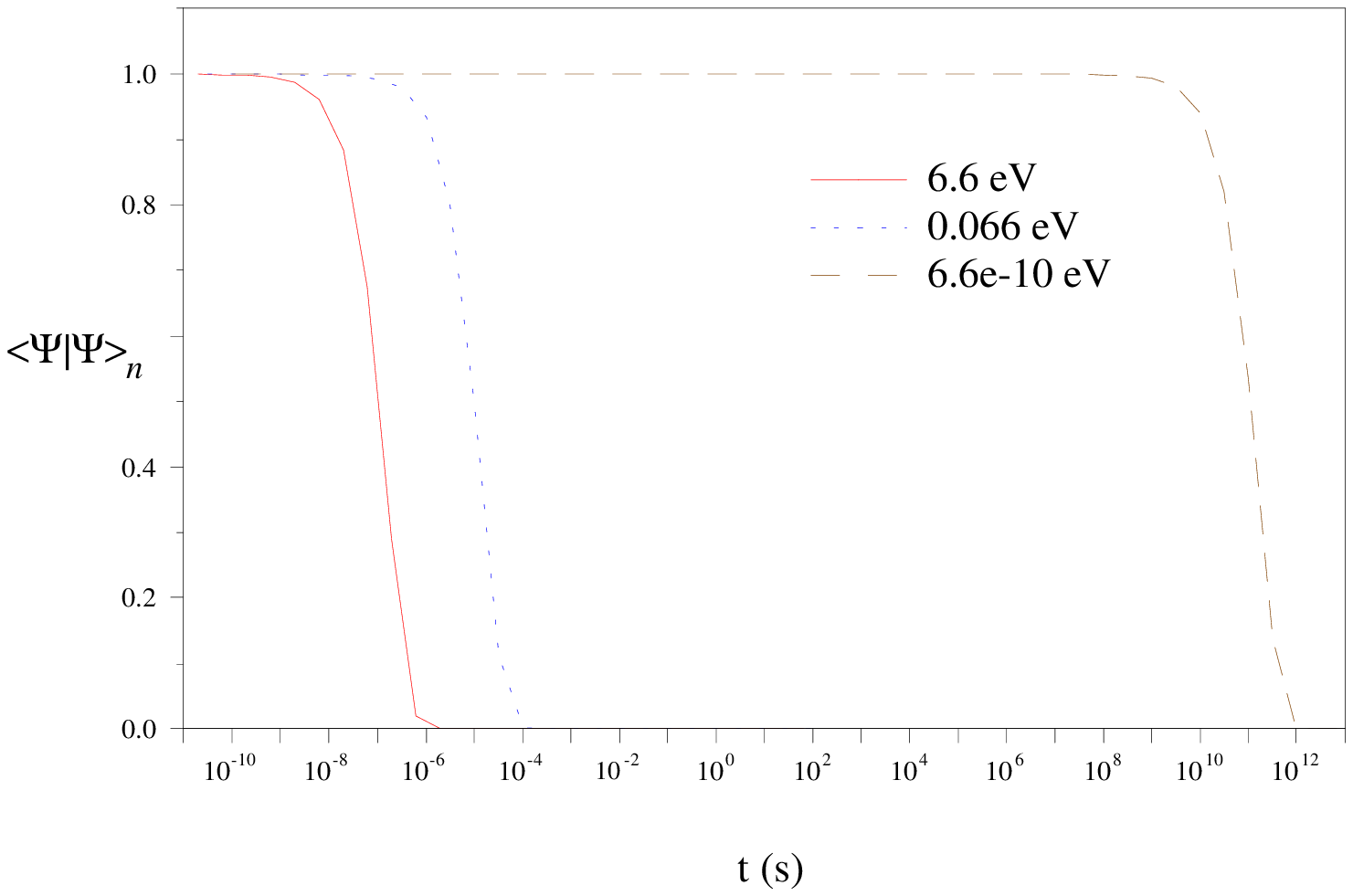}}
\caption{Typical behaviour of the damping factor associated with different
energy eigenvalues (retarded case).}
\end{center}
\label{fig:5-1}
\end{figure}

For a time independent Hamiltonian, the {\em retarded} equation always furnishes
damped solutions characteristic of radiating systems. In this case there is neither
stationary solutions nor upper limit for the energy eigenvalues. The larger the
eigenvalue the larger the damping factor and more quickly its contribution to the
state of the system tends to zero. If we write the state of the oscillator as

$$\left|{\Psi(t)}\right\rangle = \sum_{n} c_n(0) \left|{u_n}\right\rangle
\left[{1+\frac{i}{\hbar} \tau E_n}\right]^{\frac{t}{\tau}}  ,$$

\noindent
which has a norm decaying according to

\begin{equation}
 \left\langle{\Psi(t)|\Psi(t)}\right\rangle = \sum_{n} c_n(0)
\left|{u_n}\right\rangle \left[{1+\frac{\tau^2 E_n^2}{\hbar^2} }
\right]^{\frac{t}{\tau}} ,
\label{eq:oshar}
\end{equation}

\noindent
we have for an arbitrary observable that [with $\left\langle{A(t)}\right
\rangle \ \equiv \ $$\left\langle{A}\right\rangle (t)$]:

$$\left\langle{A(t)}\right\rangle = \sum_{m} \sum_{n} c_{m}^{\ast} (0)
c_n(0) A_{mn} \exp{\left[{- \frac{t}{\tau} \ln{\left[{1+\frac{\tau^2}{\hbar^2}
E_n E_m - \frac{i \tau}{\hbar}\left({E_m-E_n}\right)}\right]} }\right]}$$

\noindent
or, to the first order in $\tau$,

$$\left\langle{A(t)}\right\rangle = \sum_{m}\sum_{n} c_{m}^{\ast} (0)
c_n(0) A_{mn} \exp{\left[{\frac{i}{\hbar}\left({E_m-E_n}\right)t}\right]}
\exp{\left[{- i \left({E_m^2-E_n^2}\right) \frac{\tau}{2\hbar^2}}\right]}, $$

\noindent
so that, beside the Bohr frequencies defining the emission and absorption
frequencies of the oscillator, we obtain a damping term which causes the
average value of the observable ---which is explicitly independent of
time--- to tend to zero with time. A cursory analysis shows that even for
very small eigenvalues, smaller then $1.0 \; {\rm eV}$, the damping factor is
large, so that the decay of the average values is very fast. The damping factor
of the norm in equation (\ref{eq:oshar}) can be evaluated, and
its behaviour can be seen in figure \ref{fig:5-1}.

\subsection{Free particle}
\label{ss:5-2}

For a free particle, an electron for example, the general solution of the
symmetric equation (\ref{eq:symm}) can be obtained, in the coordinate
representation, using as an ansatz the solution for the continuous case.
Thus, a spectrum of eigenfunctions (plane waves) is obtained given by

$$\Psi_p({\bf x},t) = \left({2\pi\hbar}\right)^{-3/2}\exp{\left({
-i\alpha(|{\bf p}|)t +i\frac{({\bf p} \cdot {\bf x})}{\hbar}}\right)}. $$

Inserting this expression into the symmetric equation, we obtain for the
frequency $\alpha(|{\bf p}|)$ that

\begin{equation}
\alpha(|{\bf p}|)=\frac{1}{c}\sin^{-1}{\left({\frac{\tau}{\hbar}
\frac{{\bf p}^2}{2 m_0}}\right)} .
\label{eq:freefreq}
\end{equation}

When $\tau \rightarrow 0$, $\alpha(|{\bf p}|)\hbar$ coincides with the energy
of the particle. As has been observed for the bound particle, here we also have
an upper limit for the spectrum of eigenvalues. Thus the upper limit for the
possible values of momentum is given by

\begin{equation}
p\leq p_{\rm Max} \equiv \sqrt{\frac{2m_0\hbar}{\tau}}=10 \; {\rm MeV}/c
\end{equation}

\noindent
for the electron. In other words, there is a limit beyond which the frequencies
cease to be real.

As in the continuous case, the state of the particle is described by
a superposition of the eigenstates and can be written as

$$\Psi({\bf x},t) = \frac{1}{\left({2\pi\hbar}\right)^{3/2}} \int \drm^3 p
c({\bf p}) \exp{\left({ -i\alpha(|{\bf p}|)t +i\frac{({\bf p} \cdot
{\bf x})}{\hbar}}\right)}. $$

The coefficients $c({\bf p})$ are determined from the initial condition
$\Psi({\bf x},0)=\Psi_0({\bf x})$. From the expression for $\alpha$, it can
be observed that beyond a certain value of $p$ the expression loses
meaning. When $p\geq\sqrt{\frac{2m_0}{\tau}}$, the complete solution will
be defined only if $c({\bf p})$. From the stationary phase condition we have
that

$${\bf x}=\frac{{\bf p}}{m_0} \frac{t}{\sqrt{1-\left({\frac{\tau}{\hbar}}\right)^2
\left({\frac{p^4}{4m_0^2}}\right)}} ,$$

\noindent
and, supposing that $c({\bf p})$ corresponds to a distribution of probabilities
with a peak at ${\bf p}={\bf p}_0$~, then the wave packet will move in the
direction ${\bf p}_0$ with uniform velocity

$$v=\frac{p_0}{m_0}\left[{1-\left({\frac{\tau}{\hbar}}\right)^2
\left({\frac{p^4}{4m_0^2}}\right)}\right]^{-1/2}$$

\noindent
which coincides with the group velocity of the packet. It can be promptly
observed that when $p$ reaches its maximum value permitted, the velocity
diverges: $v\rightarrow \infty$. Thus, the introduction of a fundamental
interval of time does not bring in any restriction to the velocity of the
particle, although it results in a limit for the canonical momentum of the
eigenfunctions. Starting from the condition of stationary phase it is
possible to redefine the momentum associated with the particle, so that
this new momentum does not suffer any restriction at all. So, one can
conclude that the existence of free electrons with any
energy is possible, differently from what happens for the bound electron.

\begin{figure}
\begin{center}
 \scalebox{0.8}{\includegraphics{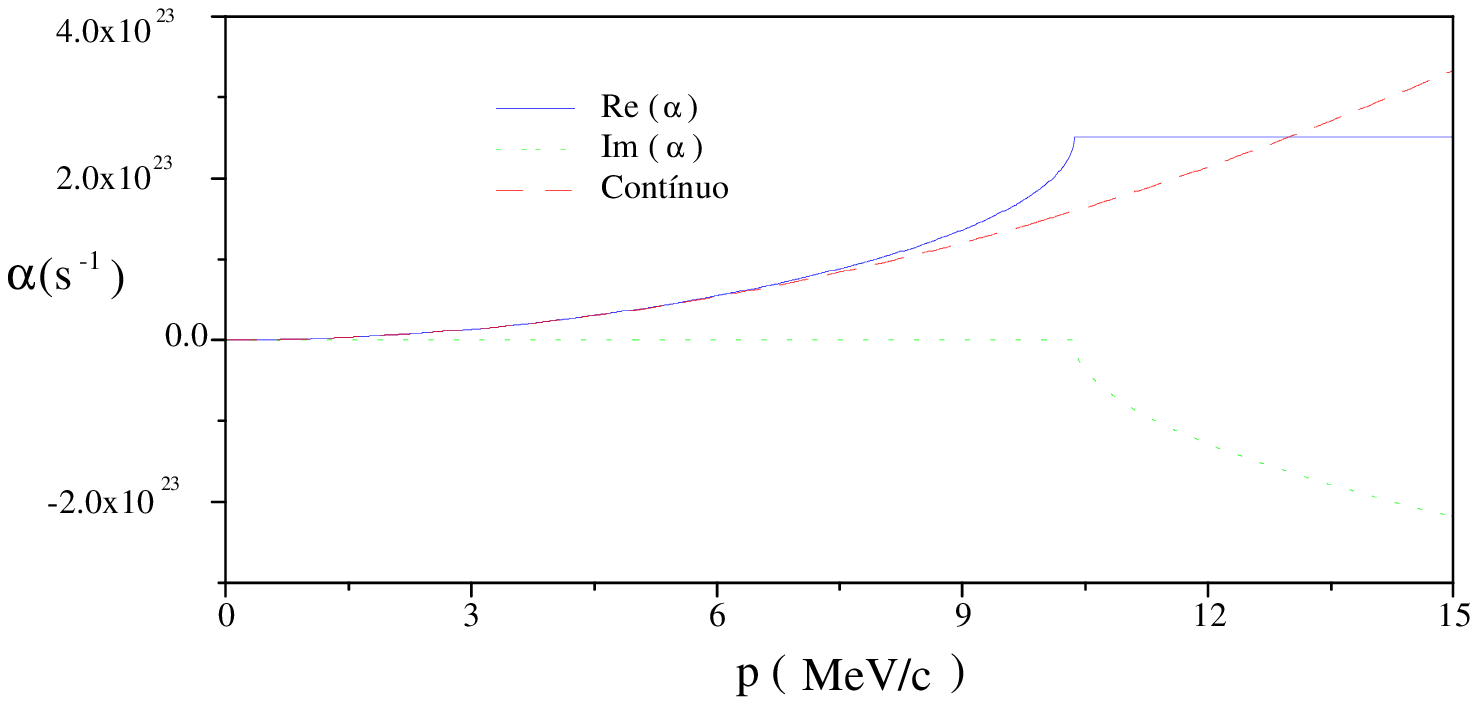}}
 \caption{Real and imaginary components of $\alpha(|{\bf p}|)$  obtained
 for the {\em symmetric} equation compared to the continuous case.}
\end{center}
\label{fig:5-2}
\end{figure}

For $p > p_{\rm Max}$ the frequency $\alpha(|{\bf p}|)$ fails to be real and its
dependence on $p$ is shown in figure {\ref{fig:5-2}}. An analysis of equation
(\ref{eq:freefreq}) shows that if $\alpha(|{\bf p}|)$ is complex then, for
$p\leq p_{\rm Max}$, the imaginary component is null and the real part is given by
expression (\ref{eq:freefreq}). When $p\geq p_{\rm Max}$, then

$${\rm Re}\left({\alpha(p)}\right)= \frac{\pi}{2 \tau} , $$

$${\rm Im}\left({\alpha(p)}\right)=-\frac{1}{\tau}\ln{\left[{\left|{\frac{\tau p^2}
{2 m_0 \hbar}}\right|+\sqrt{\left({\frac{\tau p^2} {2 m_0 \hbar}}\right)^2-1}}\right]} ,$$

\noindent
with the real part being a constant and the imaginary one tending logarithmically
to $-\infty$. Using the expressions above we can observe that, for $p>p_{\rm Max}$,
the eigenstates become unstable, with a time dependent decay term. When we look for an
equivalent Hamiltonian $\tilde{H}$ that, for the continuous Schr\"{o}dinger
equation, supplies equivalent outputs, we have that this is possible only if
$\tilde{H}$ is a non-hermitian operator. It is straightforward to see that
this is the case for $\tilde{H}={H}_1 + i {H}_2$, with ${H}_1$ and
${H}_2$ hermitian and such that ${H}_1\left|{p}\right\rangle=\hbar
\;{\rm Re}\left({\alpha(p)}\right)\left|{p}\right\rangle$ and ${H}_2\left|{p}\right\rangle=\hbar
\;{\rm Im}\left({\alpha(p)}\right)\left|{p}\right\rangle$ .

\begin{figure}
\begin{center}
 \scalebox{0.8}{\includegraphics{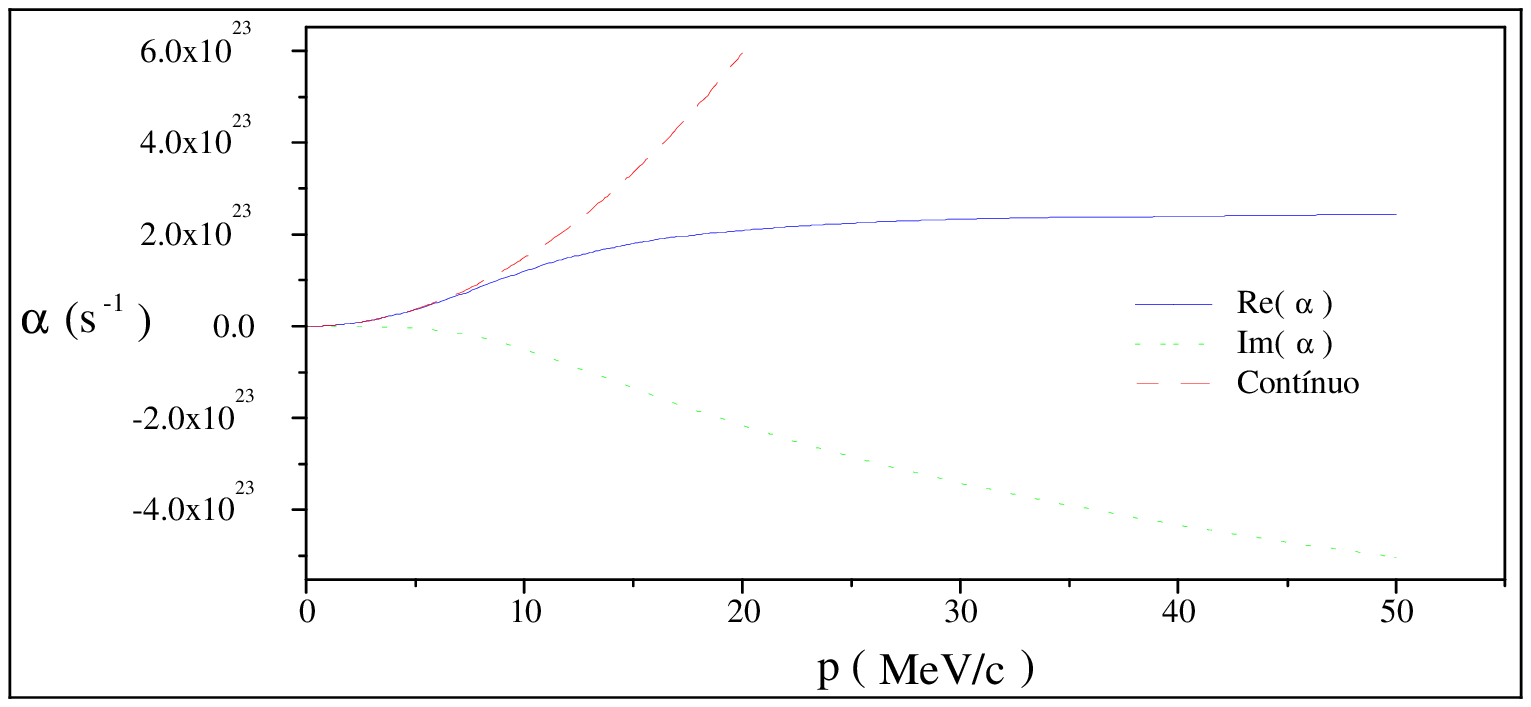}}
 \caption{Real and imaginary components of $\alpha(|{\bf p}|)$
 obtained for the
{\em retarded} equation compared with the continuous case.}
\end{center}
\label{fig:5-3}
\end{figure}

For the {\em retarded} equation, using the same {\em ansatz} of the symmetric case,
the damping factor appears for every value of $p$. There is no limitation
on the values of $p$ but, when $p\rightarrow \infty$, the real part of $\alpha(|{\bf p}|)$
tends to the same limit value observed for the symmetric case. Figure
{\ref{fig:5-3}}
illustrates the behaviour of the components of $\alpha(|{\bf p}|)$. The general expression for
an eigenfunction is found to be

$$\Psi_p(x,t)\propto \exp{\left[{\frac{ipx}{\hbar}-\frac{it}{\tau} \tan^{-1} {\frac{p^2\tau}
{2m\hbar}}}\right]} \exp{\left[{-\frac{t}{2\tau}\ln{\left[{1+\left({\frac{p^2\tau}
{2m\hbar}}\right)^2}\right]}}\right]} . $$

Performing a Taylor expansion and keeping only the terms to the first order in $\tau$
we obtain the continuous solution multiplied by a damping factor:

\begin{equation}
\Psi_p(x,t) \propto \exp{\left({\frac{ipx}{\hbar}-i\omega t}\right)}\exp{\left({\frac{1}{2}
\omega^2 \tau t}\right)}
\end{equation}

\noindent
where $\omega=p^2/2m\hbar$ is the frequency obtained for the continuous case.

It must be remarked that the damping term depends only on the Hamiltonian, through
the frequency $\omega$, and on the chronon associated with the particle. As the latter
is constant for a given particle, that term shows that for very high frequencies  the
solutions decay quite fast and, as the system evolves, the decay for smaller frequencies
also comes true.

\begin{figure}
\begin{center}
 \scalebox{0.8}{\includegraphics{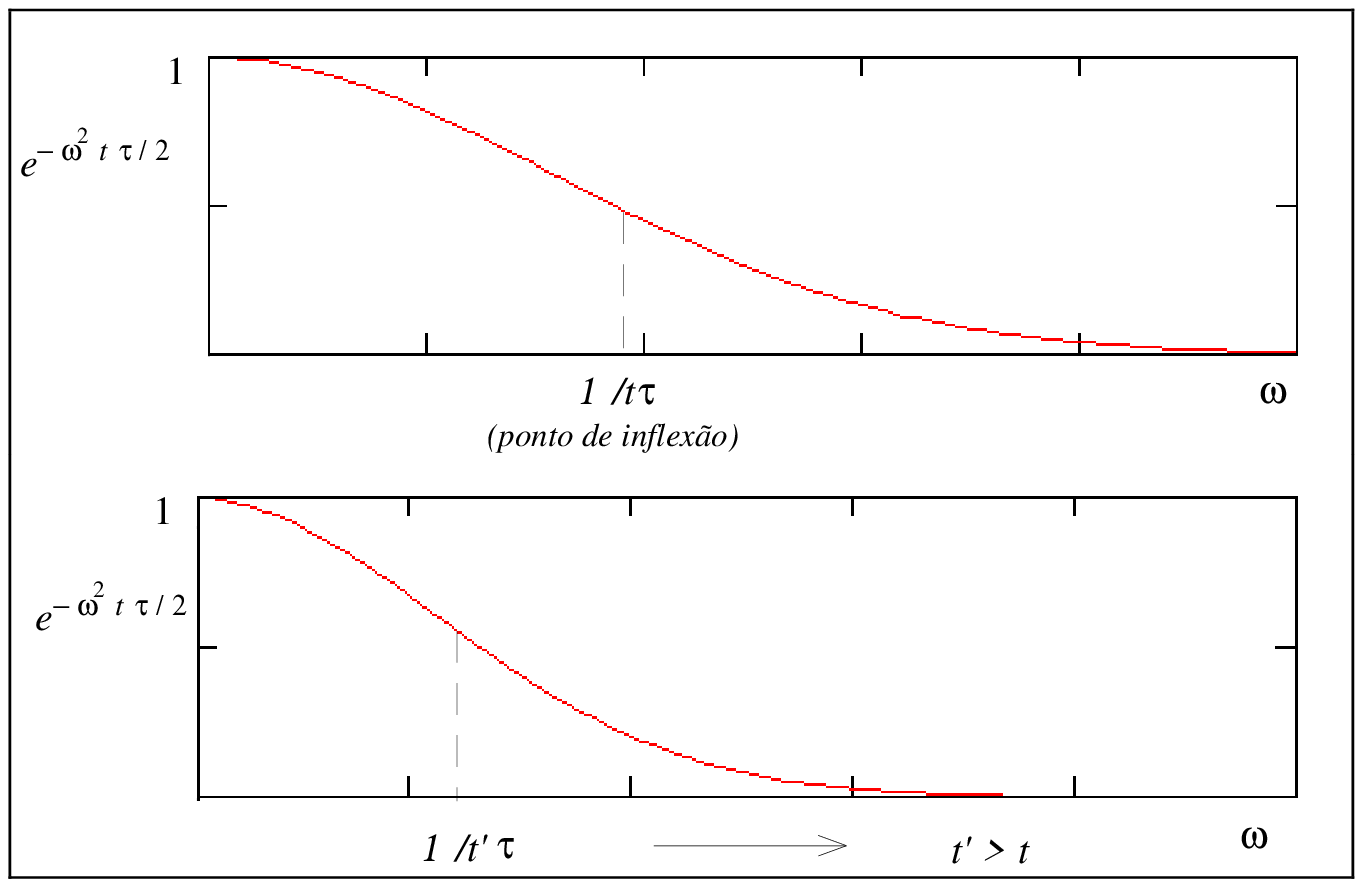}}
 \caption{Displacement with time of the inflection point.}
\end{center}
\label{fig:5-4}
\end{figure}

In figure {\ref{fig:5-4}} it can be observed that the inflection point, delimiting the region of
the spectrum where the decay is faster, moves in the direction of smaller frequencies
as time goes by. The consequence of this decay is the narrowing of the frequency
bandwidth which is relevant for the wave packet describing the particle. This is an echo
of the continuous decrease of the energy. As in the symmetric case, obtaining an equivalent
Hamiltonian is possible only if non-hermitian operators are considered.

At this point, it is worthwhile to reconsider the question of the physical meaning of the
three discretized Schr\"{o}dinger equations. Apparently, the choice of the equation to be
used in a particular situation is determined by the boundary conditions, by the restrictions
they impose on the system. The symmetric equation is used for special situations
for which the system neither emits nor absorbs radiation, or does it in a perfectly
``balanced" way: This is the case for the electrons in their `atomic orbits'.
So, the particle is stable
until a certain energy limit, beyond which the behaviour of the states is similar to that
of the retarded solutions. For energies far below that limit, the particle behaves almost
identically to the continuous case, only that the new frequencies associated with each
wave function differ from the continuous frequencies by a factor of order $\tau^2$.
The probability that a particle is found with energy larger than the limit value
decreases exponentially with time. For the bound electron, the limit is that one equivalent
to the rest mass of the muon. If a parallel with the classical approach is valid, the
symmetric equation describes an isolated system, which does not exchange energy
with the surrounding environment; or a situation of perfect thermodynamic
equilibrium, in which a perfect balance between absorbed and dissipated energies is
verified. For the classical theory of the electron the symmetric equation is only an
approximation which ignores the radiation reaction effects. In QM, however,
the existence of non-radiating states are related to the very essence of the theory. The
symmetric equation shows that, below the critical limit, the states are physically
identical to the outputs from the continuous theory: they are
{\em non-radiating} states.

The retarded equation represents a system which somehow loses energy for the
environment. The mechanism of such energy dissipation is related to the hamiltonian
of the system but also to some property of the environment  ---even the
vacuum---
as it can be inferred from the description of the free particle. From the solutions obtained
it is now observed that time has a well-defined direction of flux and that the frequency
composition of the wave packet associated with the particle depends on the instant of time
considered. It is clear that it is always possible to normalize the state in a certain instant
and consider it as being an initial state. This is permitted by the formalism. However,
in a strictly rigorous description, the frequency spectrum corresponds to a specific
instant of time which took place after the emission. This is an aspect that can be interesting
from the point of view of possible experimental verifications.

\subsection{The discretized Klein--Gordon equation}
\label{ss:5-3}

Another interesting application is the description of a free scalar particle
---a {\em scalar}
or {\em zero spin} photon,---  using a finite--difference form of the Klein--Gordon equation
for massless particles.

In the symmetric form the equation is written as

\begin{equation}
{{\square}^2} A_{\mu} = 0 \ \longrightarrow \ \frac{\Psi(t+2\tau)-2\Psi(t)+\Psi(t-2\tau)}{4 c^2 \tau^2}
-\nabla^2 \Psi(t)=0  .
\end{equation}

\noindent
Using a convenient {\em ansatz} we obtain, for this equation, in the coordinate representation,
that

$$\Psi_k(x,t)=A \exp{\left({-i\frac{t}{2\tau}\cos^{-1} ({1-2 c^2 \tau^2 k^2}})\right)}\exp{\left(
{ikx}\right)} , $$

\noindent
which can be written as

$$\Psi_p(x,t)=A \exp{\left({-i\frac{t}{2\tau}\cos^{-1} ({1-2 c^2 \tau^2 E^2/\hbar^2}})\right)}\exp{\left(
{ipx/\hbar}\right)} , $$

\noindent
since $E=p^2c^2$ and $p=\hbar k$. Expanding the time exponential in powers of $\tau$,
we find that, to the second order in $\tau$, a solution which is very similar to the
continuous expression:

$$\Psi_{\rm p}(x,t)=A \exp{\left({-\frac{i}{\hbar}\left({E' t-px}\right) }
\right)} , $$

\noindent
with

$$E' \approx E\left({1+\frac{E^2\tau^2}{6\hbar^2}}\right)  .$$

A difference of the order of $\tau^2$ is observed between the energy values of the photons
in the continuous and discrete approaches. The general solution is given by a linear combination
of the eigenfunctions found. {\em A priori}, the value of the chronon for the particle is
not known. The time dependent exponential term in the expressions above leads to
an upper limit for the allowed energy, which is given by $E\leq\hbar/\tau$. We could
suppose that the value of the chronon for this `photon' is of about the
fundamental time interval of
the electromagnetic interactions, around $10^{-9}\;{\rm s}$ , resulting in a critical value
of  approximately  $6.6 \; {\rm keV}$, which is a very low limit. A smaller chronon should
increase this limit but, if there is any generality in the classical expression obtained for
the electron, we should expect a larger value for this massless particle.

Whether, instead of a `photon' we consider a scalar `neutrino' , taking for the value of the
chronon $\tau \sim10^{-13}\;{\rm s}$ ---typical time for the weak decay---, the limit for the energy
associated with the eigenfunctions is now approximately $0.007\;{\rm eV}$. This means that in the composition
of the wave packet describing this particle the only contribution comes from eigenfunctions, the energy of which
is below that limit.

\begin{figure}
\begin{center}
 \scalebox{0.8}{\includegraphics{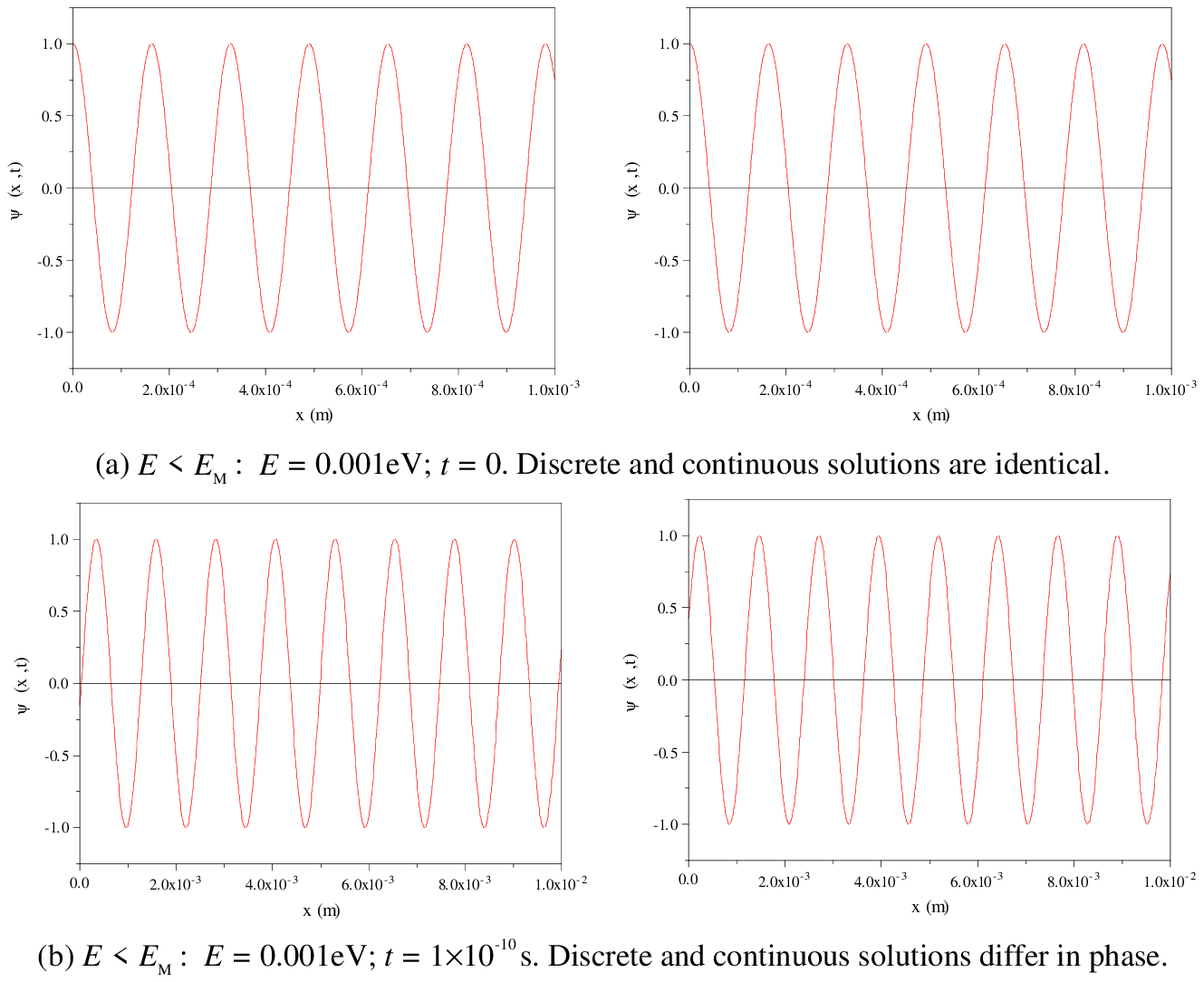}}
 \caption{Solution of the discretized Klein--Gordon equation, when the energy
is smaller than the critical limit, depicted for different values
of energy and time.}
\end{center}
 \label{fig:5-5}
\end{figure}

\begin{figure}
\begin{center}
 \scalebox{0.8}{\includegraphics{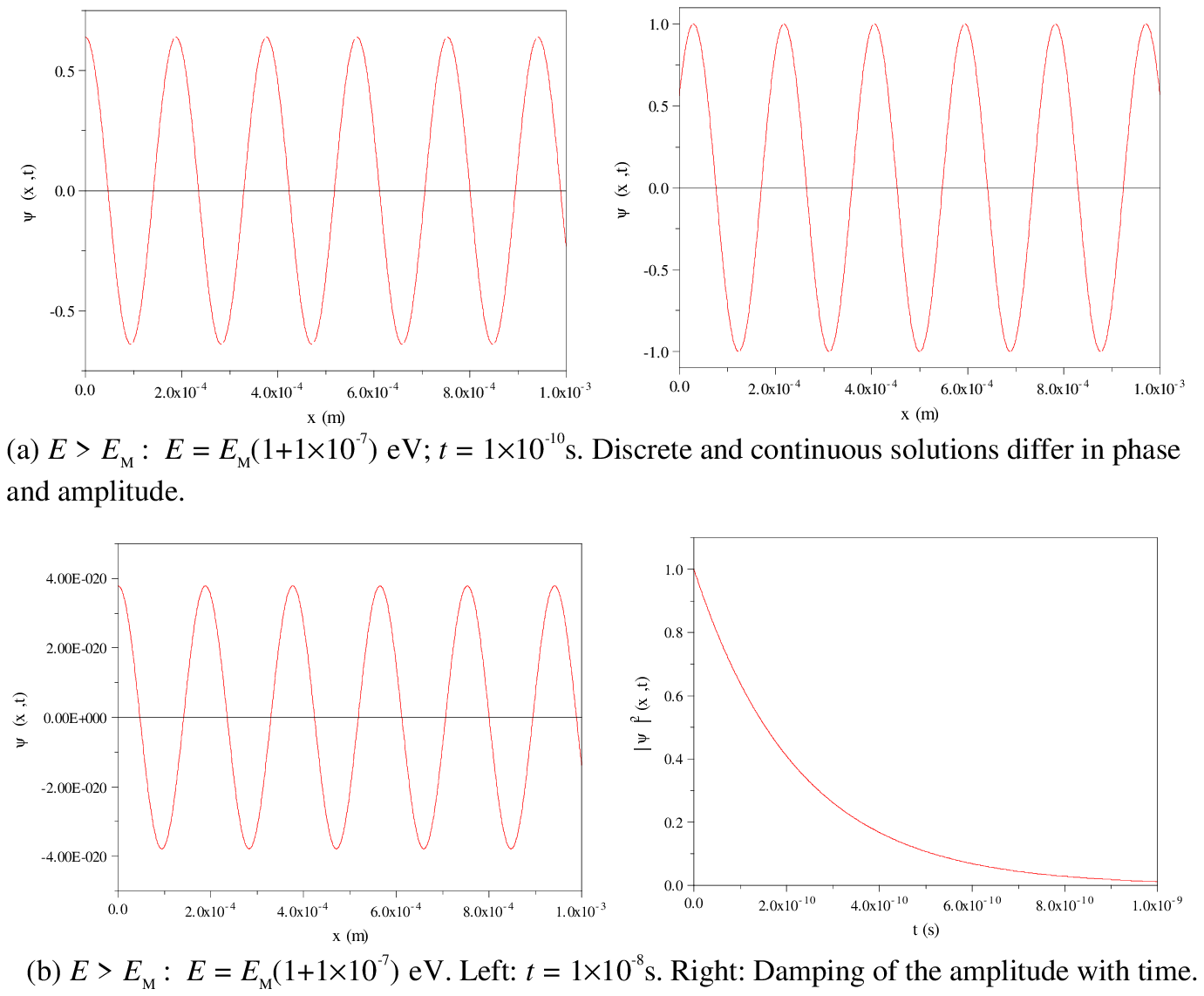}}
 \caption{Solution of the discretized Klein--Gordon equation when the energy
is larger than the critical limit, depicted for different values of energy
and time. In this case the amplitude decay is very fast.}
\end{center}
\label{fig:5-6}
\end{figure}

The eigenfunctions obtained for the Hamiltonian considered are ``plane waves"
solutions. The dependence of these solutions on energy and time is displayed
in figures {\ref{fig:5-5}} and {\ref{fig:5-6}}. For smaller values of
$\tau$ the decay of the modes with energy above the maximum is faster.

Apparently, it seems to be possible to determine a limiting value for the chronon
starting from the uncertainty relations. This could be obtained, when
describing particles, using the expression

$$\tau < \frac{\hbar}{2m_oc^2}$$

\noindent
that provides for the electron a maximum limit given by $6.4\times10^{-22}\;{\rm s}$.
However, this value is two degrees of magnitude larger than the classical
value of the chronon for the electron, which is a considerable difference.
For a complex system it is also possible to make use of this relation, as
we are going to mention ahead.

We also have to consider the conditions a `photon' must be supplied with,
in order
to be described by the symmetric equation. [For the electron, it seems clear
that not irradiating in a bound state ---which is imposed by QM---
implies the adoption of the symmetrical equation]. For the photon, as for
a free particle,
when using the retarded form of the Klein--Gordon equation, a solution is
also obtained wherein the highest frequencies decay faster than the lowest
ones. There is
always a tendency in the sense that the lowest frequencies prevail. If we are
allowed
to assign a physical meaning to such a discretized Klein--Gordon equation,
we are also
allowed to think that, the farther the light source, the more the spectrum
of the emitted
light will be shifted for the largest wavelengths, even if the source is at
rest with respect to to the observer. Thus, we could obtain a {\em red shift}
effect as a consequence
of the introduction of the chronon which could be used in the construction of
a `tired-light' theory.

Finally, we have to keep in mind that the discretization considered for the Klein--Gordon
equation does not follow exactly the same procedure which led to the discretized
Schr\"{o}dinger
equation, since it is a relativistic invariant equation. We did not change
the proper time
but the time coordinate itself into the discretized form. We considered a
discretized version of
the hamiltonian operator by applying the transformations:

$$p \hspace{1.0cm}\longrightarrow \hspace{1.0cm} \frac{\hbar}{i} \nabla , $$

$$H \hspace{1.0cm}\longrightarrow \hspace{1.0cm} i\hbar \Delta ,$$

\noindent
with $\Delta$ as defined in Subsection {\ref{sec:heisen}}, on the hamiltonian of a relativistic
free particle,

$$H = \sqrt{p^2 c^2 + m^2 c^4}$$

\noindent
as usual in the continuous case.

\subsection{Time evolution of the position and momentum\hfill\break
Operators: The harmonic oscillator}

It is possible to apply the discretized equations to determine the time evolution of the position and
momentum operators, which is rather interesting for the description of the simple harmonic
oscillator. In order to do that, we use the discretized form of the Heisenberg
equations which,
in the symmetric case, can be obtained by a direct discretization of the
continuous equation.
Starting from this equation we determine the coupled Heisenberg equations
for the two operators:

\begin{equation}
\frac{\hat{\bf p}(t+\tau)-\hat{\bf p}(t-\tau)}{2 \tau}=-m\omega^2\hat{\bf x}(t) ,
\end{equation}

\begin{equation}
\frac{\hat{\bf x}(t+\tau)-\hat{\bf x}(t-\tau)}{2 \tau}=\frac{1}{m} \hat{\bf p}(t) .
\end{equation}

Such coupled equations yield two finite--difference equations of second order,
the general
solutions of which are easily obtained. The most immediate way to determine the evolution of these
operators is to use the creation and annihilation operators. Keeping the Heisenberg equation
and remembering that for the harmonic oscillator we have $\hat{H}=\omega\left({\hat{A}^\dagger
\hat{A}+\frac{1}{2}}\right)$, we get for the symmetric case:

\begin{equation}
\frac{\hat{A}(t+\tau)-\hat{A}(t-\tau)}{2\tau}=-i\omega \hat{A}(t) ,
\end{equation}

\begin{equation}
\frac{\hat{A}^{\dagger}(t+\tau)-\hat{A}^{\dagger}(t-\tau)}{2\tau}=
i \omega \hat{A}^{\dagger}(t) ,
\end{equation}

\noindent
such that

\begin{equation}
\hat{A}(t)=\hat{A}(0) \exp{\left({-i\frac{t}{\tau}\sin^{-1}{(\omega \tau)}}\right)} ,
\end{equation}

\begin{equation}
\hat{A}^{\dagger}(t)=\hat{A}^{\dagger}(0) \exp{\left({i\frac{t}{\tau}\sin^{-1}{(\omega \tau)}}\right)} ,
\end{equation}

\noindent
where we used the fact that, for $t=0$, the Heisenberg and Schr\"{o}dinger pictures are
equivalent: $\hat{A}(t=0)=\hat{A}=\hat{A}(0)$ and $\hat{A}^{\dagger}(t=0)=
\hat{A}^{\dagger}=\hat{A}^{\dagger}(0)$, with  $\hat{A}$ and $\hat{A}^{\dagger}$
independent of time. In order to obtain these equations we considered that, for the
non-relativistic case, there is neither creation nor annihilation of particles, such that we
can impose restrictions on the frequencies in the phase term of the operators. For the
creation operators, e.g., the terms with negative frequencies ---associated
with antiparticles---  are discarded.

We can observe that the {\bf Number}, as well as the hamiltonian operator are not altered:

$$\hat{N}=\hat{A}^{\dagger}(t)\hat{A}(t)=\hat{A}^{\dagger}(0)\hat{A}(0) ,$$

$$\hat{H}=\hbar\omega\left({\hat{N}+\frac{1}{2}}\right)=
\hbar\omega\left({\hat{A}^{\dagger}(0)\hat{A}(0)+\frac{1}{2}}\right) .$$

\begin{figure}
\begin{center}
 \scalebox{0.7}{\includegraphics{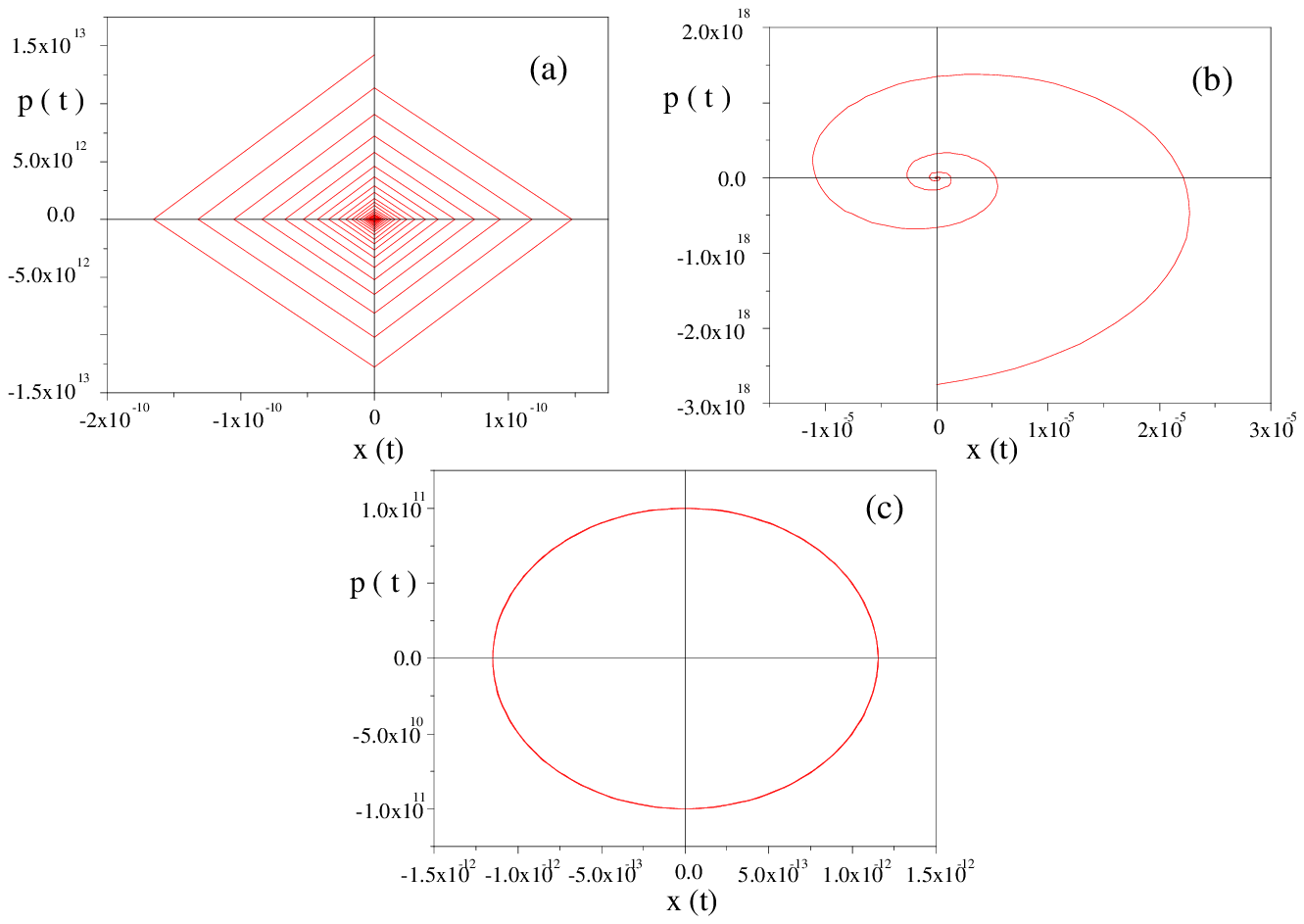}}
\caption{Phase space of the harmonic oscillator when $\omega >
\frac{1}{\tau}$ and in the discrete case, with time intervals multiples
of $\tau$: \ in case (a), time is regarded as {\em intrinsically}
discrete, so that in the picture only the points where the lines
touch one  another are meaningful, while in case (b) time is regarded as
{\em intrinsically} continuous. 
In the actually continuous case no modification is expected with respect to
the ordinary case, (c), under the present hypotheses.}
\end{center}
\label{fig:5-7}
\end{figure}

Thus, starting from these operators, we obtain for the symmetric case:

$$ \hat{\bf x}(t)=\hat{\bf x}(0) \cos{\left[{\frac{t}{\tau} \sin^{-1}{(\omega \tau)}}\right]} +
\frac{\hat{\bf p}(0)}{m\omega}\sin{\left[{\frac{t}{\tau}\sin^{-1}{(\omega \tau)}}\right]} $$

$$ \hat{\bf p}(t)=\hat{\bf p}(0) \cos{\left[{\frac{t}{\tau} \sin^{-1}{(\omega \tau)}}\right]} -
m\omega\hat{\bf x}(0)\sin{\left[{\frac{t}{\tau}\sin^{-1}{(\omega \tau)}}\right]} $$

\noindent
which differ from the continuous case since the frequency $\omega$ here is
replaced
by a new frequency $\frac{1}{\tau}\sin^{-1}{(\omega \tau)}$ that, for $\tau\rightarrow0$,
tends to the continuous one. Also, there is now an upper limit for the possible oscillation
frequencies given by  $\omega \leq {1 / \tau}$. Above this frequency the motion becomes
unstable, as it can be observed in figure {\ref{fig:5-7}}.

The existence of a maximum limit for the frequency is equivalent to an upper limit for
the energy eigenvalues given by $E_n=(n+\frac{1}{2})\hbar \omega \leq \hbar/\tau$,
which is equal to the upper limit obtained using  Schr\"{o}dinger' s picture. Since

$$\frac{t}{\tau} \sin^{-1}{(\omega \tau)} \cong \omega +\frac{1}{3!}\omega^3\tau^2+
{\rm O}(\tau^4) ,$$

\noindent
the difference expected in the behaviour of the oscillator with respect to the continuous
solution is quite small. If we take, for example, the vibration frequency of the
hydrogen molecule $(H_2)$, we have that $\omega\sim 10^{14}\;{\rm Hz}$, while the
term of the second order in $\tau$ is smaller than $10^{-3}\;{\rm Hz}$ (if the analogy with
the classical theory is valid, the chronon is expected to be smaller for more massive
systems). In terms of average values we have that, for the position operator

$$ \langle\hat{\bf x}(t)\rangle = \langle\hat{\bf x}(t)\rangle_{{\rm cont}} + \frac{\omega^2\tau^2}
{3! m} t \langle\hat{\bf p}(t)\rangle \ , $$

\noindent
in which the term of order $\tau^2$ is expected to be considerably smaller than the
mean value for the continuous case. At this point, naturally, the mean values are determined
taking for the system a state made up of a superposition of stationary states. For the
stationary states $|u_n\rangle$ themselves the mean values of $\hat{\bf x}$ and
$\hat{\bf p}$ are zero.

For the retarded case the solutions can be obtained using the time
evolution operators
for the Heisenberg equation (Appendix \ref{ap:01}). As expected, decaying terms come
out. The creation and annihilation operators obtained for this case are then given by

$$\hat{A}(t)=\hat{A}(0)\left[{1+i\omega\tau+\tau^2\omega^2\xi}\right]^
{- \frac{t}{\tau}}\approx  \hat{A}(0) \exp{\left({-i \omega \tau}\right)}
\exp{\left[{-\left({\xi+\frac{1}{2}}\right) \omega^2\tau t}\right]} , $$

$$\hat{A}^{\dagger}(t)=\hat{A}^{\dagger}(0)\left[{1-i\omega\tau+\tau^2\omega^2\xi}\right]^
{- \frac{t}{\tau}}\approx  \hat{A}^{\dagger}(0) \exp{\left({i \omega \tau}\right)}
\exp{\left[{-\left({\xi+\frac{1}{2}}\right) \omega^2\tau t}\right]} , $$

\noindent
with $\xi$ being a real positive factor. The relation $(\hat{A}^{\dagger})^{\dagger}=
\hat{A}$ continues to be valid but the {\em Number} operator and, consequently,
the Hamiltonian, is now not constant anymore:

$$\hat{H}(t)=\hat{A}^{\dagger}(0)\hat{A}(0) \left[{\left({1+\omega^2 \tau^2 \xi}\right)^2
+\omega^2 \tau^2}\right]^{- \frac{t}{\tau}}=\hat{H}(0)\exp{\left[{-2 \left({\xi+\frac{1}{2}}\right)
\omega^2\tau t}\right]} ,$$

$$\hat{H}(t)=\hbar\omega\left\{{\hat{A}^{\dagger}(0)\hat{A}(0)\left[{
\left({1+\omega^2 \tau^2 \xi}\right)^2 +\omega^2 \tau^2}\right]^{- \frac{t}{\tau}}
+\frac{1}{2}}\right\}. $$

\begin{figure}
\begin{center}
 \scalebox{0.8}{\includegraphics{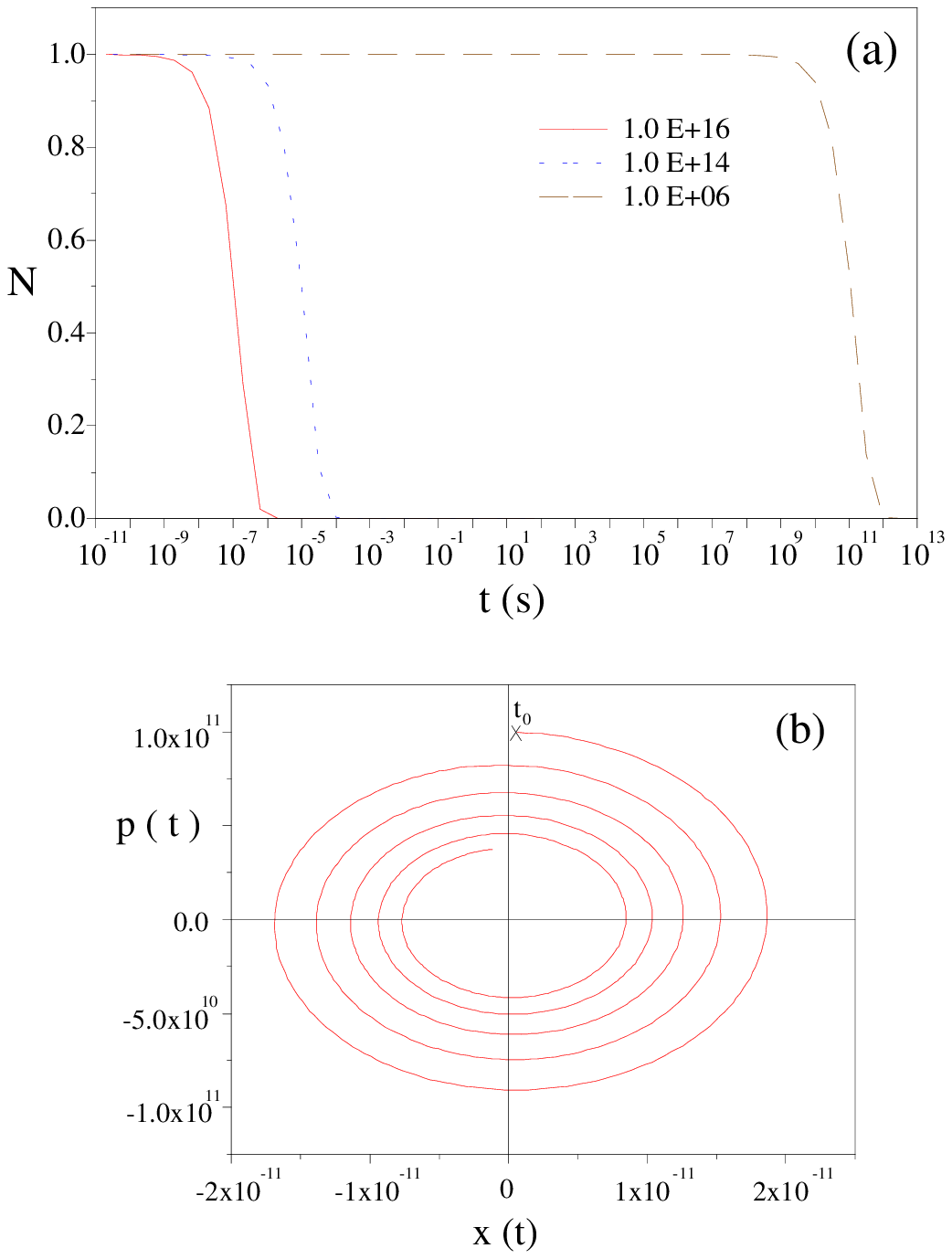}}
\caption{(a) Damping factors associated with the {\em Number}, operator
calculated for a few frequencies. \
(b) Damping of the oscillations for the harmonic oscillator described by
the {\em retarded} equation.}
\end{center}
\label{fig:5-8}
\end{figure}

Taking into account the terms to the second order in $\tau$, we get that the oscillation
frequencies also decays with time.  These results are consistent with the fact that the
system is emitting radiation, with the consequent reduction of its total energy. However,
it is remarkable that the energy of the quanta associated with the creation and annihilation
operators is not constant, even with a very tiny variation rate. In the same way, when we
calculate the position and momentum operators a damping factor is obtained.
Figure {\ref{fig:5-8}\/-a} shows the strange damping factor associated with the {\em Number}
operator. It can be observed that this damping occurs within a period of time which
is characteristic for each frequency, being slower and postponed for lower frequencies.
Figure {\ref {fig:5-8}\/-b} shows the dampening of the oscillations as described by the retarded
equation. Once the expressions for the position and momentum operators are determined,
we obtain that, to first order in $\tau$,

$$ \langle\hat{\bf x}(t)\rangle = \left\{{\hat{\bf x}(0) \cos{(\omega \tau)} + \frac{\hat{\bf p}(0)}
{m\omega} \sin{(\omega \tau)} }\right\} \exp{\left[{-\left({\xi+\frac{1}{2}}\right) \omega^2\tau t}
\right]} ,$$

$$ \langle\hat{\bf x}(t)\rangle = \langle\hat{\bf x}(t)\rangle_{{\rm cont}}
\exp{\left[{-\left({\xi+\frac{1}{2}}\right) \omega^2\tau t}\right]} .$$

Taking into account the higher order terms, we can observe a small variation in the oscillation
frequency just as observed in the symmetrical case. The introduction of time independent
perturbations does not cause any additional variations apart from those found even in the
continuous case. It must be pointed out that the results obtained with the procedure above
are in agreement with those obtained following Schr\"{o}dinger's picture.

\subsection{Hydrogen atom}

The hydrogen atom is basically a system made up of two particles attracting each other
through coulombian force, which is therefore inversely proportional to the square of the
distance between them. The basic Hamiltonian is given by

\begin{equation}
\hat{H}_0=\frac{\hat{\bf P}^2}{2 \mu}-\frac{e^2}{R},
\end{equation}

\noindent
and is composed of the kinetic energy of the atom in the centre-of-mass frame, and of the
coulombian electrostatic potential ($\mu$ is the reduced mass of the system electron-proton).
A more complete description is obtained by adding correction terms (fine structure) to the
Hamiltonian, including relativistic effects such as the variation of the electron mass with
velocity and the coupling of the intrinsic  magnetic moment of the electron with the
magnetic field due to its orbit (spin--orbit coupling). Besides, there are
also the hyperfine (hf)
corrections which appear due to the interaction of the electron with the
intrinsic magnetic
moment of the proton and, finally, the {\em Lamb shift}, due to the
interaction of the
electron with the fluctuations of the quantized electromagnetic field.
The Hamiltonian
can finally be written as\cite{COHEN}

\begin{equation}
\hat{H}_{\rm I}=m_{\rm e}c^2+\hat{H}_0-\frac{\hat{\bf P}^4}{m_{\rm e}^2 c^2 R^3}\;\hat{\bf L}
\cdot\hat{\bf S}+\hat{H}_{{\rm hf}}+\hat{H}_{{\rm Lamb}}  .
\end{equation}

The introduction of the magnetic moment of the nucleus through the hyperfine correction
causes the total angular momentum to be ${\bf F}={\bf J}+{\bf I}$. The Hamiltonian does
not depend explicitly on time such that, for the symmetric Schr\"{o}dinger equation

\begin{equation}
i{\hbar  \over {2\tau }}\left[ {\Psi \left( {\bf x,t+\tau } \right)-\Psi \left(
{\bf x,t-\tau } \right)} \right]=\hat{H}_{\rm I}\Psi \left( {\bf x,t} \right),
\end{equation}

\noindent
we obtain, using the separation of variables, the following uncoupled equations:

$$\hat{H}_{\rm I} \Phi({\bf x})=E \Phi({\bf x}) $$

$$i{\hbar  \over {2\tau }}\left[ {T \left( {t+\tau } \right)-T \left(
{t-\tau } \right)} \right]=\hat{H}_{\rm I} T \left( {t} \right), $$

\noindent
with the general solution

\begin{equation}
\Psi \left({{\bf x},t}\right)=\Phi \left({{\bf x}}\right) \exp{\left[{
-i\frac{t}{\tau} \sin^{-1}{\left({\frac{\tau E}{\hbar}}\right)}}\right]} .
\end{equation}

The difference related to the continuous case appears only in those aspects
involving  the time evolution of the states. Since the Hamiltonian is time
independent, its eigenvalues are exactly the same as those obtained in the
continuous case\cite{COHEN}

$$ E_{(n,j)}\approx m_0c^2-\frac{1}{2n^2}m_{\rm e} c^2\alpha^2-\frac{m_{\rm e}c^2}{2n^4}
\left({\frac{n}{j+\frac{1}{2}}-\frac{3}{4}}\right)\alpha^4+E_{{\rm hf}}+E_{{\rm Lamb}} . $$

\noindent
A situation where a difference between the two cases can appear  is when
taking into account the probabilities of transition between the eigenstates
for an
atom submitted to a time dependent potential. In the discrete approach it is
possible to use the method of the equivalent Hamiltonian in order to obtain the
transition probabilities. As mentioned previously (Subsection \ref{time}),
the problem is treated
using the conventional approximate methods for time dependent perturbations.

If we consider, for example, the non-relativistic interaction of an atom
with an electromagnetic field
described by the vector potential ${\bf A}({\bf x},t)$, we have for the low
intensity limit, in the Coulomb gauge, the Hamiltonian

\begin{equation}
\hat{H}(t) = \hat{H}_{\rm I} - \hat{V}(t) =  \hat{H}_{\rm I} -\frac{e}{m_{\rm e}c}
\hat{\bf A}\left({\hat{\bf R},t}\right) \cdot \hat{\bf P} ,
\end{equation}

\noindent
where the potential term is taken as being the perturbation. If we consider that the
potential describes a monochromatic field of a plane wave, then

\begin{equation}
{\bf A}\left({{\bf x},t}\right) = A_0 \; {\hat{{\epsilonbf}}} \, \left[{\exp{\left({i\omega\frac{
\hat{\bf n}\cdot{\bf x}}{c}-i\omega t}\right)}+\exp{\left({-i\omega\frac{
\hat{\bf n}\cdot{\bf x}}{c}+i\omega t}\right)}}\right]
\end{equation}

\noindent
where ${\hat{{\epsilonbf}}}$ is the linear polarization of the field and $\hat{\bf n}$ is
the propagation direction. The term depending on $(-i\omega t)$ corresponds to the
absorption of a quantum of radiation $\hbar\omega$ and the  $(i\omega t)$ term to
stimulated emission. Let us assume that the system is initially in an eigenstate
$|\Phi_{i}\rangle$ of the time independent Hamiltonian. Keeping only the perturbations
to the first order in $\hat{V}(t)$, we obtain that

$$c_n^{1}(t)=-\frac{i}{\hbar}\int_0^t \exp{\left({i\omega_{ni}t'}\right)}
V_{ni}(t') \drm t' \ ,$$

\noindent
where $\omega_{ni}$ in the discrete case is given by

$$\omega_{ni} = \frac{1}{\tau} \left[{\sin^{-1}{\left({\frac{\tau E_n}{\hbar}}\right)}
-\sin^{-1}{\left({\frac{\tau E_i}{\hbar}}\right)}}\right] . $$

\noindent
Working with the absorption term, we get by contrast that

$$c_n^{(1)}(t) = \frac{i e A_0}{m_{\rm e}^2 c \hbar} {\left\langle{\Phi_n | e^{i \omega \hat{\bf n} \cdot
{\bf x} /c} ({{\hat{{\epsilonbf}}} \cdot {\bf p}}) | \Phi_{i}}\right\rangle}
\int_0^t \exp{\left[{i(\omega_{ni}-\omega)t'}\right]} \drm t' \ .$$

\noindent
Thus, the probability of transition from the initial state $|\Phi_{i}\rangle$  to the final state
$|\Phi_{f}\rangle$  is given by

$$
P_{fi}(t)=\left|{c_{f}^{(1)}(t)}\right|^2 = \frac{e^2 |A_0|^2}{m_{\rm e}^2c^2\hbar^2}
\left| {\left\langle {\Phi_{f} | e^{i \omega \hat{\bf n} \cdot
{\bf x} /c} ({{\hat{{\epsilonbf}}} \cdot {\bf p}} ) | \Phi_{i}}
\right\rangle}\right|^2
\left| {\int_0^t \exp{\left[{i(\omega_{fi}-\omega)t'}\right]} \drm t'}
\right|^2 , $$

\noindent
or

$$P_{fi}(t)= \frac{4e^2 |A_0|^2}{m_{\rm e}^2c^2\hbar^2} \left|{\left\langle{\Phi_{f} | e^{i \omega
\hat{\bf n} \cdot {\bf x} /c} ({{\hat{{\epsilonbf}}} \cdot {\bf p}} ) | \Phi_{i}}\right\rangle}\right|^2
\frac{\sin^2{[(\omega_{fi}-\omega)t/2]}}{(\omega_{fi}-\omega)^2}  ,$$

\noindent
so that the determination of the matrix elements of the spacial term, using the electric dipole
approximation, provides the selection rules for the transitions. What is remarkable in this
expression is the presence of a {\em resonance} showing a larger probability for the
transition when

\begin{equation}
\omega = \omega_{fi} = \frac{1}{\tau}  \left[{\sin^{-1}{\left({\frac{\tau E_{f}}{\hbar}}\right)}
-\sin^{-1}{\left({\frac{\tau E_{i}}{\hbar}}\right)}}\right] .
\end{equation}

This expression is formally different from the one obtained for the continuous approach.
When we expand this expression in powers of $\tau$, we get that

\begin{equation}
\omega \approx \frac{E_{f} - E_{i}}{\hbar}+\frac{1}{6}\frac{E_{f}^3-E_{i}^3}{\hbar^3}\/\tau^2 .
\end{equation}

\noindent
The first term supplies the Bohr frequencies as in the continuous case; the second,
the deviation in the frequencies caused by the introduction of the time discretization:

$$\Delta\omega_{fi}=\frac{1}{6}\frac{E_{f}^3-E_{i}^3}{\hbar^3}\/\tau^2 .$$

If we consider the chronon of the classical electron, $\tau \approx 6.26\times10^{-24}\;{\rm s}$,
it is possible to estimate the deviation in the frequency due to the time discretization. Then,
for the hydrogen atom,

$$\Delta\omega_{\rm fi} \approx 2.289\times10^{-2} \left({E_{\rm f}^3 - E_{\rm i}^3}\right) .  $$

\noindent
If we take into account, for example, the transitions corresponding to the first lines of
the series of Lyman and Balmer, i.e., of the non-disturbed states $n=n_{\rm i}\rightarrow n=n_{\rm f}$,
we have

\begin{center}
\begin{tabular}{|c|c|c|c|c|} \hline \hline
${\bf n}_{\rm i}$ & ${\bf n}_{\rm f}$ & ${\Delta E}\;({\rm eV})$  & ${\nu}\;({\rm Hz})$ & ${\Delta \nu}_{\rm D}\;({\rm Hz})$ \\ \hline
1 & 2 & 10.2 & $2.465\times10^{15}$ & $ \sim 10$ \\
1 & 3 & 12.1 & $2.922\times10^{14} $ & $ \sim 10 $ \\
1 & 4 & 12.75 & $3.082\times10^{14} $ & $ \sim 10$ \\
2 & 3 & 1.89 & $4.566\times10^{14} $ & $ < 1$ \\ \hline \hline
\end{tabular}
\end{center}

\noindent
where $\Delta E$ is the difference of energy between the states, $\nu$ is the frequency of the
photon emitted in the transition and $\Delta\nu_{\rm D}$ is the frequency
deviation due to the discretization.
Such deviation is always very tiny. We must remember that the hyperfine corrections
and those due to the Lamb shift are of order of a Gigahertz. For the transition $n=1 \rightarrow n=2$,
e.g., the correction due to the Lamb shift is approximately 1.06 GHz.

Larger deviations caused by the discretization occurs for mono-electronic
atoms with larger
atomic numbers. For the first transition the deviation is of approximately 90 Hz for the
$^2He$, 1.1 {\rm kHz} for the $^3Li$, and 420 {\rm kHz} for the $^6C$.  However, these deviations
are still quite smaller than that one due to the Lamb shift. That is also the case for the
{\em muonic} atoms. For a muonic atom with a proton as nucleus, using for the
chronon a value derived from the classical expression for the electron
($\tau_{\mu}=3.03\times10^{-26} \; {\rm s}$) the deviation is of about 1.4 kHz for the
transition $n=1\rightarrow n=2$. For that transition the frequency of the emitted radiation
is approximately $4.58\times10^{17}$ Hz.

For the {\em retarded} equation, a difference with respect to the symmetrical
case is present in the
time evolution of the states. The procedure is identical to the one used
above and the general solution is now given by

$$
\Psi{\left({{\bf x},t}\right)}=\Phi({\bf x})\left[{1+i\frac{\tau E}{\hbar}}
\right]^{-t/\tau} , $$

\noindent
so that the transitions now occur with frequencies given by

\begin{equation}
\omega = \omega_{fi} = \frac{-i\hbar}{\tau}  \left[{\ln{\left({1+\frac{i\tau E_{f}}{\hbar}}\right)}
-\ln{\left({1+\frac{i\tau E_{i}}{\hbar}}\right)}}\right] .
\end{equation}

As results from the characteristics of the retarded equation, this is a
complex frequency. The real component of such frequency can be
approximated by

$$
{\rm Re}(\omega_{fi})\approx \frac{E_{f}-E_{i}}{\hbar}+\frac{1}{3}
\frac{E_{\rm f}^3-E_{\rm i}^3}{\hbar^3} \tau^2, $$

\noindent
where the first term is the expression for the continuous case. For the
particular transition $n=1\rightarrow n=2$ we have that the deviation
due to the discretization is of about 18 Hz.

The imaginary component, on the other hand, can be approximated by

$$
{\rm Im}(\omega_{\rm fi})\approx -\frac{i}{2} \frac{E_{\rm f}^2-E_{\rm i}^2}{\hbar^2} \tau .
$$

\begin{figure}
\begin{center}
 \scalebox{0.8}{\includegraphics{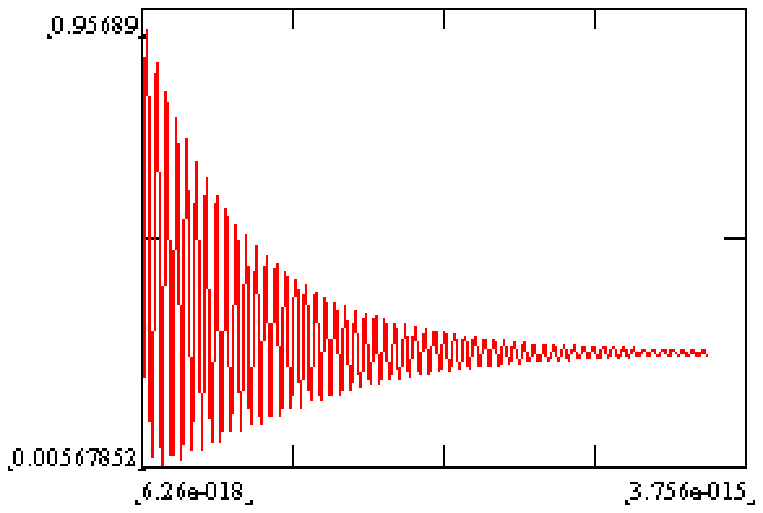}}
\caption{Behaviour of the time dependent component of the transition
probability.}
\end{center}
\label{fig:5-9}
\end{figure}

In the expression for the probability of transition we have the module
of an integral involving the time dependency of the general solution.
In this case, the characteristic damping causes the probability to tend
to a fixed, non zero value. An example of such behaviour is shown in
figure {\ref{fig:5-9}}, which shows the variation of the time dependent
term between an initial instant $t_0=0$ and some hundred chronons later.
In order to observe the decay of the amplitude factor we have used a larger
value for the chronon, of about $10^{-18}$ s. When the chronon is of the
order of the one we have been considering for the electron the decay
is slower.

As it can be observed, the effect of the time discretization on the
emission spectrum of the hydrogen is extremely small. Using the expressions
obtained above we can estimate that, in order that the effect of the time
discretization is of the same order of the Lamb shift, the chronon
associated with the electron should be of about $10^{-18}$~s, far above
the classical value (but close to the typical interval of the
electromagnetic interactions). \ In any case, it should be remembered that
the Lamb shift measurements do not seem to be in full agreement\cite{LP} with
quantum electrodynamics.

\vspace{1.in}

Concluding this Subsection, it is worthwhile to remark that, for a time
independent Hamiltonian, the outputs obtained in the discrete formalism
using the symmetric equation are very similar to those from the
continuous case. For such Hamiltonians, the effect of the discretization
appears basically in the frequencies associated with the time dependent
term of the wave function. As already observed, the difference in the
time dependency is of the kind

$$
\exp{\left[{-iE_n(t-t_0)/\hbar}\right]} \hspace{0.5cm} \longrightarrow
\hspace{0.5cm} \exp{\left[{-i\sin^{-1}{\left({\frac{\tau E_n}{\hbar}}\right)}(t-t_0)/\tau}\right]}
$$

The discretization causes a change in the phase of the eigenstate, which can
be quite large. The eigenfunctions individually describe stationary
states, so that the time evolution appears when we have a linear combination
of such functions, this way describing the state of the system. This
state evolves according to

$$
\left|{\Psi(t)}\right\rangle = \sum_n c_n(0) \exp{\left[{-i\sin^{-1}{\left({\frac{\tau E_n}{\hbar}}
\right)}(t-t_0)/\tau}\right]}\left|{\phi_n}\right\rangle ,
$$

\noindent
considering that ${H}|\phi_n\rangle = E_n |\phi_n\rangle$ is the
eigenvalue equation associated with the Hamiltonian.

When the stationary states of a particle under, e.g.,  one-dimensional squared
potentials are studied, the same reflection and transmission coefficients
and the same tunnel effect are obtained, since they are calculated starting from the
stationary states.
When we consider a linear superposition of these stationary states,
building a wave packet, the time dependent terms have to be taken into account,
resulting in some differences with respect to the continuous case. Some attempts
have been carried out in order to find out significant measurable differences
between the two formalisms\cite{WOL87,WOL89} but no encouraging
case has been found yet.

\section{Density Operators and the ``Coarse Graining" Hypothesis}
\label{sec:5}

\subsection{The ``coarse graining" hypothesis}

First of all, it is convenient to present a brief review of some topics
related to the introduction of the {\em coarse grained} description of a
physical system. This hypothesis is then going to be used to obtain a
discretized form of the Liouville equation, which represents the evolution law
of the density operators in the usual QM.

An important point to be remarked is that the introduction of a fundamental
interval of time is perfectly compatible with a coarse grained description.
The basic premise of such description, in statistical physics, is the impossibility
of a precise determination of the position and momentum of each particle
forming the system, in a certain instant of time. Let us consider, for the sake of
simplicity, a system composed by $N$ similar pointlike particles, each of
them with three degrees of freedom described by the coordinates $(q_1, q_2,q_3)$.
We can associate with this ensemble of particles an individual phase space (named
$\mu$\/-\/space) defined by the six coordinates $(q_1, q_2, q_3; p_1, p_2, p_3)$
so that the system as a whole is represented by a crowd of points in this space.

Since the macroscopic observation is unable to precisely determine the six coordinates
for each particle, let us assume that it is possible only to know if a given
particle has its coordinates inside the intervals $(q_i+dq_i)$ and $(p_i+dp_i)$,
with $i=1,2,3$. In order to describe the state of the system in the $\mu$\/-\/space,
we divide it into cells corresponding to the macroscopic uncertainties $\delta q_i$
and $\delta p_i$, each one occupying in the $\mu$ -space a volume

\begin{equation}
w_i = \delta q_1\; \delta q_2\; \delta q_3\; \delta p_1\; \delta p_2\; \delta p_3\; .
\end{equation}

These cells must be sufficiently small related to the macroscopically measurable
dimensions but also sufficiently large to contain a great number of particles.

When considering the system as a whole, its macroscopic state is given by a collection
of points ${n_i}$ corresponding to the number of particles inside each cell. Now,
if we take into account the 6N-dimensional phase space $\Gamma$, in which each of the states
assumed by the system is represented by a point, to each configuration ${n_i}$ corresponds
in $\Gamma$ a cell with volume given by

$$
\left({\delta V}\right)_{\Gamma} = \prod_{n=1}^N \left({w_i}\right)^{n_i}  .
$$

Considering that the permutation of the particles inside the cells of the $\Gamma$ space
does not change the macroscopic state of the system, then to each collection of numbers
${n_i}$ corresponds a volume $\Omega_n$ in the $\Gamma$ space
\footnote{Jancel\cite{JANC} calls it a {\em star}.}
given by

$$
W\left({\Omega_n}\right)=\frac{N!}{\prod_i n_i!}\prod_i \left({w_i}\right)^{n_i}
\hspace{1.in} \left({\sum_i n_i=N}\right) .
$$

The state of the system is determined by the star occupied by the representative
point of the system in the $\Gamma$ space. This way, macroscopically, it is only
possible to distinguish in which star the system is, such that any point in this star
corresponds to a same macroscopic state. When we consider a system which is not
in equilibrium, a change in its macroscopic state can only be observed when the
point describing the system changes star. The crossing time is small but finite.
During this period of time the macroscopic state of the system does not change
notwithstanding its microscopic state is continuously changing.

Thus, from the point of view of statistical physics, the introduction of a fundamental
interval of time appears in a very natural way. That is still more significant when
we remember that the predictions of QM are always obtained as
mean values of observables. The uncertainty relations, according to the usual
interpretation of QM ---the Copenhagen interpretation---, are
independent of the arguments above. If we accept that they play a fundamental
role in the microscopic world ---and this is postulated by Copenhagen---, then
the concept of chronon, as a fundamental interval of time, must be related to
them.

\subsection{Discretized Liouville equation and\hfill\break
the time--energy uncertainty relation}

 An attempt to set up a relationship between the chronon and the time--energy
uncertainty relation has been put forward by Bonifacio (1983)\cite{BONI},
extending the {\em coarse graining} hypothesis to the time coordinate. In the
conventional QM the density operator evolves according to
the Liouville--von~Neumann equation

\begin{equation}
\frac{\partial \hat{\rho}}{\partial t}=-i{\cal L} \hat{\rho}(t) =
-\frac{i}{\hbar}\: \left[{\hat{H},\hat{\rho}}\right] ,
\label{eq:liouv}
\end{equation}

\noindent
where ${\cal L}$ is the Liouville operator. One can immediately observe that,
if ${H}$ is time independent, the solution is given by

\begin{equation}
\hat{\rho}(T)=\exp{\left({-i\frac{{H}}{\hbar}}\right)}
 \; \hat{\rho}(0) \; \exp{\left({i\frac{{H}}{\hbar}}\right)} ,
\end{equation}

\noindent
which gives the time evolution of the density operator starting from an initial
time $t_0$, such that $T=t-t_0$ is the evolution time.

When we build a coarse grained description of the time evolution, by introducing
a graining of value $\tau$ such that the evolution time is now given by
$T=k\tau\; \left({k=1,2,\ldots,\infty}\right)$, we have that the resulting
density operator $\rho$ does not satisfy the continuous equation (\ref{eq:liouv})
but a discretized form of it given by

\begin{equation}
\frac{\hat{\rho}(t)-\hat{\rho}(t-\tau)}{\tau} \ = \ -i{\cal L}\hat{\rho}(t) ,
\label{eq:liouvd}
\end{equation}

\noindent
with $t=k\tau$, which reduces to the Liouville--von~Neumann equation when
$\tau \rightarrow 0$. In the energy representation ${|n\rangle}$, once satisfied
certain conditions which ensure that $\rho (k)$ is a density operator, we
have that (\ref{eq:liouvd}) rules for $\rho$ an evolution which preserves trace,
obeys the semigroup law and is an irreversible evolution towards a stationary
diagonal form. In other words, we observe a {\em reduction of state} in the same
sense as in the measurement problem of QM. This {\em reduction}
is not instantaneous and depends on the characteristic value $\tau$:

$$ \rho (t) \vspace{1.0cm} \stackrel{t\rightarrow 0}{\rightarrow} \vspace{1.0cm}
\sum_n \rho_{nn}(0) |n\rangle \langle n| \ .$$
It is important to observe that the non diagonal terms tend exponentially to zero
according to a factor which, to the first order, is given by

$$ \exp{\left|{\frac{-\omega_{nm}^2 \tau t}{2}}\right|} .$$

\noindent
Thus, the reduction to the diagonal form occurs provided we have a finite value
for $\tau$, no matter how small, and provided we do not have $\omega_{nm} \tau \ll 1$
for every $n$,$m$, where $\omega_{nm} = (E_n-E_m)/\hbar$ are the transition
frequencies between the different energy eigenstates. This latter condition is always
satisfied for systems not bounded.

These results, together with an analysis of the discrete Heisenberg equation
defined in terms of the average values of observables

$$\bar{A}(t)={\rm Tr} \left({{\rho}(t)\;{\hat A}}\right )$$

\noindent
in the {\em coarse grained} description, suggest an interpretation of $\tau$
in terms of the uncertainty relation $\Delta E\Delta t\geq \hbar/2$ such that
$\tau$ is a characteristic interval of time satisfying the inequality

\begin{equation}
\tau \geq \tau_E \equiv \frac{\hbar}{2 \Delta E} \hspace{1.cm}{\rm with}
\hspace{1.0cm} \Delta E = \sqrt{\left\langle{H^2}\right\rangle - \left\langle
{H} \right\rangle^2} \; ,
\end{equation}

\noindent
so that the mathematical meaning of the time-energy uncertainty relation
is that of fixing a lower limit for the time interval within which the time
evolution can be described. Thus, ``\ldots\ the coarse grained
irreversibility would become
a necessary consequence of an intrinsic impossibility to give an instantaneous
description of time evolution due to the time--energy uncertainty relation".

Since the density operator, in the energy representation, tends to a diagonal
form, it seems to be tempting to apply it to the measurement problem.
We can also observe that, even without assuming any {\em coarse graining}
of time, namely,
without using the statistical approach adopted by Bonifacio, the reduction to
a diagonal form results straightforwardly from the discrete Liouville equation
and some asymptotic conditions regarding the behaviour of the solution, once
satisfied\cite{CABO} the inequality $\omega_{nm} \tau \ll 1$. \ See also
\cite{GHIR}.

The crucial point, from which derives both the decay of the non-diagonal terms
of the density operator and the very discrete Liouville equation, is that
the time
evolution operator obtained from the {\em coarse grained} description
{\bf is not a unitary operator}. This way, the operator

\begin{equation}
\hat{V}\left({t=k\tau,t=0}\right)= \frac{1}{\left({1+\frac{i \tau
\hat{\cal L}}{\hbar}
}\right)^k}   ,
\label{eq:evoliou}
\end{equation}

\noindent
as all the non-unitary operators, does not preserve the probabilities
associated with each
of the energy eigenstates that make up the expansion of the initial state
in that basis
of eigenstates. We must recall that the appearance of non-unitary time
evolution operators is not associated with the {\em coarse grained} approach
only, since they also result from the discrete Schr\"{o}dinger equations.

\subsection{The measurement problem in quantum\hfill\break
mechanics}

Let us apply the discrete formalism introduced in the previous Subsection to the measurement
problem. Using a quite general formalization, we can describe the measurement process
taking advantage of the properties observed for the evolution of the density operator as
determined by the discrete Liouville--von~Neumann equation.\footnote{We follow closely
the description exhibited in Ballantine.\cite{BALL}}

When speaking of measurement, we have to keep in mind that, in the process,
an object ${\cal O}$, of which we want to measure a dynamic variable  $R$,
and an apparatus ${\cal A}$, which is used to perform such measurement, are
involved. Let us suppose that ${\hat R}$ is the operator associated with the
observable $R$, with an eigenvalue equation given by ${\hat R} |r\rangle =
r |r\rangle$ and defines a complete basis of eigenstates. Thus, considered
by itself, any possible state of the object can be expanded in this basis:

\begin{equation}
|\Psi\rangle_0 = \sum_r c_r |r\rangle_0 .
\end{equation}

As regards the apparatus ${\cal A}$, we are interested only in its observable
$A$, whose eigenvalues $\al$ represent the possible values indicated by a
pointer.  Besides, let its various internal quantum numbers be labelled by
an index $n$.
These internal quantum numbers are useful to specify a complete basis of
eigenvectors associated with the apparatus:

\begin{equation}
\hat{A} \left|{\alpha,n}\right\rangle_A= \alpha \left|{\alpha,n}\right\rangle_A  .
\end{equation}

Now, let us suppose that the apparatus is prepared in an initial state given
by $|0,n \rangle_A$, i.e., in the initial state the value displayed is zero.
The interaction between the two systems is introduced by means of the time evolution
operator and is such that there is a correlation between the value of {\em r}
and the measure $\alpha_r$. In order to deal with the measurement process
itself we consider a quite general situation. First of all, let us consider
the following pure state of the system {\em object} + {\em apparatus}
({\cal O} + {\cal A}):

\begin{equation}
|\Psi_n^{\rm i}\rangle = |\Psi\rangle_0 \; |0,n\rangle_A  .
\end{equation}

The evolution of this state, in the continuous description, using the evolution
operator, is given by

\begin{equation}
\hat{U}(t,t_0)|\Psi\rangle_0 |0,n\rangle_A=\sum_r c_r  \left|
{\alpha_r;r,n}\right\rangle = |\Psi_n^{\rm f}\rangle
\label{eq:pure}
\end{equation}

\noindent
which is a coherent superposition of macroscopically distinct eigenstates,
each one corresponding to a different measure $\alpha_r$. The great problem
for the Copenhagen interpretation results from the fact that it considers
the state $|\Psi_n^{\rm i}\rangle$  as associated with a single system: a pure state
provides a complete and exhaustive description of an individual system. Thus,
the coherent superposition above describes a single system so that, at the end
of the interaction which settles the measurement, the display should not
show a well-defined output since (\ref{eq:pure}) describes a system
which is a superposition of all its possible states.

However, we know from experience that the apparatus always displays
a single value as the output of the measurement. It is this disagreement
between observation and the description provided by the formalism, when
interpreted according to Copenhagen, which results in the necessity of
introducing the postulate of the reduction of the vector state

$$ |\Psi_n^{\rm f}\rangle \hspace{1.cm} \longrightarrow \hspace{1.cm}
|\alpha_{r_0}; r_0, n\rangle $$

\noindent
where $r_0$ is the value displayed by the apparatus.

This fact has been considered by many as being a problem for the usual
interpretation of quantum mechanics.\cite{WIGN,BALL}  The attempts
to find a solution, in the context of different interpretations, have been
numerous,
from the {\em Many-Worlds interpretation},
proposed by Everett and Wheeler,\cite{EVERW} to the measurement
theory by Daneri, Loinger and Prosperi,\cite{DLP,ROS} in which the reduction
of the quantum state is described as
a process triggered by the appearance of aleatory phases in the state
of the apparatus, just because of its interaction with the elementary object. The
approach introduced here is ---by contrast--- rather simpler.

As an initial state in the measurement process, let us consider a mixed
state for the composite system ${\cal O} + {\cal A}$,

\begin{equation}
\rho^{\rm i} = \sum_n C_n \left| {\Psi_n^{\rm i}} \right\rangle\left\langle {\Psi_n^{\rm i}} \right| ,
\end{equation}

\noindent
where $C_n$ is the probability associated with each of the states
$ \left|{\Psi_n^{\rm i}}\right\rangle $. Such probability is, as in the classical
physics, an `ignorance' probability, i.e., it is not intrinsic to the
system. In the continuous case, when we apply the time evolution operator
to that density operator we get a final state given by

\begin{equation}
\rho^{\rm f} = \hat{U}\rho^{\rm i}\hat{U}^{\dagger} \sum_n C_n
\left| {\Psi_n^{\rm i}} \right\rangle\left\langle {\Psi_n^{\rm i}} \right| =
\end{equation}

\begin{equation}
\rho^{\rm f} = \sum_{r_1,r_2} c_{r_1}^{\ast} c_{r_2} \sum_n C_n
\left\{ {\left| {\alpha_{r_1};r_1,n} \right\rangle\left\langle {\alpha_{r_1};r_1,n} \right|  } \right\} ,
\end{equation}

\noindent
so that the presence of non-diagonal terms corresponds to a coherent
superposition of states. In this case, the postulate of the reduction of
the quantum state is connected with the non-diagonal terms of the density
operator. It is postulated that when a measurement is carried out on the
system, the non-diagonal terms tend instantaneously to zero. Since in the
continuous case the time evolution of the state results from the application
of a unitary operator, which preserves the pure state condition $\hat{\rho}^2
=\hat{\rho}$, it is impossible to obtain the collapse of the pure state
from the action of such operator. In the diagonal form the density operator
describes an incoherent mixture of the eigenstates of  ${\hat A}$,
and the indetermination regarding the output of the measurement is a sole
consequence of our ignorance about the initial state of the system.

In the discrete case, which has the time evolution operator given by
(\ref{eq:evoliou}), with the interaction between apparatus and object
embedded in the Hamiltonian ${H}$, the situation is quite different. The main
cause of such difference lays in the fact that the time evolution operator
is not unitary. Let us consider the energy representation, describing the
eigenvalue equation of the Hamiltonian as ${H}|n\rangle=E_n |j\rangle$
so that the eigenstates $|n  \rangle$ are the states with defined energy.
From the formalism of the density matrices we know that when the operator
${\hat R}$ is diagonal in the energy representation then, when calculating
the expected value of the observable, we do not obtain the interference
terms describing the {\em quantum beats} typical of a coherent superposition
of the states $|n\rangle$.

As the time evolution operator is a function of the Hamiltonian and, therefore,
commutes with it, the basis of the energy eigenstates is also a basis for this
operator. We can now use a procedure identical to the one applied by
Bonifacio, and consider the evolution of the system in this representation.
Thus, we have that the operator $\hat{V}\left({t=k\tau,t=0}\right)$
takes the initial density operator $\rho^{\rm i}$ to a final state for which
the non-diagonal terms decay exponentially with time:

\begin{equation}
\rho_{rs}^{\rm f} = \langle r|V \left({t=k\tau,t=0}\right)|s\rangle =
\frac{\rho_{rs}^{\rm i}}{\left({1+i\omega_{rs} \tau}\right)^{t/\tau}} ,
\label{eq:decay}
\end{equation}

\noindent
with

\begin{equation}
\omega_{rs} = \frac{1}{\hbar} \left({E_r - E_s}\right) =
\frac{1}{\hbar} \Delta E_{rs} .
\end{equation}

The expression (\ref{eq:decay}) can be written as

\begin{equation}
\rho_{rs} (t) = \rho_{rs} (0) e^{-\gamma_{rs}t} e^{-i\nu_{rs}t}
\end{equation}

\noindent
such that,

\begin{equation}
\gamma_{rs} = \frac{1}{2\tau} \ln{\left({1+\omega_{rs}^2 \tau^2}\right)} ,
\end{equation}

\begin{equation}
\nu_{rs} = \frac{1}{\tau} \tan^{-1}{\left({\omega_{rs} \tau}\right)} .
\end{equation}

\begin{figure}
\begin{center}
 \scalebox{0.8}{\includegraphics{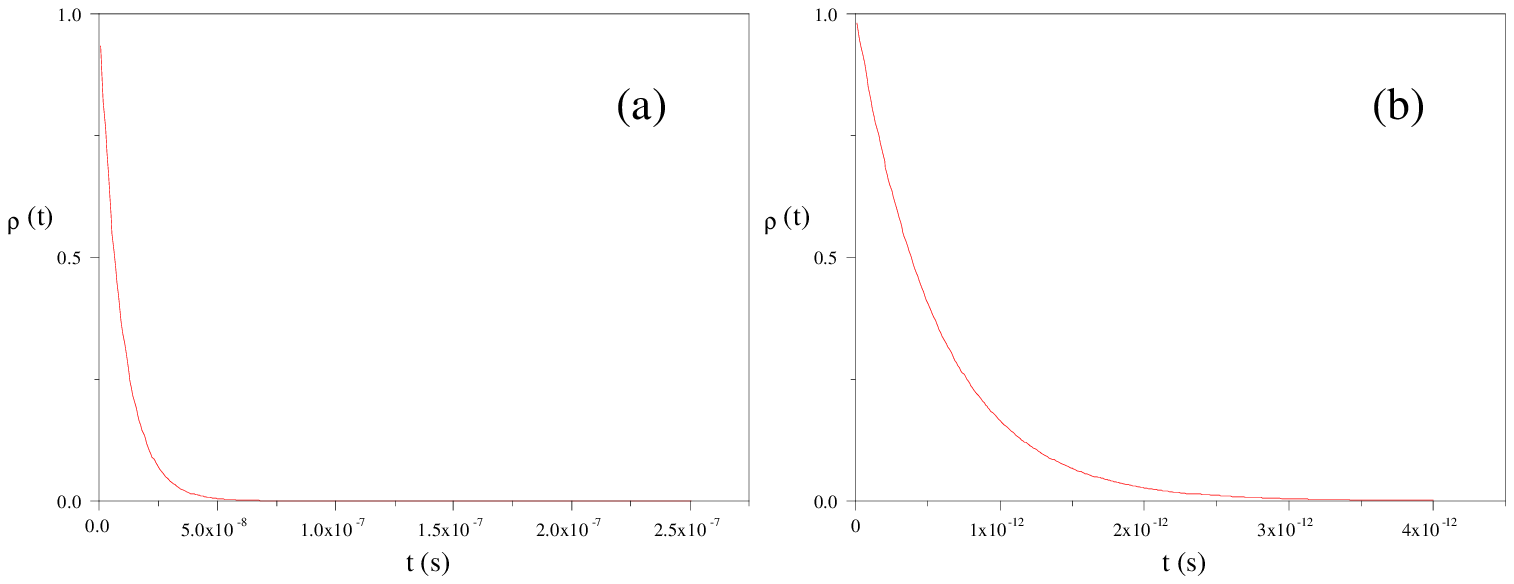}}
\caption{Vanishing in time of the non-diagonal terms of the density operator
for two different
values of $\tau$. For both cases we have used $\Delta E=4$ eV. \ One gets: \
(a) Slower
damping for $\tau=6.26\times10^{-24}$ s; \ (b) faster damping for
$\tau= \times 10^{-19}$ s.}
\end{center}
\label{fig:6-1}
\end{figure}

We can observe right away that the non-diagonal terms tend to zero with time
and the decay is faster the larger the value of $\tau$, which here is an
interval of time related to the whole system ${\cal O} + {\cal A}$. If
we keep in mind that in the {\em coarse grained} description the value of
the time interval $\tau$ originates from the impossibility to distinguish
between two different states of the system, we must remember that the system
${\cal O} + {\cal A}$ is not an absolutely quantum system. That means that
$\tau$  {\em could be significantly larger, implying an extremely faster damping
of the non-diagonal terms of the density operator} (figure \ref{fig:6-1}). We
then arrived at a process like the one of the reduction of the quantum state, even
if in a very elementary formalization. This result seems to be very encouraging
regarding  future researches on such important and controversial subject.

Some points must be pointed out from this brief approach of the measurement
problem. First of all, we must emphasize that  this result does {\bf not} occur
when we use the time evolution operators obtained directly from the retarded
Schr\"{o}dinger equation. The dissipative character of that equation causes the norm
of the state vector to decay with time, leading also to a non-unitary evolution operator.
However, this operator is such that, in the definition of the density operator we get
damping terms which are effective even for the diagonal terms. This point, as well
as the question of the compatibility between Schr\"{o}dinger's picture and
the formalism of the density matrix are going to be analysed in Appendix \ref{ap:01}.
As the composite system ${\cal O} + {\cal A}$ is a complex system, it is suitably
described by the {\em coarse grained} description, so that  the understanding of
the relationship between the two pictures is necessary in order to have a deeper
insight on the processes involved.

Notwithstanding the simplicity of the approach we could also observe the intrinsic
relation between {\em measurement process} and {\em irreversibility}. The time
evolution operator $\hat{V}$  meets the properties of a semigroup, so that it does
not necessarily possess an inverse: and non-invertible operators are related to
irreversible processes. In a measurement process, in which the object is lost just after
the detection, we have an irreversible process that could very well be described by
an operator such as $\hat{V}$.

Finally, it is worth mentioning that the measurement problem is controversial even
regarding its mathematical approach. In the simplified formalization introduced above,
we did not include any consideration beyond those common to the quantum formalism,
allowing an as clear as possible individualization of the effects of the introduction
of a fundamental interval of time in the approach to the problem.

\vspace{0.5in}

The introduction of a fundamental interval of time in the description of the measurement
problem makes possible a simple but effective formalization of the state
reduction
process. Such behaviour is only observed for the retarded case. When we take into account
a symmetric version of the Liouville--von~Neumann equation the solution is given by

$$
\rho_{nm}(t) = \rho_{nm}(0) \exp{\left\{ {- \frac{i t}{\tau}\sin^{-1}{ \left[ {\frac{\tau}
{\hbar} (E_n-E_m)} \right] } } \right\} },  $$

\noindent
where the diagonal elements do not change with time and the non-diagonal elements
have an oscillatory behaviour. This means that the symmetric equation is not  suitable
to describe a measurement process, and this is an important distinction between the two
descriptions.

It is important to stress that the retarded case of direct discretization of the
Liouville--von~Neumann equation results in the same equation obtained via the
{\em coarse grained} description. This lead us to the consideration of this equation
as the basic equation to describe complex systems, which is always the
case when a measurement process is involved.

\section{Conclusions}
\label{sec:6}

In this review we tried to get a better insight into the applicability of the various
distinct formalisms obtained when performing a discretization of the continuous
equations. For example, what kind of physical description is provided by the
retarded,
advanced and symmetric versions of the Schr\"{o}dinger equation? This can be achieved
by observing the typical behaviour of the solutions obtained for each case and,
particularly, attempting to the derivation of these equations from Feynman's approach.
Then we get that the advanced equation describes a system that absorbs energy
from the environment. We can imagine that, in order to evolve from one instant to another,
the system must absorb energy, and this could justify the fact that, by using Feynman's
approach with the usual direction of time, we can only obtain the advanced equation.
The propagator depends only on the Hamiltonian, being independent of the wave function
which describes the initial state.  So, it describes a transfer of energy
to the system.

The retarded equation is obtained by a time reversion, by an inversion of the
direction of the propagator, i.e., by inverting the flux of energy. The damping factor
characteristic of the retarded solutions refers to a system continuously releasing
energy into the environment.
Thus, both the retarded and the advanced equations describe open systems.

Finally, the
symmetric equation describes a system in an energy equilibrium with the environment. Thus,
the only way to obtain stationary states is by using the symmetric equation.

Regarding the
nature of such an energy, it can be related to the very evolution of the system. It can be
argued that a macroscopic time evolution is possible only if there is some energy flux
between system and environment. The states described by the symmetric equation are basically
equilibrium states, without nett dissipation or absorption of energy by the system as a
whole.  We can also conceive the symmetric equation as describing a closed system, which does not
exchange energy with the external world.

On the other hand, when a comparison is made with the classical approach, we can speculate
that the symmetric equation ceases to be valid when the interaction with the environment
changes fastly within a chronon of time. Thus, phenomena such as the collision of highly
energetic particles require the application of the advanced or retarded equations.
The decay of the norm associated with the vector states described by the retarded equation
would indicate the very decay of the system, i.e., of a system abandoning its initial
``equilibrium state". The behaviour of the advanced equation would indicate the
{\em transition} of the system to its final state. This speculation suggests another
interpretation, closer to the quantum spirit. We could consider the possible
behaviour of the system as being described
by all the three equations.  However, the ordinary QM works with averages
over ensembles, which is a description of an ideal, purely mathematical
reality. The question is that, if we accept the ergodic hypothesis, such
averages over ensembles are equivalent to {\em averages over time}. Anyway,
the quantum formalism always deals with average values when
tackling with the real world. When the potentials involved vary slowly
with respect to the value of the chronon of the system, which means a long
interaction time, we have that the contributions due to the transient factors
coming from the retarded and advanced equations compensate each other, and
cancel out. Then, in the average, the system behaves according to the
symmetric equation. \ On the contrary, when the potentials vary strongly
within intervals of time of the order of the chronon, we do {\bf not} have
stationary solutions. {\em The discrete formalism describes such a situation
by making recourse, during the interaction, to the transient solutions}, which
will yield the state of the system after the interaction.
Afterwards, the system will be described again by a symmetric solution.

The most conservative quantum interpretation would be that of believing that
only the
symmetric equation describes a quantum system. During the interaction process
the theory does not provide any description of the system, pointing only
to the possible states of the system after the transient period. The
description of the interaction would demand one more ingredient: the
knowledge of the interaction process (which would imply an additional
theoretical development, as, for example, the working out of an interaction
model).

Besides the question of the physical meaning of the discretized equations,
i.e., of the type of physical description underlying it, there is also the
question of the time evolution of the quantum states.  The Schr\"{o}dinger
equations describe the evolution of a wave function, with which
an amplitude of probability is associated.  An analogy with the electron
theory makes us suppose that this wave function does not react instantaneously
to the external action, but reacts after an interval of time
which is  characteristic of the described system. In discrete QM, the
justification of the non-instantaneous reaction comes from the fact that the
uncertainty principle prevents a reaction arbitrarily close to the
action application instant.\cite{WOL90,WOL94}  Such uncertainty could be
related to the very perturbation caused by the Hamiltonian on the state of
the system, resulting in an uncertainty relation like the Mandelstam and
Tamm\cite{FOCK} time--energy correlation. What we meet is a time evolution
in which the `macroscopic' state of the system {\em leaps discontinuously}
from one instant to the other. Therefore, the {\em quantum jumps} appear not
only in the measurement process, but are an intrinsic aspect of the time
evolution of the quantum system. The difference, in our case, is that the jump
does not take the system suddenly out of the quantum state it was endowed
with, but only determines the {\em evolution} of that state.

Another aspect characteristic of the discrete approach is the existence of an
{\em upper limit} for the eigenvalues of the Hamiltonian of a bounded system. \
In the description of a free particle it has been observed the existence of
an upper limit for the energy of the eigenfunctions composing the wave packet
which describes the particle, but this limit does not imply an upper value
for the energy of the particle. The existence of this limiting value determines
the Hamiltonian eigenvalue spectrum within which a normalization condition
can hold. Once exceeded that value, a transition to the internal
excited states of the system takes place. This allowed us, e.g., to obtain
the {\em muon} as an excited internal state of the electron.

It must be noticed the {\em non-linear} character of the relation
between energy and oscillation frequency of a state, and the fact that the
theory is intrinsically {\em non-local}, as can be confirmed by looking at the
discretized equations. It must also be stressed that the theory described in this
paper is non-relativistic.

Finally, it must be remembered that the symmetric form of the discrete
formalism reproduces grosso modo the results of the continuous theory. The
effects of the
introduction of a fundamental interval of time are evident in the evolution
of the quantum systems, but they are -- -in general---  extremely tiny.
There have been some attempts to find physical situations in which measurable
differences between the two formalisms can be observed, but till now with
little success.\cite{WOL87,WOL89,WOL90}   Maybe this could be afforded by
exploiting the consequences of the phase shifts caused by the discretization,
that we saw in Subsections (\ref{ss:5-2}) and (\ref{ss:5-3}). Regarding the
justifications for introducing a fundamental interval of time, let us for
instance
recall what Bohr\cite{BOHR} replied to the famous 1935 paper by Einstein,
Podolski and Rosen\cite{EPR}: \ ``The extent to which an unambiguous meaning
can be attributed to such an expression as {\em physical reality} cannot of
course be deduced from {\em a priori} philosophical conceptions, but $\ldots$ must be founded on a direct
appeal to experiments and measurements": Considering time as continuous may
be regarded as a criticizable philosophical position since, at the level of
experiments and measurements, nature seems to be discrete.

More important is to recall that, as already mentioned, the new formalism
allows not only the description of the stationary states, but also a space-time
description of transient states: The Retarded Formulation yields a
natural quantum theory for dissipative systems. \ It is not
without meaning that it leads to a simple solution of the measurement
problem in QM. \ Since the present review is still in a preliminary form, we shall
come back to such interesting problems also elsewhere.

\vfill
\newpage

\centerline{\bf{\large{APPENDICES}}}

\

\appendix
\section{Evolution Operators in the Schr\"{o}dinger\hfill\break
and Liouville--von~Neumann Discrete\hfill\break
Pictures}
\label{ap:01}

When we think of applying the formalism introduced in the previous Sections to the
measurement problem, the requirement of the existence of a well-defined evolution
operator comes out. By well-defined we mean, as in the continuous case, a unitary
operator satisfying the properties of a group.

In the continuous case, when the Hamiltonian is independent of time, the time evolution
operator has the form

$$ \hat{U}(t,t_0)=\exp{\left({-i(t-t_0) \hat{H}/\hbar }\right)} $$

\noindent
and is a unitary operator which satisfies the condition that $\hat{H}$ be
hermitian. In the
continuous case, by definition, every observable is represented by a hermitian operator.
An operator is unitary when its hermitian conjugate is equal to its inverse, such that

$$ \hat{A}^{\dagger} \hat{A} = \hat{A} \hat{A}^{\dagger} = 1 . $$

Another important aspect regarding a unitary operator is related to the probability
conservation. In other words, if the initial state is normalized to one,  it will keep
its norm for all subsequent times. The evolution operator does not change the
norm of the states it operates on.  Thus, we know beforehand that the evolution
operators associated with the retarded and advanced discretized Schr\"{o}dinger equations
are not unitary operators.

\subsection{Evolution operators in the Schr\"{o}dinger\hfill\break
picture}

For the discretized Schr\"{o}dinger equation the discrete analogue of the time evolution
operator can easily be obtained. Let us initially consider the symmetric equation that,
as already remarked, is the closest to the continuous description. After some algebraical
handling we get the evolution operator as being

$$
\hat{U}(t,t_0) = \exp{\left[{-\frac{i(t-t_0)}{\tau} \sin^{-1}{\left({
\frac{\tau \hat{H}}{\hbar}}\right)}}\right]} ,  $$

\noindent
so that

$$
\left|{\Psi (x,t)}\right\rangle = \hat{U}(t,t_0) \left|{\Psi (x,t_0)}\right\rangle =
\exp{\left[{-\frac{i(t-t_0)}{\tau} \sin^{-1}{\left({
\frac{\tau \hat{H}}{\hbar}}\right)}}\right]}  \left|{\Psi (x,t_0)}\right\rangle  . $$

Thus, if the eigenvalue equation of the Hamiltonian is given by

$$ \hat{H} \left|{\Psi (x,t_0)}\right\rangle = E  \left|{\Psi (x,t_0)}\right\rangle $$

\noindent
we have that

$$  \left|{\Psi (x,t)}\right\rangle =  \exp{\left[{-\frac{i(t-t_0)}{\tau} \sin^{-1}{\left({
\frac{\tau E}{\hbar}}\right)}}\right]}  \left|{\Psi (x,t_0)}\right\rangle  . $$

As   $\hat{H}$ is a hermitian operator, the evolution operator for the symmetric
equation is also hermitian. However, the existence of a limit for the possible
values of the eigenvalues of $\hat{H}$ implies that, beyond such threshold,
{\em the evolution operator is not hermitian anymore}. In fact, if we consider
that beyond the threshold the operator $\hat{H}$ has the form

$$ \hat{H}=\hat{\nu} + i \hat{\kappa}  , $$

\noindent
where $\hat{\nu}$ and $\hat{\kappa}$ are hermitian operators, we obtain
in the  continuous approach the same results obtained in the discrete case. One of
the characteristics of a non-hermitian operator is just the fact that it does not conserve
the norm of the state it acts on.

For the retarded equation, the evolution operator is given by

\begin{equation}
\hat{U}(t,t_0) = \left[{1+\frac{i}{\hbar} \tau \hat{H}}\right]^{-(t-t_0)/\tau}  ,
\label{eq:a-5}
\end{equation}

\noindent
such that, in the limit $\tau \rightarrow 0$,

$$\lim_{\tau \rightarrow 0}  \left[{1+\frac{i}{\hbar} \tau \hat{H}}\right]^{-(t-t_0)/\tau} =
\; e^{-\frac{i}{\hbar} (t-t_0) } \hat{H} , $$

\noindent
which is an expression known as the Trotter equality. Taking the conjugate hermitian operator
$\hat{U}^{\dagger}$ we can verify that this operator is not unitary.  In the basis of
eigenstates of $\hat{H}$ we can verify that

$$ \langle n | \hat{U}^{\dagger} \hat{U} | n \rangle = \left[{1+\frac{\tau^2 E_n^2}{\hbar^2}
}\right]^{-(t-t_0)/\tau} , $$

\noindent
is not equal to 1. Thus, that is the reason why the probabilities are not conserved for the
solutions of the retarded equation. Besides, as the evolution operator for the advanced
equation is given by

$$\hat{U}(t,t_0) = \left[{1- \frac{i}{\hbar} \tau \hat{H}}\right]^{(t-t_0)/\tau}  ,$$

\noindent
it can be verified that the formal equivalence between the two equations is obtained by the
inversion of the time direction and of the sign of the energy. In the relativistic case, this
is understandable if we remember that, if a transformation changes the sign of the time
component of a coordinate 4-vector, then it also changes the sign of the energy, which is
the corresponding element of the energy-momentum 4-vector. Then the retarded equation
describes a  particle endowed with positive energy travelling forward in time, and the
advanced equation describes an object with negative energy travelling backwards in
time, i.e., an
{\em antiparticle}.\cite{REC,RECW1,RECW2,PAVREC}

\subsection{Evolution Operator in the density matrix\hfill\break
picture}

For the sake of simplicity let $|\psi (t)\rangle$ be a pure state. The density
of states operator is defined as

$$ \hat{\rho}(t) = |\psi (t)\rangle \langle \psi(t) | . $$

It can be shown that such operator evolves according to the following dynamic
laws. For the retarded case,

$$\Delta_{\rm R} \hat{\rho}(t) = \frac{1}{i \hbar} \left[{\hat{H}(t),
\hat{\rho}(t)}\right]-\frac{\tau}{\hbar^2}\hat{H}(t) \hat{\rho}(t)
\hat{H}(t); $$

\noindent
for the advanced case,

$$\Delta_A \hat{\rho}(t) = \frac{1}{i \hbar} \left[{\hat{H}(t),
\hat{\rho}(t)}\right]+\frac{\tau}{\hbar^2}\hat{H}(t) \hat{\rho}(t)
\hat{H}(t); $$

\noindent
and, finally, for the symmetric case,

$$\Delta \hat{\rho}(t) = \frac{1}{i \hbar} \left[{\hat{H}(t),
\hat{\rho}(t)}\right] .$$

We can thus observe that the retarded and the advanced equations cannot
be obtained by a direct discretization of the continuous Liouville--von~Neumann
equation. Such formal equivalence occurs only for the symmetric case.
Taking into account the retarded case, we can obtain the equivalent time
evolution operator as being

\begin{equation}
\hat{V}(t,t_0) = \frac{1}{\left[{1+\frac{i \tau}{\hbar}\hat{\cal L}
+\frac{\tau^2}{\hbar^2}\hat{H} \ldots \hat{H}}\right]^{(t-t_0)\tau}} .
\label{eq:a-13}
\end{equation}

We must remark that this operator is different from the one obtained from
the {\em coarse grained} approach,

\begin{equation}
\hat{V}_{\rm CG}(t,t_0) = \frac{1}{\left[{1+\frac{i \tau}{\hbar}\hat{\cal L}
}\right]^{(t-t_0)\tau}} .
\label{eq:a-14}
\end{equation}

\noindent
and that it is {\bf not} unitary as well. Quantity $\hat{V}_{\rm CG}$ is so defined
to have the properties of a semigroup: without having necessarily
an inverse, but possessing the other group properties such as commutativity
and existence of an identity (besides the translational invariance of the
initial condition).

What we can conclude from such a difference between the two operators is that,
apparently, the descriptions clash with each other. In the {\em coarse
grained} approach the starting point was the continuous Liouville--von~Neumann
equation and, by introducing the graining of the time coordinate, an evolution
operator was obtained satisfying the retarded equation

$$\Delta_{\rm R} \hat{\rho}(t) = \frac{1}{i \hbar} \left[{\hat{H}(t),
\hat{\rho}(t)}\right] .$$

The second path consisted in starting from the definition of the density operator, in order
to determine the dynamical equation it satisfies, and then obtaining the
evolution operator.

For the symmetric case the evolution operator is given by

\begin{equation}
\hat{V}(t,t_0) = \exp{\left[{-\frac{i(t-t_0)}{\tau} \sin^{-1}{\left({
\frac{\tau \hat{\cal L}}{\hbar}}\right)}}\right]} ,
\end{equation}

\noindent
which is  similar to the operator obtained for the continuous case.

\subsection{Compatibility between the previous pictures}

We thus have two distinct evolution operators for the retarded Schr\"{o}dinger
and Liouville equations so that, once established a connection between them,
we arrive at the compatibility of the two descriptions. We try to set up a
relation between those operators by observing their action on the
density operator. So, we expect that both operators satisfy the expression

$$\hat{V}(t,t-\tau) \hat{\rho_0} = \hat{U}(t,t-\tau) \hat{\rho_0}
\hat{U}^{\dagger}(t,t-\tau) , $$

\noindent
where the different action of the operators is basically due to the bilinearity
of the operator $\hat{V}$ given by (\ref{eq:a-13}), while $\hat{U}$, given by
(\ref{eq:a-5}), is linear. This relation is valid in the continuous case,
where the evolution operators act on the density operator according to

$$
\hat{\rho}(t) = \exp{\left[{-i{\cal L}(t-t_0)/\hbar}\right]} \hat{\rho}_0
= \exp{\left[{-i{H}(t-t_0)/\hbar}\right]} \hat{\rho}_0
\exp{\left[{i{H}(t-t_0)/\hbar}\right]} .$$

\noindent
Considering the basis of hamiltonian eigenstates ${|n\rangle}$, we have

$$
\langle n| \hat{\cal L} \hat{\rho} (0) |m \rangle = (E_n-E_m)\rho_{nm}(0),
$$

\noindent
so that

\begin{equation}
\exp{(- i \hat{\cal L} t)} \hat{\rho}(0)=\exp{\left[{-i t (E_n - E_m)
}\right]} \rho_{nm}(0),
\label{eq:a-18}
\end{equation}

\begin{equation}
\exp{(- i \hat{H} t)} \hat{\rho}(0) \exp{(i \hat{H} t)}=\exp{\left[{-i t (E_n - E_m)
}\right]} \rho_{nm}(0).
\label{eq:a-19}
\end{equation}

\noindent
The question is to know if the same is valid for the discrete case. For the
retarded approach we must check whether the relation

$$\frac{1}{\left[{1+\frac{i \tau}{\hbar}\hat{\cal L}
+\frac{\tau^2}{\hbar^2}\hat{H} \ldots \hat{H}}\right]^{(t-t_0)/\tau}}\;\hat{\rho}_0 \, = \,
\frac{1}{\left[{1+\frac{i}{\hbar} \tau \hat{H}}\right]^{(t-t_0)/\tau} } \: \hat{\rho_0} \:
\frac{1}{\left[{1-\frac{i}{\hbar} \tau \hat{H}}\right]^{(t-t_0)/\tau} }
$$

\noindent
is valid.  What we get is that, if we consider equations such as
(\ref{eq:a-18}) and (\ref{eq:a-19}) to continue to be valid in the discrete
case, then the above relation is valid. For a generic element of the operator
we then get

$$\frac{1}{\left[{1+\frac{i \tau}{\hbar} (E_n-E_m)
+\frac{\tau^2}{\hbar^2} E_n E_m }\right]^{t/\tau}}\;\rho_{nm}(0) =
\frac{1}{\left[{1+\frac{i}{\hbar} \tau E_n}\right]^{t/\tau} }\: \rho_{nm}(0) \:
\frac{1}{\left[{1-\frac{i}{\hbar} \tau E_m}\right]^{t/\tau} }  .
$$

Such equivalence can be also observed for the other cases. However, when
we consider the evolution operator obtained from the {\em coarse grained}
approach we find an incompatibility with the operator deriving from the
Schr\"{o}dinger one. For the operator (\ref{eq:a-5}) we have

$$
\langle n \left| {\frac{1}{\left[{1+\frac{i}{\hbar} \tau \hat{\cal L}}\right]^{t/\tau} }
\hat{\rho}(0)} \right| m\rangle = \frac{1}{\left[{1+\frac{i}{\hbar} \tau (E_n - E_m)}\right]^{t/\tau} }
\rho_{nm}(0) .
$$

The question is to know what is the fundamental difference between the two
descriptions: if both are valid and under what conditions. Some points must
be emphasized.
First of all, we must remember that the {\em coarse grained} description is
a semi-classical
approach which assumes a system with a certain degree of complexity, while the vector
state description is a fundamentally quantum approach without any imposition, in principle,
on the number of degree of freedom of the system described. There is a basic difference even
in the way of conceiving the chronon. In the {\em coarse grained} approach it is
understood as a magnitude inwardly connected to the experimental limitations or, for
an ideal measurement device, to the limitations imposed by the uncertainty relations.
For the Schr\"{o}dinger equation, the value of the chronon is taken as a fundamental interval of
time associated with interaction processes among the components of the system,
and of the system as a whole with some external potential; i.e., it is
associated with the internal processes of the system (as it has been
conceived for the classical electron).
In this way, the absence of the mixed term in the evolution operator obtained
with the semi-classical procedure is comprehensible, as well as its
 incompatibility with the pure quantum description
provided by the Schr\"{o}dinger equations. As a semi-classical  approach, the range of
applicability of the {\em coarse grained} formalism extends to the cases where the system
to be studied is not purely microscopic, particularly in the measurement processes.
We have to stress the fact that, in this formalization, only the retarded equation was
obtained. Thus, the system described dissipates energy, i.e., it is an open system.
This is just the characteristic that allows us to have access to the output of a
measurement.

In connection with the operator obtained directly in the Schr\"{o}dinger picture for the
retarded case, we have that all the elements of the density matrix, even the diagonal ones,
are damped with time. Besides, there is also the controversy linked to the non-existence,
in QM, of an applicability limit of the theory due to the number of
degrees of freedom involved. The formalism does not distinguish between a microscopic and
a macroscopic system, so that it should reproduce what is obtained with the
{\em coarse grained} formalism. This means that the measurement problem also appears
in the discrete formalism through the non-equivalence of the
evolution operators (\ref{eq:a-13})
and (\ref{eq:a-14}).

\section{Non-Hermitian Operators in the Discrete\hfill\break  Formalism}
\label{ap:02}

 One of the features we have been stressing through this work is the {\bf non-hermitian}
character of the discrete formalism. In the Schr\"{o}dinger representation, for example,
the continuous equation can reproduce the outputs obtained with the discretized equations
once we replace the conventional Hamiltonian by a suitable non-hermitian Hamiltonian
we have called `equivalent Hamiltonian'. One of the characteristics of a non-hermitian
operator is that its eigenvalues are defined over the complex number field. A linear
non-hermitian operator can always be considered as being composed by a hermitian part,
which supplies the real component of the eigenvalues, and by an anti-hermitian part,
which gives the complex component.

In the continuous case, let us take the Hamiltonian as being a non-hermitian operator
given by

$$ \tilde{H} = \hat{\nu} + i \hat{\kappa} ,$$

\noindent
where $\hat{\nu}$ and $\hat{\kappa}$ are hermitian. Then we have, in the
Schr\"{o}dinger picture, that the time evolution operator is given by

\begin{equation}
 \hat{U}_{{\rm cont}} (t,t_0) = \exp{\left[{\frac{1}{\hbar} \left( {\hat{\kappa} - i \cdot
\hat{\nu} } \right) (t-t_0)}\right]} .
\label{eq:b-2}
\end{equation}

For the discrete case, we get from Appendix \ref{ap:01} that the
evolution operator for
the retarded states is given by (\ref{eq:a-5})

\begin{equation}
\hat{U}(t,t_0) = \left[{1+\frac{i}{\hbar} \tau \hat{H}}\right]^{-(t-t_0)/\tau}  ,
\end{equation}

\noindent
where $\hat{H}$ is the hermitian operator associated with the conventional hamiltonian. This
evolution operator can be written as

\begin{eqnarray}
\hat{U}_{{\rm ret}} (t,t_0) \ =\nonumber \\
 = \ \exp{\left[{-\frac{(t-t_0)}{2 \tau} \ln{\left({1+\frac{\tau^2
\hat{H}^2}{\hbar^2}}\right)}}\right]}  \exp{\left[{-\frac{i (t-t_0)}{\tau} \tan^{-1}{
\left({\frac{\tau \hat{H}}{\hbar}}\right)}}\right]} .
\label{eq:b-3}
\end{eqnarray}

Comparing (\ref{eq:b-2}) and (\ref{eq:b-3}) we obtain the equivalence of the hamiltonians
once $\hat{\nu}$ and $\hat{\kappa}$ are given by

$$\hat{\nu} = \frac{\hbar}{\tau} \tan^{-1}{\left({\frac{\tau \hat{H}}{\hbar}}\right)} ,$$

$$\hat{\kappa} = -\frac{\hbar}{2 \tau} \ln{\left({1+\frac{\tau^2
\hat{H}^2}{\hbar^2}}\right)} . $$

For the advanced case we obtain the same expressions except by a minus sign for $\hat{\kappa}$.
For the symmetric case, below the critical limit, we have

$$\hat{\nu} = \frac{\hbar}{\tau} \sin^{-1}{\left({\frac{\tau \hat{H}}{\hbar}}\right)} ,$$

$$\hat{\kappa} = 0 . $$

Above that limit $\hat{\nu}$ ceases to be hermitian and, in this case, the
evolution operator can be written as

$$
\hat{U}_{{\rm sym}}(t,t_0) = \exp{\left[{-\frac{i \pi}{2 \tau}(t-t_0)}\right]} \;
\exp{\left\{ {-\frac{(t-t_0)}{\tau} \ln{\left[ {\left| {\frac{\tau \hat{H}}{\hbar}}
\right| +\sqrt{\left( {\frac{\tau \hat{H}}{\hbar}} \right)^2 - 1 }} \ \right]}} \right\}}
$$

\noindent
so that

$$ \hat{\nu} = \frac{\hbar \pi}{2 \tau} ,$$

$$ \hat{\kappa} = -\frac{\hbar}{\tau} \ln{\left[ {\left| {\frac{\tau \hat{H}}{\hbar}}
\right| +\sqrt{\left( {\frac{\tau \hat{H}}{\hbar}} \right)^2 - 1 }} \ \right]} , $$

\noindent
with $\hat{\nu}$ being now independent of the Hamiltonian and $\hat{\kappa}$ ceases to be
zero.

The expressions obtained above show the characteristics that $\hat{\nu}$ and $\hat{\kappa}$
must fulfil in order that the continuous equation reproduces the outputs of the
discretized equations. By observing the continuous evolution operator we have that the
anti-hermitian part of $\tilde{H}$ shows a non-stationary behaviour, resulting in a
damping or amplifying term associated with the evolution of the quantum state. Thus, the
stationary solutions appear only for the symmetric case below the critical limit.
In all the other cases the transient term always appears.

In QM, the non-hermitian operators have been used only as mathematical
shortcuts, as in the case of the Lippmann--Schwinger equation in the scattering theory.
It has already been observed that the introduction of such operators could make possible
 the description of unstable states, by phenomenologically linking the transient factor
to the lifetime of the considered states.\cite{COHEN}  If in a certain instant $t_0=0$
the system is in one of the eigenstates $|n\rangle$ of the Hamiltonian $\hat{H}$
then, if such state is unstable, the probability of  the system to be found in the same
state in a later instant {\em t} is

$$P_n(t) = \left|{\langle n | \hat{U}^{\dagger} \hat{U} | n \rangle }\right|
= \exp{(-t/\tau_L)} , $$

\noindent
and that allows us to specify a lifetime $\tau_L$, for the retarded case, as being

\begin{equation}
\tau_L = \frac{\tau}{\ln{\left( {1+ \frac{\tau^2 E_n^2}{\hbar^2}} \right) }} ,
\label{eq:b-10}
\end{equation}

\noindent
and for the symmetric case, above the critical energy,

$$
\tau_L = \frac{\tau}{2 \ln{\left( { \left| {\frac{\tau E_n}{\hbar}} \right| +
\sqrt{\frac{\tau^2 E_n^2}{\hbar^2}-1 } } \right) } } .$$

Such lifetimes are connected with states that, in the discretized formalism, are intrinsically
unstable. Only the retarded equation seems to be associated with quantum states which decay
with time. If that is really valid, we have an expression which could be used for
phenomenologically determining the value of the chronon. Finally, we can conclude that
the time discretization brings forth a formalism which, even if only hermitian Hamiltonians
are involved, is equivalent to the introduction of non-hermitian operators in the continuous
QM.

\

{\bf Acknowledgements --} \ The
authors are grateful to Y.Akebo, V.Bagnato, A.P.L.Barbero, J.Bellandi,
R.Bonifacio, G.Degli Antoni,
C.Dobrigkeit-Chinellato, S.Esposito, F.Fontana, G.Giuffrida, D.Grilli,
P.Leal-Ferreira, D.Mugnai, A.Natale,
E.C.Oliveira, F.Pisano, I.Radatti, S.Randjbar-Daemi, A.Ranfagni, A.Salanti,
I.Torres Lima Jr., C.Ussami, M.Zamboni-Rached and
in particular to G.C.Ghirardi, H.E.Hern\'andez-Figueroa, L.C.Kretly,
J.W.Swart and
D.Wisniweski, for stimulating discussions or kind collaboration. \ At last,
one of us [RHAF] acknowledges a former PhD fellowship from FAPESP.

\end{document}